# A review of ptychographic techniques for ultrashort pulse measurement

Daniel J. Kane, Ph.D.*
Andrei B. Vakhtin, Ph.D.
Mesa Photonics, LLC, 1550 Pacheco St., Santa Fe, NM 87505
djkane@mesaphotonics.com

**Abstract:**
The measurement of optical ultrafast laser pulses is done indirectly because the required bandwidth to measure these pulses exceeds the bandwidth of current electronics. As a result, this measurement problem is often posed as a 1-D phase retrieval problem, which is fraught with ambiguities. The phase retrieval method known as ptychography solves this problem by making it possible to measure ultrafast pulses in either the time domain or the frequency domain. One well known algorithm is the principal components generalized projections algorithm (PCGPA) for extracting pulses from Frequency-Resolved Optical Gating (FROG) measurements. Here, we discuss the development of the PCPGA and introduce new developments including an operator formalism that allows for the convenient addition of external constraints and the development of more robust algorithms. A close cousin, the ptychographic iterative engine will also be covered and compared to the PCGPA. Additional developments using other algorithmic strategies will also be discussed along with new developments combining optics and high-speed electronics to achieve megahertz measurement rates.

## Table of Contents

# I. Introduction

Since the development of mode-locked lasers, optical pulses could be created that exceeded the capability of electronic measurement. The created ultrafast laser pulses have time durations that span from picoseconds to attoseconds. While such short pulses might seem to be more in line with only scientific applications, all applications—scientific, medical, commercial and industrial have exploded. It is extremely likely that every day you use and own something that has been built with the help of these lasers. From resistor trimming to machining antennas in smart phones, eye surgery, precision clocks, precise spectroscopy, applications continue to grow. Because of the extreme speed and bandwidth of these lasers, characterization of the intensity and phase of these pulses has been a difficult problem that has always been out of the reach of electronic measurement technology. However, the need for measurement and characterization has been important. For example, pulse measurement determined that the pulse limitation of mode-locked titanium sapphire oscillators was caused by dispersion and not coherent ringing [1]–[3]. Pulse measurement was also important for analysis, design and alignment of chirped-pulse amplifiers [4]–[6] as well as analysis of pulses shortened using nonlinear methods [7], [8].

Without detectors that match the speed of the pulse, direct intensity measurement is not sufficient [9], [10] and the optical methods were developed. Initial optical methods such as intensity autocorrelation only provides an approximate idea of the pulse duration, and no information about the phase, or chirp, of the pulse. Interferometric autocorrelation provides some, but not complete information about the chirp [11]. Initial work on time-frequency measurements of the ultrafast pulses started with Treacy [12]. Later, full characterization of ultrafast pulses through a group delay measurement was accomplished by Chilla and Martinez [13]. Frequency-resolved optical gating (FROG) [14], [15], a spectrographic technique [16], was the first technique to completely characterize ultrafast laser pulses without assumptions required by group velocity measurement and in a single-shot [14], [15], [17], [18]. FROG spectrally resolves temporal slices of the pulse which are formed by the pulse gating itself. The pulse is extracted from the resulting spectrogram via a 2-D phase retrieval algorithm [19]. Because FROG was initially hampered by a slow, iterative phase retrieval inversion, pulse measurement methods requiring no iterative inversion algorithms were sought. One full characterization method based on interferometric techniques, called Spectral Phase Interferometry for Direct Electric-field Reconstruction (SPIDER) was developed [20] that used a direct (non-iterative) reconstruction algorithm. The experimentally simpler Spectrally and Temporally Resolved Upconversion Technique (STRUT), providing an approximate, real-time measure of the pulse phase was also developed [21].

The period throughout the 1990's and into the millennium saw the development of many other pulse measurement techniques besides the spectrographic and interferometric methods introduced above, which was motivated by the need for real-time feedback, simplicity, and specific needs such as measurement at a focus. Examples of pulse measurement techniques developed include tomographic methods [22], [23], temporal imaging [24]–[29], time domain interferometry [30]–[32] as well as frequency domain interferometry utilizing electronics detection [33]. Additional techniques were also developed to use parallel or in line filtering methods including Dispersion Propagation Time Resolved Optical Gating (DP-TROG) [34], [35], Multiphoton Intrapulse Interference Phase Scan (MIIPS) [36]–[39], Spectral Phase and Amplitude Retrieval and Compensation (SPARC)[40] and d-scan[41]. (For a more complete and excellent review, see [16].)

The speed of the inversion of FROG spectrograms was greatly improved by the development of the principal components generalized projections algorithm (PCGPA) [42] and real-time FROG [43], [44] making FROG one of the most popular pulse measurement techniques. With the inversion speed no longer the issue, the primary drawback to FROG was the sampling requirement of maintaining a Fourier transform relationship between the time and the frequency axis. This made, for example, measuring high bandwidth, highly chirped pulses difficult. Later, it was found that a new algorithm called the ptychographic iterative engine (PIE) could be applied to pulse measurement applications, using the same type of data used by FROG [45]–[48]. The only difference is that the sampling along the time axis could be nearly arbitrary as long as temporally gated portions are overlapped in time (see for example [49]). (This is a weaker requirement than traditional FROG because in FROG, the gated samples must overlap, and the time sampling is determined by the reciprocal of the span of the frequency axis; e.g., Nyquist sampling.)

Indeed, ptychography is a phase retrieval process where an underlying unknown function (i.e., field) is multiplied by a second function (field) that is translated along an independent axis of the field such as position or delay, and the intensity of the product field propagated to another plane (or transformed to another domain) is measured [50]. Perhaps one of the earliest forms of ptychography is synthetic aperture radar where a traveling antenna monitors radio waves reflected off a target. The spatial location maps to time. Ordering the received signals builds an aperture much larger than the physical antenna length. Later, ptychography, in addition to other forms of data diversity, were used to improve convergence of 2-D phase retrieval problems [51]–[53] in optics, and algorithms to include applications in acoustics [54]. Historically, ptychography has been applied to imaging applications where the transverse translation is spatial [49], [51], [55]. Ptychography is probably best known for imaging applications where a diffraction pattern (electron and optical) is recorded as a function of position [56], and the ptychographic iterative engine (PIE) extracts the image from the diverse data [49].

It is well known that ultrafast laser pulse measurement is a 1-D phase retrieval problem. That is, to obtain the pulse from its spectrum, the phase of the spectrum must be recovered. Unconstrained, 1-D phase retrieval problems are unsolvable [57]–[59]. Unintuitively, 2-D phase retrieval problems are amenable to solutions [60]–[62]. One solution tactic for 1-D phase retrieval problems has been to use data diversity [51]–[53], or multiple related measurements to pose the 1-D ultrafast laser pulse measurement problem as a 2-D phase retrieval problem. Experimentally, this may be accomplished by creating a spectrogram, or time-frequency measurement of the pulse. The first pulse measurement technique to utilize this methodology was frequency-resolved optical gating (FROG)[14], [15], [19]. Later, the ptychographic iterative engine (PIE) gained popularity because the retrieval algorithm did not have FROG's Nyquist sampling requirement, requiring less data, and making the retrieval of highly chirped pulses easier[45]–[48].

In this paper, we review the class of ptychographic pulse measurement techniques applied to ultrafast laser pulse measurement: both FROG[14], [15], [19], which traditionally requires a Nyquist sampling of the time axis relative to the frequency axis, and what has been differentiated in the pulse measurement literature as ptychography[45]–[48], which relaxes the time sampling requirement when some *a priori* information of the gating function is known (i.e., the duration of the gate is long enough not to require Nyquist sampling). Both techniques are in the strictest sense ptychographic techniques and use data diversity to convert an unsolvable 1-D phase retrieval problem into a solvable 2-D phase retrieval problem[19]; gated spectra are recorded as a function of time delay between two pulses. Indeed, data diversity and ptychography [49], [55], [63] has been found to improve 2-D phase retrieval as well [51]–[53]. We classify both as spectrogram/sonogram methods even though the sampling methodology for ptychography is not as constrained as for a spectrogram measured by FROG. In the case of gating in time, we refer to the obtained data as a spectrogram; when gating in the frequency, or equivalently in the spectral, domain, we refer to the obtained data as a sonogram. The acquisition of a sonogram may also be called time-resolved optical gating (TROG)[13], [64], [65], where the signal is frequency gated and the temporal profile of the gated signal is measured (sonogram). Sonogram measurements have been less popular than spectrogram measurements because of the difficulty of measuring the time profile of the frequency filtered pulse. However, in this paper, we show that as the speed of electronics have increased to include optically relevant bandwidths (>~ 30 GHz), sonogram pulse measurement techniques may become more common and effective. In addition, there is a little known technique called Dispersion Propagated (DP) TROG which is a method to convert autocorrelations after varying amounts of dispersion is applied, to a sonogram[34], [35], [66], [67]. The data acquisition is close to the popular d-scan technique but fits the description of ptychographic pulse measurement once mathematical transformations are applied.

In the sections that follow, we first introduce the mathematical quantities we will measure, then introduce the spectrogram[68] before reviewing FROG and ptychography. We follow with a discussion on data acquisition, which defines the pulse measurement techniques. From the data acquisition discussion, we move into phase retrieval starting with the ptychographic iterative engine (PIE) [49] then continuing with the Principal Components Generalized Projections Algorithm (PCGPA)[42]–[44], [69]. We choose to primarily limit the algorithm discussion to these two algorithms because the two have similar structure yet differ in approach. The PIE is a steepest decent algorithm [45] while the PCGP algorithm is a generalized projection algorithm.

We also include new work regarding the development of the PCGPA into a fully operator-based algorithm, which became possible after the development of the cross-correlation FROG (X-FROG) PCGPA algorithm where one of the fields is known and fixed. The development of the new operator in this algorithm allowed for the completion of the constrained PCGPA (C-PCGPA). The resulting algorithm is also robust enough for most situations, and allows both external (e.g., spectral constraints) and internal constraints (e.g., pulse-gate relationships) to be added to the PCPGA [70]. From there, we continue with an error analysis that utilizes constructs from the PCGPA to provide insight into the performance of phase retrieval algorithms. We end the phase retrieval section with a comparison of the PCGPA and the PIE algorithms.

Next, we discuss sonogram methods that can be summarized by the name Time-Resolved Optical Gating (TROG). These techniques have not had much application in the past, largely because the temporal waveform needs to be measured with high-speed detection systems such as streak-cameras [71] or cross-correlation methods [64]. Two more recent developments stand out: dispersion propagation TROG (DP-TROG) [34], [35], [67] and detection using very high-speed electronics. In our discussion, we present for the first-time initial experiments demonstrating that TROG using high-speed electronics for detection of the temporal waveform can effectively increase the bandwidth of the measurement electronics by up to 40X while providing single-shot measurement rates up to one megahertz. We conclude with a discussion on some recent developments on uniqueness of ptychographic measurements along with a summary of work on both new algorithms and data acquisition schemes.

## II. Mathematical Background

The pulse electric field can be written as:

$$\mathrm{e}(t) = \frac{1}{2}\left(\sqrt{I(t)}e^{-i[\omega_0 t+\phi(t)]} + c.c.\right) \tag{1}$$

where *I(t)* is the temporal intensity of the pulse given by $|e(t)|^2$, $\phi(t)$ is the temporal phase of the pulse, and $\omega_0$ is the center frequency. c.c. is the complex conjugate, which is added to obtain the real part. However, to simplify the mathematical discussion, we will use the analytic representation (no negative frequencies):

$$e(t) = \sqrt{I(t)}e^{-i\phi(t)}\ . \tag{2}$$

Similarly, the frequency domain, or equivalently, the spectral domain representation is given by

$$E(\omega) = \sqrt{S(\omega)}e^{-i\psi(\omega)}, \tag{3}$$

where $S(\omega)$ is the spectral intensity (as a function of frequency) and $\psi(\omega)$ is the spectral phase. Thus, $E(\omega)$ is the Fourier transform of $e(t)$. In general, the pulse measurement techniques described here do not retrieve constant terms or the linear terms of $\phi(t)$ and $\psi(t)$. Indeed, the linear terms are just translations in the corresponding Fourier domains and the constant terms are only phase offsets that contain no information about the duration and shape of the recovered pulse.

The data obtained in ptychographic measurements is an unknown field being gated by a second function, either known or unknown, that is displaced in some variable across the unknown field, and the result is Fourier transformed. Mathematically, it is written as:

$$\boldsymbol{\Psi}(\mathrm{\tau},\mathrm{t}) = \mathrm{p}(\mathrm{t})\mathrm{g}(\mathrm{t}-\mathrm{\tau}), \tag{4}$$

where p is the unknown pulse, g is the gate, and $\tau$ is the delay variable. The spectrum of $\Psi(\tau, t)$ is taken and only the intensity can be measured (slow detector limit):

$$I(\tau, \omega) = \left| \int_{-\infty}^{\infty} \mathrm{p(t)g(t - \tau)} e^{-i\omega t} dt \right|^2 , \tag{5}$$

where $I(\omega, \tau)$ is the ptychographic data. If the data is obtained for enough time delays, then we can refer to the result as a spectrogram of the pulse. If this pulse is sampled at the Nyquist limit such that there is a Fourier transform relationship between the frequency and the time axes ($\Delta\omega = 1/(N\Delta\tau)$, where N is the number of points, $\Delta\tau$ is the spacing between points on the time axis and $\Delta\omega$ is the spacing between frequency points along the frequency axis), then it has been historically referred to as a FROG trace. In the following review, we will see that the lines delineating ptychography as applied to pulse measurement and FROG become blurred. We are not advocating a change in names of these techniques, however. We simply desire to promote an appreciation for the historical perspective of the nomenclature.

### *A. Properties of Spectrograms*

As stated above, a spectrogram is obtained by the Fourier transform of a time-windowed signal. In the equation below, we see that the pulse, denoted by $\mathrm{p(t)}$, is gated by the gate function $\mathrm{g(t)}$, and is scanned across time with delay $\tau$:

$$S_{\tau\omega}(\tau, \omega) = \left| \int_{-\infty}^{\infty} \mathrm{p(t)g(t - \tau)} e^{-i\omega t} dt \right|^2 , \tag{6}$$

$S_{\tau\omega}(\tau, \omega)$ is the spectrogram and is equivalent to $I(\tau, \omega)$ in Equation 5 [68], [72]. Interestingly, the spectrogram can also be obtained as a convolution between the pulse spectrum, $P(\omega)$ and the gate spectrum, $G(\omega)$ as:

$$S_{\tau\omega}(\tau,\omega) = \left| \int_{-\infty}^{\infty} \mathrm{P}(\Omega)\mathrm{G}(\omega - \Omega) e^{i\Omega\tau} d\Omega \right|^2, \tag{7}$$

where $P(\omega)$ and $G(\omega)$ are convolved rather than correlated. We should note that a convolution is not as experimentally convenient as a correlation; in the case of the convolution, the direction of the gate must be reversed, in time or frequency, depending on the domain. In fact, this means that if we apply the PCGPA on the rows of a FROG trace (as opposed to the columns), we will obtain frequency domain representations of the pulse and gate where the spectrum of the gate is frequency reversed.

While the many properties of spectrograms are defined elsewhere, we state the most pertinent to pulse measurement here. The time marginal of a spectrogram is defined as:

$$\int_{-\infty}^{\infty} S_{\tau\omega}(\tau,\omega) d\omega = \int_{-\infty}^{\infty} |p(t)|^2 |g(t-\tau)|^2 dt, \tag{8}$$

which is the intensity cross-correlation of the pulse and the gate [68], [72]. In second harmonic generation FROG, it would be the autocorrelation of the input pulse. The frequency marginal of $S_{\tau\omega}(\tau,\omega)$ is:

$$\int_{-\infty}^{\infty} S_{\tau\omega}(\tau,\omega) d\tau = \int_{-\infty}^{\infty} |P(\Omega)|^2 |G(\omega-\Omega)|^2 d\Omega, \tag{9}$$

Which is the convolution of the spectrum of the pulse, $p(t)$, denoted $P(\omega)$ and the spectrum of the gate $g(t)$, denoted $G(\omega)$. In the case of second harmonic generation FROG, it would be the autoconvolution of the pulse spectrum [15].

Therefore, for this review, we will define sonograms by how the data are obtained rather than how the data are represented. That is, sonograms are correlations in the frequency domain, and correspondingly as convolutions in the time domain:

$$S_{\tau\omega}^{sonogram}(\tau,\omega) = \left|\int_{-\infty}^{\infty} \mathrm{P}(\Omega)\mathrm{G}(\Omega-\omega)e^{i\Omega\tau}d\Omega\right|^2, \tag{10}$$

$$S_{\tau\omega}^{sonogram}(\tau,\omega) = \left|\int_{-\infty}^{\infty} \mathrm{p(t)g(\tau-t)}e^{-i\omega\mathrm{t}}d\mathrm{t}\right|^2. \tag{11}$$

Thus, the sonogram, as defined here, correlates two frequency domain functions. That is, a pulse spectrum $P(\omega)$ is gated in the frequency domain by $G(\omega)$ and the gated result is inverse Fourier transformed to the time domain. In the case of the sonogram, integrating $S_{\tau\omega}^{sonogram}(\tau,\omega)$ over time results in the correlation of the pulse and gate spectra while integrating over frequency obtains the convolution of the temporal functions of the pulse and the gate.

## III. Data Acquisition for Ptychographic Ultrafast Pulse Measurement

The goal of the data acquisition is to obtain sufficient data to extract the intensity and phase of the pulse. Data can be obtained in multiple straightforward ways. One commonly used method is spectrally resolving a second harmonic generation autocorrelation. In the case of traditional FROG, the data set required has a time spacing of the reciprocal of the frequency span. This can be accomplished either by sampling the data or interpolation. The Ptychographic Iterative Engine (PIE) [49] allows more freedom in the sampling at the expense of convergence. For highly chirped pulses, the reduced sampling requirements can be invaluable, and convergence is still excellent provided the gate is known.

Shown in Figure 1 is a schematic diagram of a cross-correlator. Two different pulses are sent into the cross-correlator. One pulse is delayed relative to the other, and the pulse and the delayed reference pulse are combined in a second harmonic generation (SHG) crystal to produce a signal proportional to $I_{unk}(t)I_{ref}(t-\tau)$, where $I_{ref}(t)$ is the intensity of the reference pulse as a function of time and $I_{unk}(t)$ is the intensity of the unknown pulse as a function of time. The slow detector integrates the signal and the recorded function as a function of delay is given by:

$$A^{(2)}(\tau) = \int_{-\infty}^{\infty} I_{unk}(t) I_{ref}(t-\tau) dt \qquad (12)$$

where $A^{(2)}(\tau)$ is the cross-correlation between $I_{ref}(t)$ and $I_{unk}(t)$ and only the signal of interest is shown. The SHG crystal provides the needed multiplication process. It is important to note that while this is identical to the gating process used in ptychography, it is not a ptychographic data acquisition because the resulting intensity is not converted to another domain. Fortunately, converting $A^{(2)}(\tau)$ to another domain is straightforward. All that needs to be done is to spectrally resolve $A^{(2)}(\tau)$ in a spectrometer (See Figure 2). In this case, we obtain the familiar spectrogram:

$$S_{\tau\omega}(\tau,\omega) = \left|\int_{-\infty}^{\infty} \mathrm{E}_{unk}(\mathrm{t}) \mathrm{E}_{ref}(\mathrm{t}-\tau) e^{-i\omega t} dt\right|^2, \qquad (13)$$

where $I_{unk}(t) = |E_{unk}(t)|^2$ and $I_{ref}(t) = \left|E_{ref}(t)\right|^2$ .

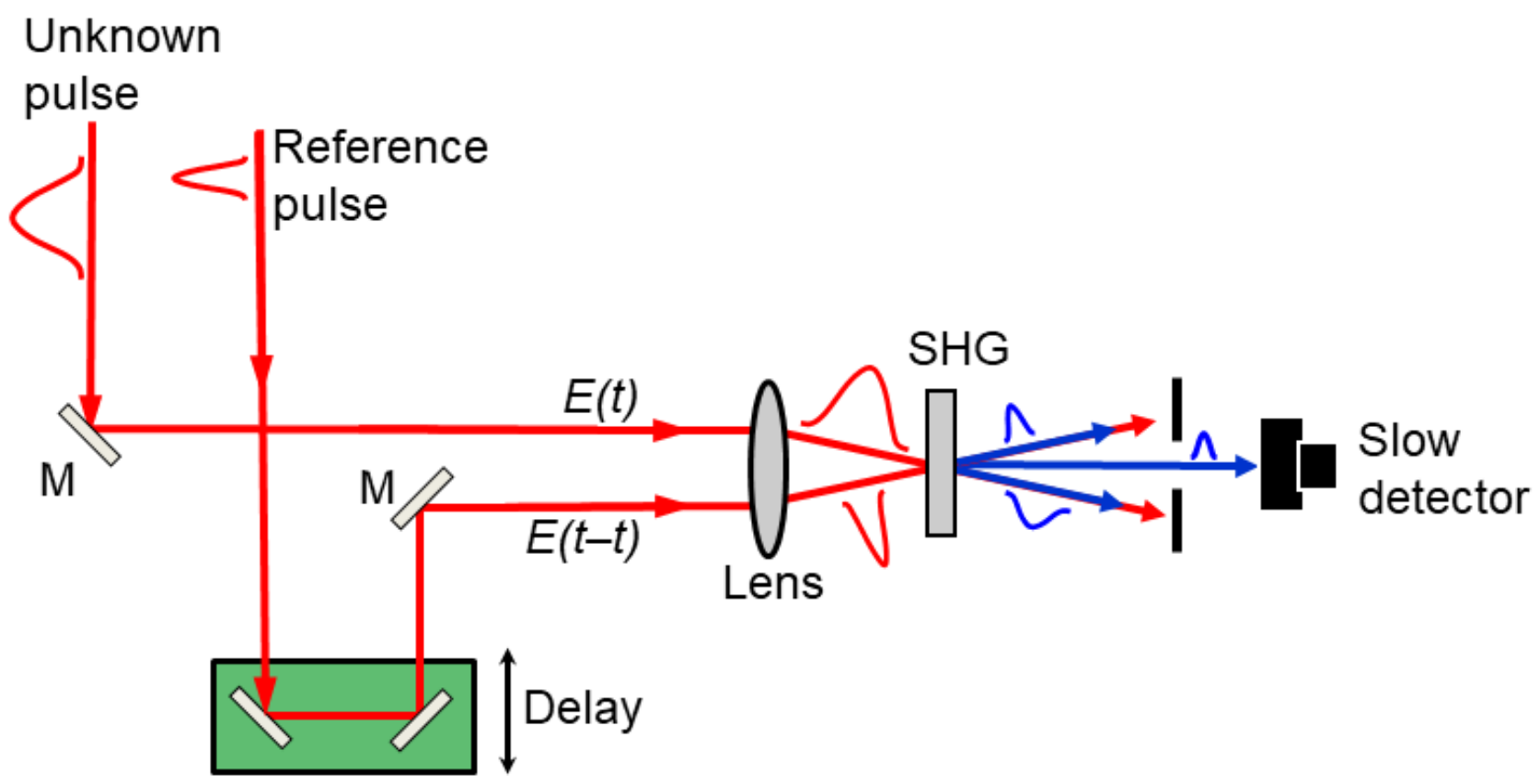

Figure 1. Schematic of an optical cross-correlator. A second harmonic generation crystal acts as a multiplier for gating.

Shown in Figure 2 is the experimental apparatus used for both ptychography and X-FROG. An unknown pulse is cross-correlated with a known reference pulse in a nonlinear medium (an SHG crystal in this case). For ptychographic data acquisition, the temporal step size only has to be about the width of the gate. In the case of X-FROG, the temporal step is determined by the Fourier relationship between the time and frequency axis the of the spectrogram; that is the time spacing, $\Delta t = 1/(N\Delta f)$, where N is the number of points and $\Delta f$ is the frequency axis spacing.

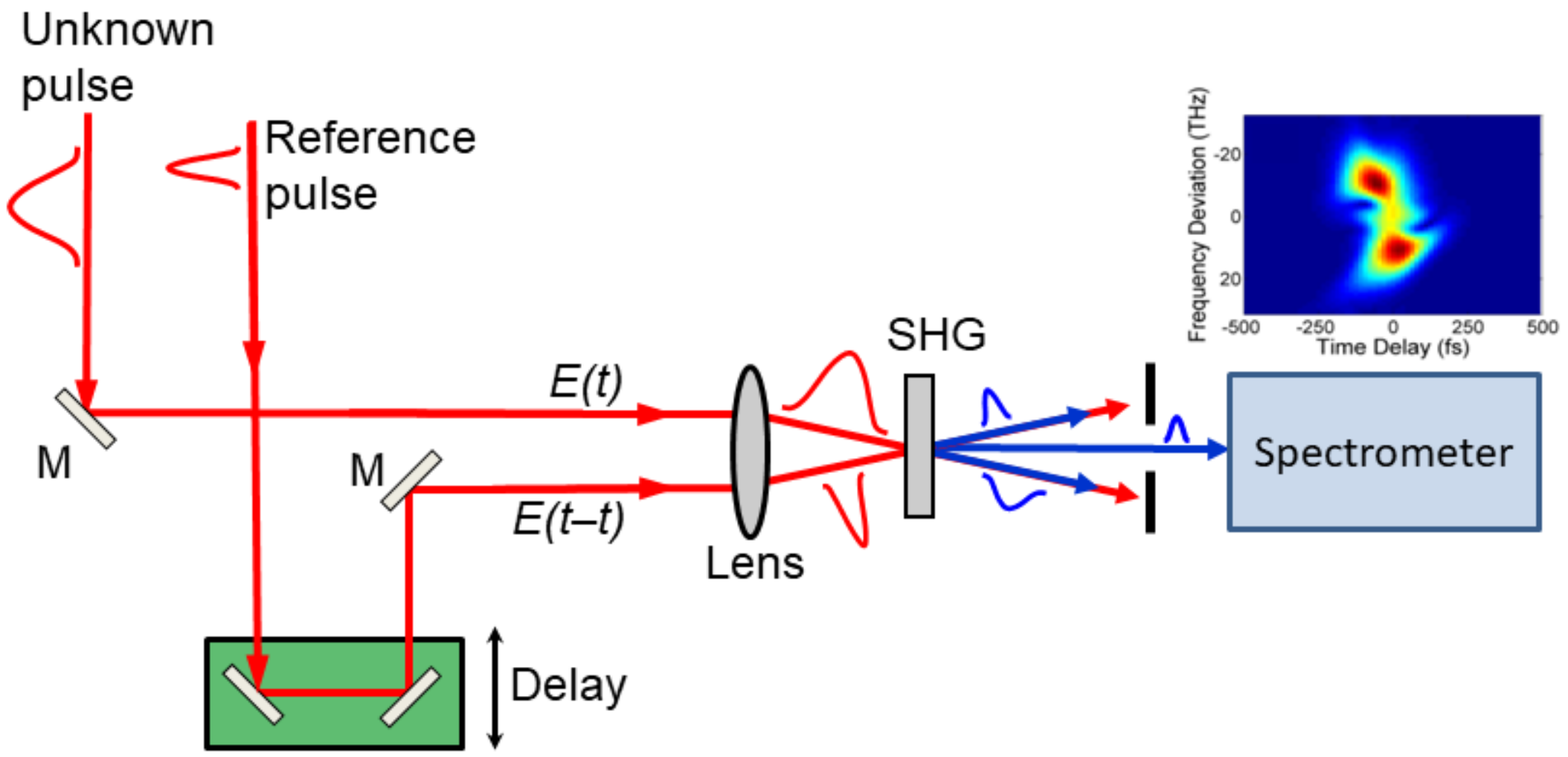

Figure 2. Schematic of a ptychographic pulse measurement system for ultrafast optical pulse measurement. In this case, a known pulse interrogates an unknown pulse. The inset shows an example of a spectrogram.

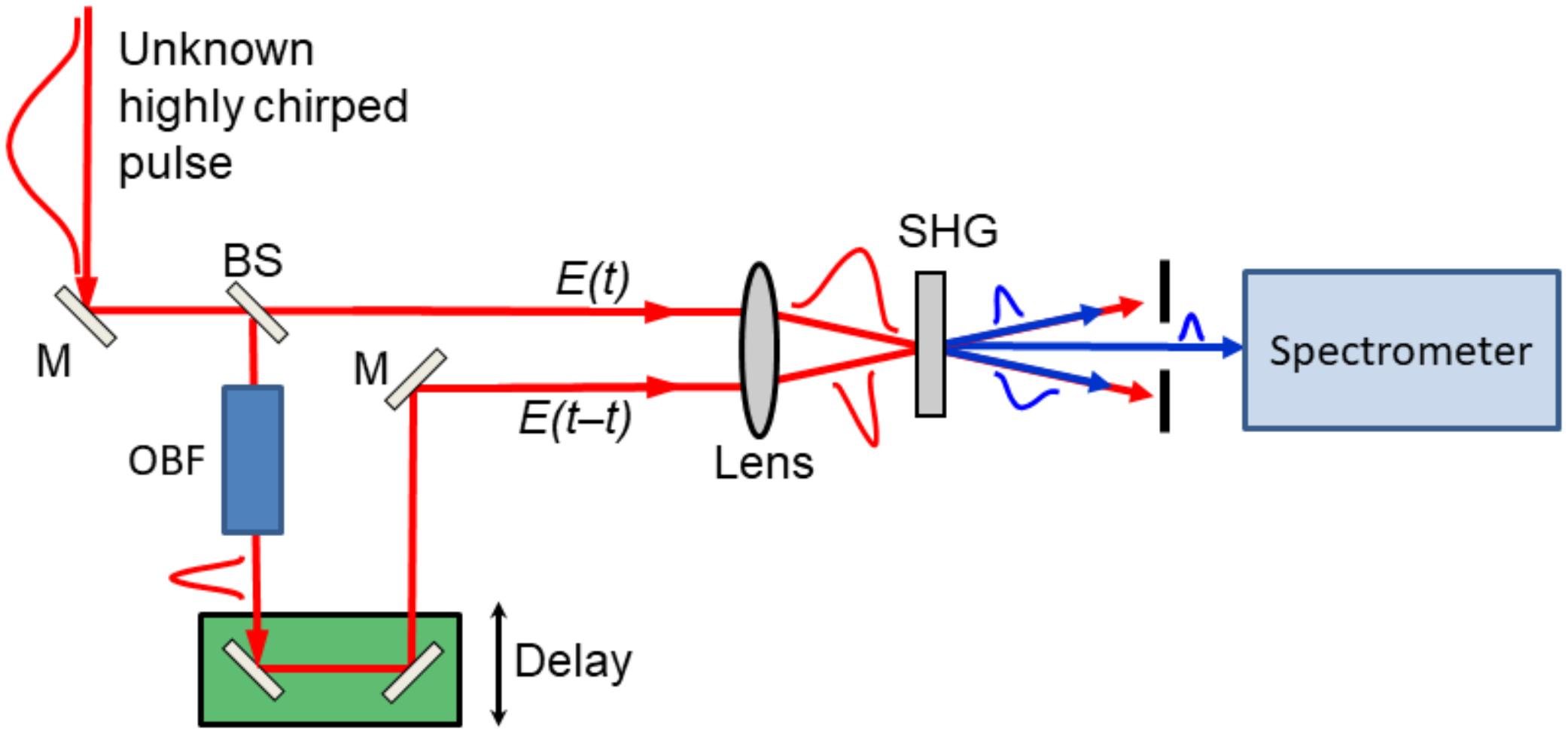

Figure 3. Schematic of a ptychography system for measuring highly chirped optical pulses. The Optical Bandpass Filter (OBF) slices out a spectral portion of the pulse, which may be shorter than the highly chirped pulse.

Another type of ptychography, which is reminiscent of a technique called Spectrally and Temporally Resolved Upconversion Technique (STRUT) [21], is shown in Figure 3. This method has been found to be very useful for measuring highly chirped pulses. Like standard pulse measurement techniques, the highly chirped input pulse is split into two identical pulses. The difference is that one replica is bandpass filtered and the filtered pulse is cross-correlated with the chirped pulse [45]. The characteristics of the optical bandpass filter (OPF) are known and can be used to update the computed gate as the algorithm iterates. Often, the input pulse is so highly chirped that the filtered pulse is actually shorter than the unfiltered pulse. In this regime, traditional FROG measurements are difficult because the Nyquist sampling requirement makes the grid size very large, greatly increasing the computational requirements for the inversion algorithm.

### *A. Making the gate a function of the pulse: FREQUENCY RESOLVED OPTICAL GATING*

For most pulse measurement applications, a known pulse is unavailable, and the unknown pulse is often not very highly chirped. In this case, the pulse can be used to gate itself, which is the premise behind Frequency Resolved Optical Gating (FROG). A schematic of a second harmonic generation FROG device is shown in Figure 4. As with an autocorrelation, the pulse is split into two replicas of itself. These replicas are time delayed relative to each other and focused into a nonlinear medium. The nonlinear medium can be thought of as a mixer, or switch, allowing one pulse to gate a portion of the other pulse. The gated portion is spectrally resolved (i.e., the power spectrum is recorded) as a function of delay between the pulse to be measured and the gate.

The resulting spectrogram, or FROG trace, provides immediate, qualitative information about the pulse.

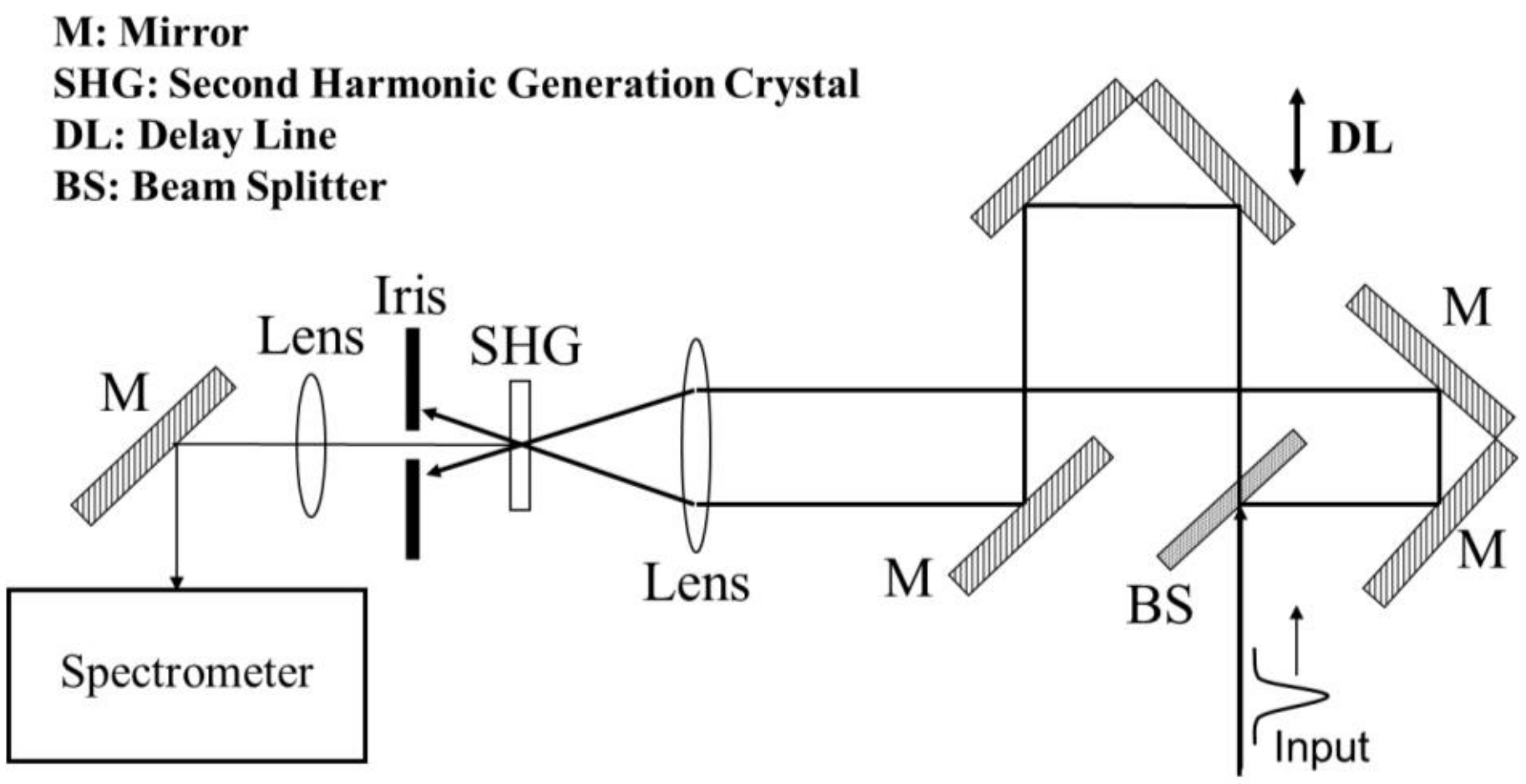

Figure 4. Schematic of an SHG FROG device. A beam splitter splits the input E(t) into probe and gate beams. The two beams are focused into an SHG crystal. The spectrum of the second harmonic is collected as a function of delay.

FROG measurements are not just limited to optical nonlinearities such as $\chi^{(2)}$ (second harmonic generation [73], [74]) and $\chi^{(3)}$ (optical Kerr effect[14], [75]) based nonlinearites. Other techniques such as phase modulation either by another optical pulse or the electro-optical effect, or intensity modulation via the electro-absorptive effect can generate pulse spectrograms that can be inverted using phase retrieval algorithms [76], [77]. Conveniently, autocorrelation-based techniques use a gate that is a function of the pulse itself while correlation-based techniques use a gate that is either known or unknown. Quantitative pulse characteristics require extraction from the spectrogram via a 2-D phase retrieval [19], [42], [43], [78].

## IV. Extracting the Pulse from Ptychographic Data

In this section, we will describe the mathematical background for ptychographic pulse measurement starting with the mathematical formulation of the data acquisition and moving on to develop the formalisms for extracting the pulse from the ptychographic data.

A pulse input into a FROG device (see Figure 4) can be represented by the equation:

$$E(t) = Re\left[\sqrt{I(t)}\,exp(-\,i\omega_0 t - i\phi(t))\right], \tag{14}$$

where $I(t)$ and $\varphi(t)$ are the time dependent intensity and phase, respectively, and $\omega_0$ is the carrier frequency. Upon entering the FROG device, the pulse is split into two identical pulses, via a beam splitter. The identical pulses are combined in a nonlinear material producing a signal with the mathematical form:

$$E_{sig}(t,\tau) \propto E(t)\Gamma[E(t-\tau)] \tag{15}$$

where E(t) is referred to as the *probe*, and Γ is the gate function that converts the pulse into the gate, which depends on the nonlinear interaction used. For the FROG device depicted in Figure 4, Γ[E(t-τ)] = E(t-τ), because an SHG crystal multiplies E(t) by its time delayed replica generating the second harmonic that is collected. Another nonlinear interaction commonly used in FROG devices is the optical Kerr effect in the polarization-gate (PG) geometry [14], [17]. In that case, $\Gamma[E(t-\tau)] = |E(t-\tau)|^2$.

A spectrometer spectrally resolves the signal; that is modeled in the algorithm via a Fourier transformation into the frequency domain. A detector, such as a CCD array, obtains the FROG trace by recording spectral intensity of the signal at each time delay. These data can be represented as the magnitude squared of the Fourier transform of $E_{sig}(t-\tau)$:

$$I_{FROG}(\omega,\tau) \propto \left|\int_{-\infty}^{\infty} E(t)\Gamma[E(t-\tau)]\,exp(-\,i\omega t)dt\right|^2 \tag{16}$$

$I_{FROG}(\omega, \tau)$ is a real quantity; therefore, it has no direct phase information. The goal of the FROG inversion algorithm is to determine the phase by solving the equation,

$$\sqrt{I_{FROG}(\omega,\tau)}\varphi(\omega,\tau) \propto \int_{-\infty}^{\infty} E(t)\Gamma[E(t-\tau)]\,exp(-\,i\omega t)dt \tag{17}$$

for $\phi(\omega,\tau)$ which is a complex function of unity magnitude.

Thus, Equations 15 and 17 define the two constraints common to all ptychographic algorithms that must be satisfied [15], [19]. Equation 15 is the physical constraint and is often called the mathematical form constraint. It is used in ptychographic (and FROG) algorithms both to obtain the next guess for E(t) and to construct the new signal field and is applied in the time domain. Equation 17 defines the intensity constraint for the spectrogram, FROG trace, or ptychographic data, which is applied in the frequency domain. The goal of the algorithm is to minimize the difference between the measured data and the results calculated from the current pulse, E(t) (see Equation 18) [19].

Regardless of the numerical method used, the process of extracting the pulse from ptychographic data is phase retrieval. The phase retrieval approach to extracting signals from their spectrograms seems to have originated in the acoustics field. One notable paper by Griffin and Lim derived a generalized projections-based phase retrieval algorithm to extract a signal from its spectrogram when the gate is known [54]. Later in this review, we derive a similar algorithm using a linear algebra approach. They also first reported what Trebino *et al.* [19] reported as the "Vanilla" algorithm. In general, the Vanilla algorithm, while fast, had convergence problems especially for compound pulses. Later, nonlinear solvers were utilized for robust retrieval, but they could be incredibly slow, and sensitive to the initial guess [78].

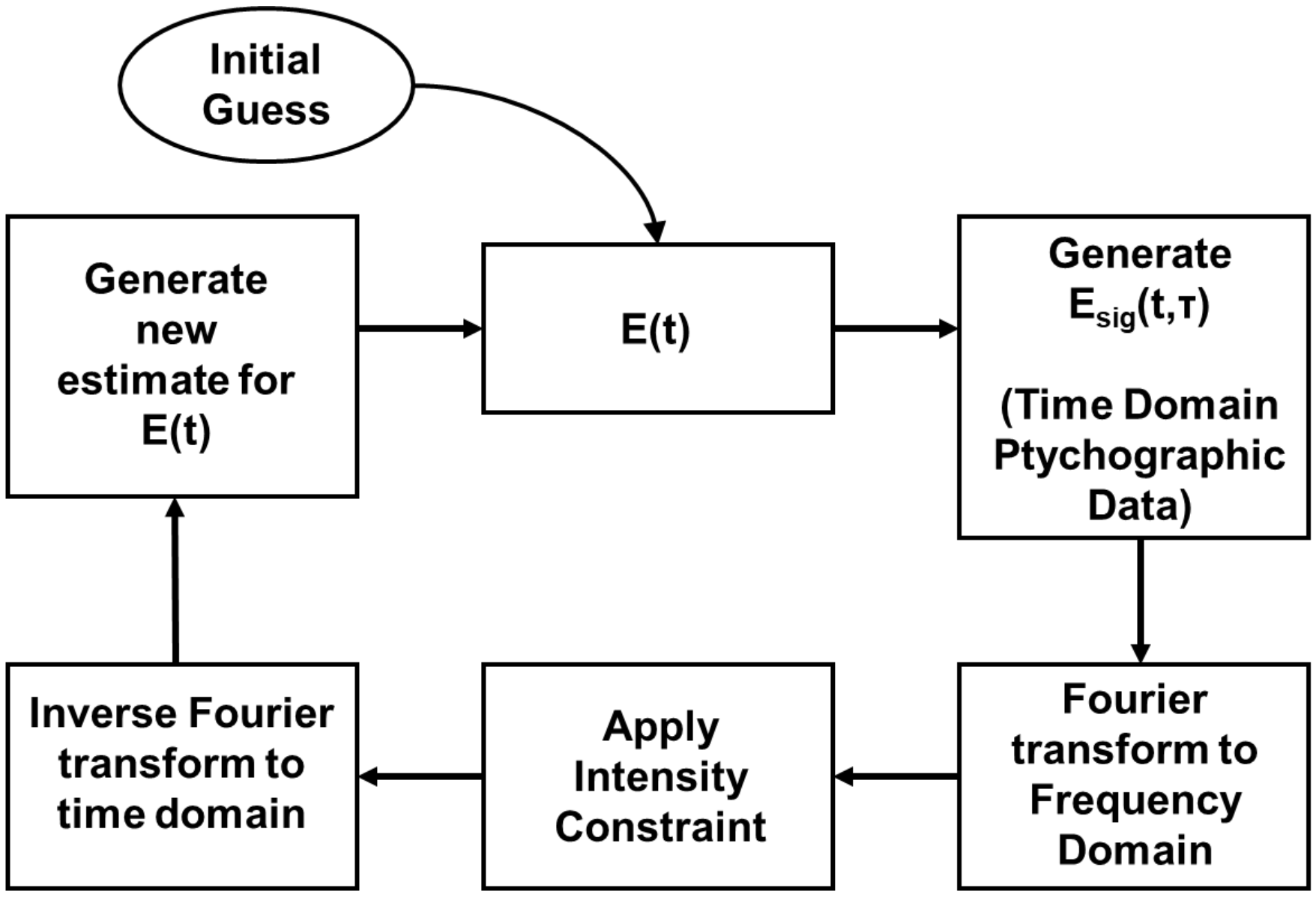

Figure 5. A flow chart of the general steps a ptychographic retrieval algorithm takes. Most all the differences are in how the new estimate for the retrieved field is computed. An initial guess starts the iterative algorithm, and the time domain data is generated. The intensity constraint is applied in the inverse domain.

Figure 5 shows the general form of most ptychographic phase retrieval algorithms. An initial guess is provided for E(t) to get the algorithm started [19]. $E_{calc}(t, \tau)$, a guess for $E_{sig}(t, \tau)$, is calculated and Fourier transformed into $\sqrt{I_{calc}}(\omega, \tau)\phi_{calc}(\omega, \tau)$. $\sqrt{I_{calc}}(\omega, \tau)$ is replaced by the square root of the measured FROG trace, $\sqrt{I_{FROG}}(\omega, \tau)$. The next guess for E(t) is then calculated from $\sqrt{I_{FROG}}(\omega, \tau)\phi_{calc}(\omega, \tau)$. Consequently, the phase retrieval problem looks much like a recursive equation with $N^2$ variables (where N is the number of time points and frequency points).

Once the new estimate for the E(t) is obtained, a new spectrogram is constructed. The process is repeated (see Figure 5) until the spectrogram error, $\varepsilon_{TF}$, (equivalent to the FROG trace error) reaches an acceptable minimum:

$$\varepsilon_{FROG} = \left[\frac{1}{N^2}\sum_{i=1}^{N}\sum_{j=1}^{N}\left[I_{CALC}(\omega_i,\tau_j) - I_{FROG}(\omega_i,\tau_j)\right]^2\right]^{1/2} \tag{18}$$

where $\varepsilon_{FROG}$ represents the rms error per element of the spectrogram, $I_{CALC}(\omega_i, \tau_j)$ is the current iteration of the spectrogram, $I_{FROG}(\omega_i, \tau_j)$ is the measured spectrogram, and $\omega_i$ and $\tau_j$ are the $i^{th}$ frequency and $j^{th}$ delay in the frequency and delay vectors, respectively.

The various FROG algorithms differ in how E(t) is calculated from $\sqrt{I_{FROG}}(\omega, \tau)\phi_{calc}(\omega, \tau)$ [19], [42], [43], [78], [79]. Ideally, after each iteration of the algorithm, a slightly better estimate for the phase of the spectrogram is obtained until convergence (Figure 5). Originally, integrating the *time domain* FROG trace, $E_{sig}(t, \tau)$, with respect to $\tau$, the time delay, was used to obtain subsequent guesses for E(t) (so called "vanilla" or "basic" algorithm) [19], [54]. While fast, this algorithm stagnates easily and fails to invert spectrograms of double pulses [19], [78]. To overcome stagnation problems and increase robustness, the vanilla algorithm was used to provide an initial guess to a brute force minimization of the root-mean square difference between the retrieved FROG trace and the experimental FROG trace [78]. While this method is robust, converging in most cases, its iteration speed, and hence the convergence time, is slow.

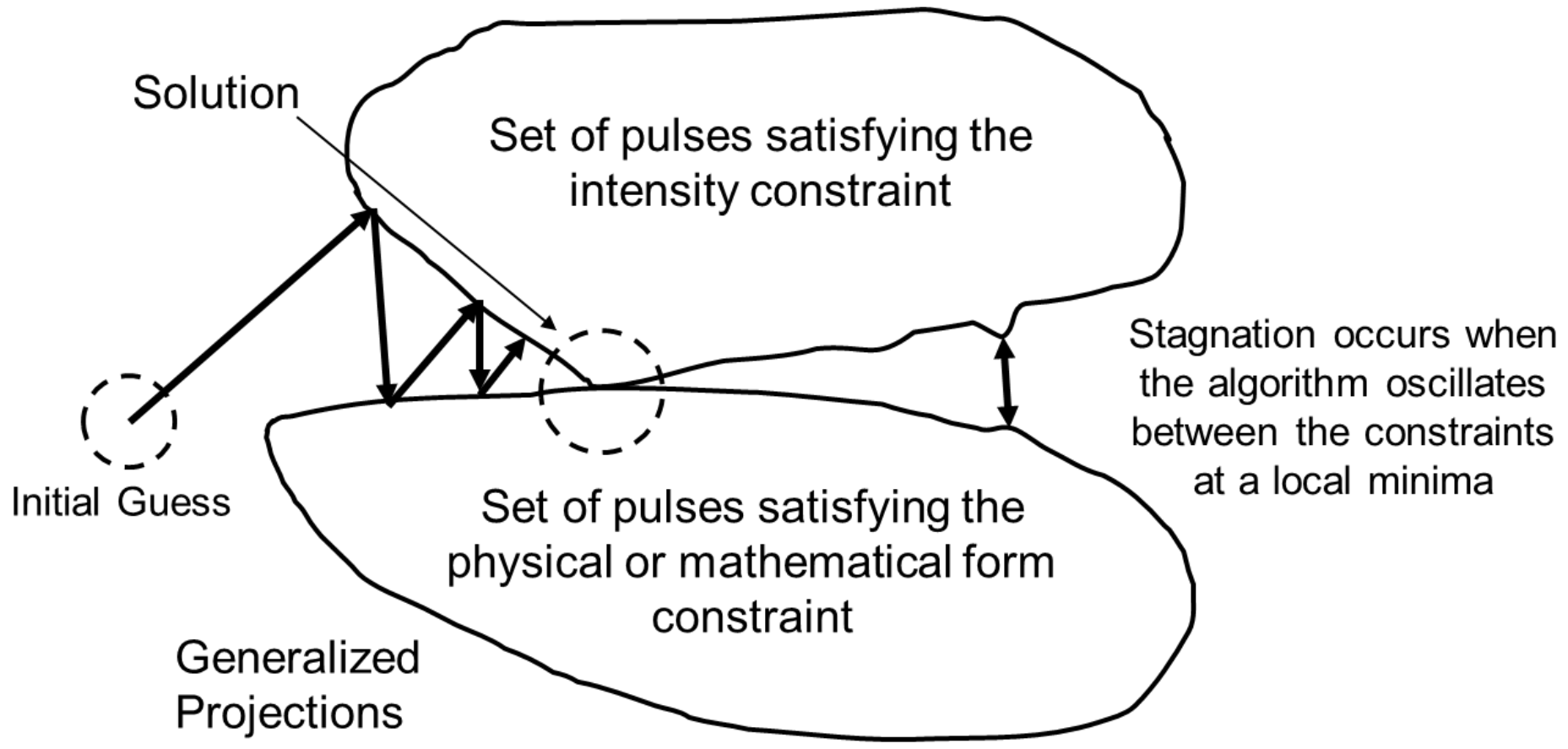

Figure 6. Schematic of how a generalized projections algorithm works. The idea is that by minimizing the distance between the set of pulses satisfying the intensity constraint and the set of pulses satisfying the physical constraint, the algorithm will converge to the solution—where the two sets join. Local minima can confound the process, however.

The development of a FROG inversion algorithm that applied the generalized projections concepts to the inversion algorithm first shown by Griffin and Lim [54] on spectrograms was published by DeLong *et. al.* [79] greatly improving the inversion speed. First, it virtually guarantees that the error always decreases for each iteration [79], [80]. Second, the generalized projections algorithm is very robust [42], [43], [79]. Last, it converges well even in the presence of noise. A complete discussion, along with derivatives used in the minimization, appears in a review article by Trebino, *et al* [15].

Like the previous algorithms, generalized projections works by alternating between two (or more) sets, $S_1$ and $S_2$ [79]. (See Figure 6; for FROG trace inversion, $S_1$ is the set of all $E_{sig}(t, \tau)$'s satisfying the nonlinear material response[81], including absorption [82], and $S_2$ is the set of all complex functions with a magnitude $I_{FROG}(\omega, \tau)$.) Unlike the previous algorithms, generalized projections finds the next guess by ensuring that the distance between $S_1$ and $S_2$ is minimized for each iteration of the algorithm. The generalized projections algorithm for FROG used a minimization algorithm to find the E(t) that minimizes the Euclidian distance between the signal field constructed from E(t) and the signal field that satisfies the intensity constraint:

$$Z = \sum_{t,\tau=1}^{N} \left| E'_{sig}(t,\tau) - E(t)\Gamma[E(t-\tau)] \right|^2 \tag{19}$$

$Z$ is minimized with respect to E(t) to obtain the next estimate of E(t), and Γ[] is the function that converts E(t) into the gate function [15], [79].

The generalized projections approach first demonstrated on spectrograms by Griffin and Lim [54] is very powerful; it can be used for any FROG geometry and can include material response [81], [82], but applied to FROG retrievals, it requires a time-consuming minimization process, and is not as easy to use for blind retrievals [42], [73]. For common FROG geometries, such as second harmonic generation (SHG), polarization gate (PG) and self-diffraction (SD) as well as blind retrievals, a generalized projections algorithm, called Principal Component Generalized Projections (PCGP) [42]–[44], [69], [70], [83] has been developed that does not require a minimization step, increasing the iteration rate by over an order of magnitude. Because the PCGPA code is compact, it easily optimized for digital signal processing techniques as well as multiprocessor approaches using graphics processor units (GPU) making simultaneous data acquisition and FROG trace inversion easily obtainable. Using such a scheme, this new algorithm has been used to invert FROG traces in real-time [43].

## V. PTYCHOGRAPHIC ITERATIVE ENGINE

Unlike the generalized projections algorithms, the ptychographic iterative engine (PIE) uses a steepest descent algorithm to find an unknown function that is multiplied by, and scanned transversely over, a known function [46]. The work of Fienup *et al.* [84] prompted others to examine using the PIE and the extended PIE (ePIE) for the 1D measurements and, therefore, the measurement of ultrafast laser pulses [45], [46] and [47], [48], respectively.

The flow chart of the PIE and the ePIE is shown in Figure 7. Using the previous guess for the gate, $G_j(t)$, and the probe, $P_j(t)$, the next guess for the signal is formed by multiplying the probe by the shifted gate, $G_j(t-R_{s(j)})$. The product is Fourier transformed, and the modulus is replaced by the measured intensity. The result is inverse Fourier transformed, and $P_j(t)$ is updated using:

$$P_{j+1}(t) = P_j(t) + \alpha \frac{G_j^*\left(t-R_{s(j)}\right)}{\left|G_j\left(t-R_{s(j)}\right)\right|_{max}^2} \left(\Psi_j'(t) - \Psi_j(t)\right), \tag{20}$$

where $P_j(t)$ and $G_j(t)$ are the pulse and gate defined previously, $R_{s(j)}$ are the set of translations, $\Psi_j(t)$ is the product of the object function and translated probe before replacing the intensity with the measured data and $\Psi'_j(t)$ is $\Psi_j(t)$ after replacing the modulus with the measured data [49]. $\alpha$ is a constant that adjusts the step-size of the update and is often set to one, but that may be varied to improve convergence [49]. An iteration is complete once all the data are used to complete the update of the object function. The order of the updating is random. The PIE assumes that one of the functions is known and updates only the unknown function.

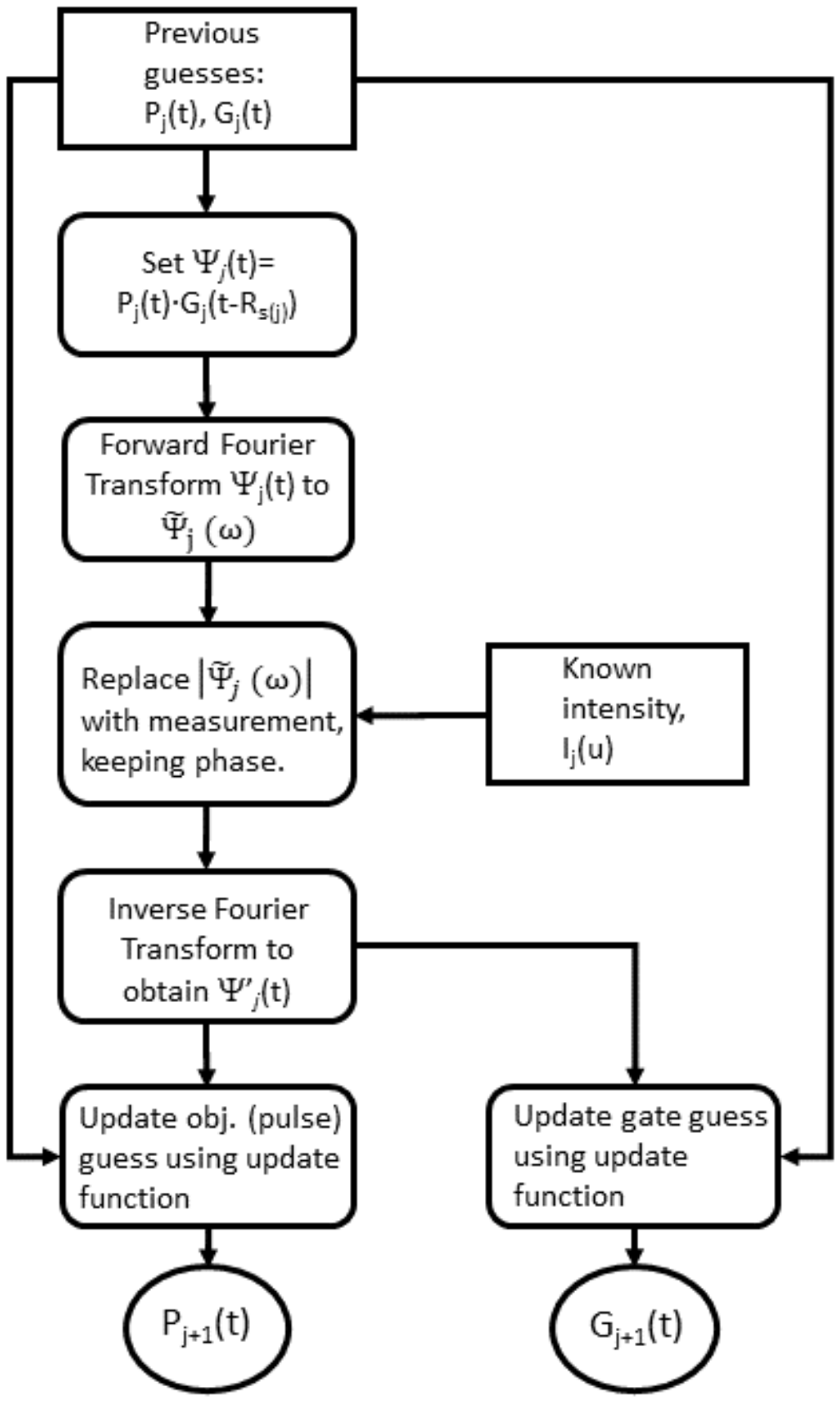

Figure 7. Flow chart for the Ptychographic Iterative Engine and the extended Ptychographic Iterative Engine (ePIE) adapted for pulse measurement.

In the case of ePIE, both functions are unknown, and the function updates alternate between the unknown functions where each update uses all the measured data. It is the ePIE algorithm that is sometimes applied to FROG-type data where the gate is a function of the pulse. In the case of SHG FROG-type data, the pulse is updated first, with the gate held fixed. The gate is then set equal to the pulse and the gate is updated while the pulse is held fixed. In general, this converges well, but not as well as the full PCPGA even when a full FROG data set is used. However, it has the advantage that the time spacing (or frequency spacing) can be arbitrary. The Fourier transform relationship is not required ($\Delta t = \frac{1}{N\Delta f}$, where N is the number of points).

# VI. Principal Components Generalized Projections Algorithm

The principal components generalized projections algorithm (PCGPA) takes advantage of a linear algebra framework for the construction of both the spectrogram and its inversion. There are several advantages brought about by this framework including both algorithm simplicity, and algorithm speed because the minimization step needed in projections algorithm is replaced by matrix vector multiplications. A new advantage that we present for the first time in this work is the construction of a full operator-based algorithm that provides simple methods for adding constraints to the algorithm without causing stagnation problems that are common for previous methods [85]. This new operator-based formulation is called the constrained PCGPA or C-PCGPA. In the sections that follow, we describe the steps for the formulation of this powerful algorithm structure. First, we describe the formulation of the spectrogram using linear algebra methods followed by the inversion. We then describe methodologies for converting the blind PCGPA into a FROG PCGPA, greatly increasing its utility. However, perhaps the most important step in the evolution of the PCPGPA to the operator formulation of the constrained PCGPA is the development of a linear algebra derivation of the PCGPA version of the PIE, where one of the fields is known. This is also known as the cross-FROG or X-FROG algorithm [86]–[89]. The section following the X-FROG PCGPA derivation shows the utility of this operator to create the constrained PCGPA.

## *A. Forming the spectrogram*

For the derivation of the PCGPA, we start with the general form of the spectrogram assuming no relationship between the pulse and the gate:

$$I_{FROG}(\omega,\tau) = \left| \int_{-\infty}^{\infty} E(t)G(t-\tau)\, exp(-i\omega t)dt \right|^2 \tag{21}$$

where $I_{FROG}(\omega,\tau)$ is the spectrogram, *E(t)* is the pulse and *G(t)* is the gate and the integration performs the Fourier transform with respect to *t*. Virtually all-practical data collection methods rely on discretizing τ and t. Suppose *E(t)* and *G(t)* are discrete samples in time with a constant spacing of Δt. Then *E(t)* and *G(t)* are column vectors of length N that can be written as:

$$E_{probe} = \begin{bmatrix} E_1 \\ E_2 \\ E_3 \\ \vdots \\ E_N \end{bmatrix}, \; G_{gate} = \begin{bmatrix} G_1^* \\ G_2^* \\ G_3^* \\ \vdots \\ G_N^* \end{bmatrix} \tag{22}$$

where each point in the vector is complex and $G_{gate}$is the complex conjugate of the gate.

The outer product of $E_{Probe}$ and $E_{Gate}$, $O_{kj} = E_k G_j^*$ written as a matrix is:

$$\begin{bmatrix} E_1G_1 & E_1G_2 & E_1G_3 & E_1G_4 & \dots & E_1G_{N-1} & E_1G_N \\ E_2G_1 & E_2G_2 & E_2G_3 & E_2G_4 & \dots & E_2G_{N-1} & E_2G_N \\ E_3G_1 & E_3G_2 & E_3G_3 & E_3G_4 & \dots & E_3G_{N-1} & E_3G_N \\ E_4G_1 & E_4G_2 & E_4G_3 & E_4G_4 & \dots & E_4G_{N-1} & E_4G_N \\ . & . & . & . & \dots & . & \\ . & . & . & . & \dots & . & \\ E_NG_1 & E_NG_2 & E_NG_3 & E_NG_4 & \dots & E_NG_{N-1} & E_NG_N \end{bmatrix}, \tag{23}$$

which is referred to as the *outer product form (OPF*) matrix.

By rotating the elements of the rows in the outer product to the left by the row number minus one, we obtain:

$$\begin{bmatrix} E_1G_1 & E_1G_2 & E_1G_3 & E_1G_4 & \dots & E_1G_{N-1} & E_1G_N \\ E_2G_2 & E_2G_3 & E_2G_4 & E_2G_5 & \dots & E_2G_N & E_2G_1 \\ E_3G_3 & E_3G_4 & E_3G_5 & E_3G_6 & \dots & E_3G_1 & E_3G_2 \\ E_4G_4 & E_4G_5 & E_4G_6 & E_4G_7 & \dots & E_4G_2 & E_4G_3 \\ . & . & . & . & \dots & . & \\ . & . & . & . & \dots & . & \\ E_NG_N & E_NG_1 & E_NG_2 & E_NG_3 & \dots & E_NG_{N-2} & E_NG_{N-1} \end{bmatrix} \tag{24}$$

$$\tau = 0 \quad \tau = -\Delta t \quad \tau = -2\Delta t \quad \tau = -3\Delta t \qquad \tau = 2\Delta t \quad \tau = \Delta t$$

which produces all the columns contained in the time domain FROG trace.

The τ=0 column is the first column, where τ is the time delay in increments of Δt, a point-by-point multiplication of the probe by the gate with no time shift between them. The next column is the τ= -1 column where the gate is delayed relative to the probe by one resolution element, Δt. The gate appears to be shifted "up" by one resolution element with the first element wrapped around to the other end of the vector. Column manipulation places the most negative τ on the left and the most positive on the right; thus, yielding the *time-domain representation of the spectrogram formed by the multiplication of the probe and gate functions*; a discrete version of the product $E_{Probe}(t)E_{Gate}(t-\tau)$. The columns are constant in τ (delay) while the rows are constant in *t* (time). This obtains the same result as calculating the time domain FROG trace directly by shifting the gate in time and multiplying the shifted gate by the probe. However, the matrix manipulation process provides a reversible way to move between the outer product form and the time domain FROG trace. By Fourier transforming each column, the Fourier transform of $E_{Probe}(t)E_{Gate}(t-\tau)$ is obtained as a function of t at each delay increment τ. The final step of taking the magnitude of the complex result produces the FROG trace. This fact that a FROG trace can be constructed from the outer product of two vectors, the pulse, and the gate, using a reversible process is the basic premise of the PCGPA. For this formulation, the pulse and the gate are assumed to be completely independent. For SHG FROG, for example, the gate is set equal to the pulse, for polarization gate FROG, the gate is set to the magnitude squared of the pulse, etc.

### *B. PCGPA-Inversion*

To extract the pulse and gate from a spectrogram, we start with an initial guess for the pulse and gate to construct a spectrogram. Most often the guesses have Gaussian amplitude with random phase. The magnitude of the constructed spectrogram is replaced with the measured magnitude obtaining:

$$\begin{bmatrix} \tilde{E}_{1,1}^{spec} & \tilde{E}_{1,2}^{spec} & \tilde{E}_{1,3}^{spec} & \tilde{E}_{1,4}^{spec} & \cdots & \tilde{E}_{1,N-1}^{spec} & \tilde{E}_{1,N}^{spec} \\ \tilde{E}_{2,1}^{spec} & \tilde{E}_{2,2}^{spec} & \tilde{E}_{2,3}^{spec} & \tilde{E}_{2,4}^{spec} & \cdots & \tilde{E}_{2,N-1}^{spec} & \tilde{E}_{2,N}^{spec} \\ \tilde{E}_{3,1}^{spec} & \tilde{E}_{3,2}^{spec} & \tilde{E}_{3,3}^{spec} & \tilde{E}_{3,4}^{spec} & \cdots & \tilde{E}_{3,N-1}^{spec} & \tilde{E}_{3,N}^{spec} \\ \tilde{E}_{4,1}^{spec} & \tilde{E}_{4,2}^{spec} & \tilde{E}_{4,3}^{spec} & \tilde{E}_{4,4}^{spec} & \cdots & \tilde{E}_{4,N-1}^{spec} & \tilde{E}_{4,N}^{spec} \\ . & . & . & . & \cdots & . & \\ . & . & . & . & \cdots & . & \\ \tilde{E}_{N,1}^{spec} & \tilde{E}_{N,2}^{spec} & \tilde{E}_{N,3}^{spec} & \tilde{E}_{N,4}^{spec} & \cdots & \tilde{E}_{N-1,N}^{spec} & \tilde{E}_{N,N}^{spec} \end{bmatrix}, \tag{25}$$

Inverse Fourier transforming each column we obtain:

$$\begin{bmatrix} E_{11}^{spec} & E_{1,2}^{spec} & E_{1,3}^{spec} & E_{1,4}^{spec} & \cdots & E_{1,N-1}^{spec} & E_{1,N}^{spec} \\ E_{2,1}^{spec} & E_{2,2}^{spec} & E_{2,3}^{spec} & E_{2,4}^{spec} & \cdots & E_{2,N-1}^{spec} & E_{2,N}^{spec} \\ E_{3,1}^{spec} & E_{3,2}^{spec} & E_{3,3}^{spec} & E_{3,4}^{spec} & \cdots & E_{3,N-1}^{spec} & E_{3,N}^{spec} \\ E_{4,1}^{spec} & E_{4,2}^{spec} & E_{4,3}^{spec} & E_{4,4}^{spec} & \cdots & E_{4,N-1}^{spec} & E_{4,N}^{spec} \\ . & . & . & . & \cdots & . & \\ . & . & . & . & \cdots & . & \\ E_{N,1}^{spec} & E_{N,2}^{spec} & E_{N,3}^{spec} & E_{N,4}^{spec} & \cdots & E_{N,N-1}^{spec} & E_{N,N}^{spec} \end{bmatrix}, \quad (26)$$

or it can be written as $\tilde{E}_{i,j}^{spec}$ where the Fourier transform is performed over constant $j$, yielding $E_{i,j}^{spec}$. The next step is to convert from the time domain spectrogram to the outer product form matrix $E_{i,j}^{OPF}$ by reversing the one-to-one transformation that converted the outer product to the time domain spectrogram. We then need to minimize the following:

$$\varepsilon^2 = \sum_{i,j=1}^{N} \left| E_{i,j}^{OPF} - E_i^{Pulse} \overline{E_j^{Gate}} \right|^2 \quad (27)$$

where $E_i^{Pulse} \overline{E_j^{Gate}}$ is the outer product and ε, is the error, the bar accent indicating the complex conjugate [44]. The primary issue is that because the magnitude of the constructed spectrogram is replaced with the measured spectrogram, $E_{i,j}^{OPF}$ is no longer a single outer product; it is a sum of outer products. Our goal from the least squares minimization is to find the best rank 1 approximation of $E_{i,j}^{OPF}$. Minimizing the sum of the squares is also minimizing the distance, which computes a projection finding the best rank 1 approximation of $E_{i,j}^{OPF}$ [90]. A simple way to solve this minimization is to use a singular value decomposition (SVD) of $E_{i,j}^{OPF}$, which decomposes $E_{i,j}^{OPF}$, written here as $\boldsymbol{O}$, into three matrices:

$$\boldsymbol{O} = \boldsymbol{U\Sigma V^*}. \quad (28)$$

where $\boldsymbol{U}$ and $\boldsymbol{V^*}$ are orthogonal square matrices and $\boldsymbol{\Sigma}$ is a square, real diagonal matrix. $\boldsymbol{V^*}$ is the conjugate transpose of $\boldsymbol{V}$. Thus, the matrix $\boldsymbol{O}$, the outer product form, is decomposed into a superposition of outer products between various "pulse" vectors (columns of $\boldsymbol{U}$) and various "gate" vectors (rows of $\boldsymbol{V^*}$). The diagonal values in $\boldsymbol{\Sigma}$ (the only non-zero elements of $\boldsymbol{\Sigma}$) determine the relative weights of each outer product and, therefore, how much each outer product contributes to matrix $\boldsymbol{O}$ [44], [80]. If we keep the outer product pair with the largest weighting factor, or *principal* *component*, for the next iteration of the algorithm, we minimize Equation 27; therefore, the principal component is the generalized projection.

The SVD approach is convenient because many commercially available mathematical libraries contain routines to compute SVDs. However, the SVD recovers much more information than the principal component. It would be much faster if only the principal outer product was recovered.

In the decomposition process, the SVD computes eigenvectors that form an orthonormal basis—not just the eigenvectors of ***O***. Consequently, an SVD finds the eigenvectors of $\boldsymbol{OO}^*$ (given as columns of ***U***) and $\boldsymbol{O}^*\boldsymbol{O}$ (given as columns of ***V***) which are orthonormal [80], [90]. If the columns of U are written as $P_i$ and the columns of ***V*** are written as $G_i$, and the diagonal elements of the matrix $\boldsymbol{\Sigma}$ are denoted as $\sqrt{\lambda_i}$, then ***O*** can be constructed as a sum of outer products:

$$\boldsymbol{O} = \sum_{i=1}^{N} \sqrt{\lambda_i} P_i G_i^* \tag{29}$$

where $\lambda_i$, $P_i$, and $G_i$ are provided by the SVD, but we only need the $P_i$ and $G_i$ corresponding to the largest $\lambda_i$, or the principal eigenvectors. Suppose we multiply an arbitrary nonzero vector $x_0$ by $\boldsymbol{OO}^*$. Then

$$\boldsymbol{OO}^* x_0 = \sum_{i=1}^{N} \kappa_i \lambda_i P_i \tag{30}$$

where $P_i$ are the eigenvectors of $\boldsymbol{OO}^*$, $\lambda_i$ the eigenvalues, and $\kappa_i$ a set of constants. $\boldsymbol{OO}^*$ can be thought of as an operator that maps $x_0$ onto a superposition of eigenvectors. The process can be repeated resulting in $\boldsymbol{OO}^* \kappa_i \lambda_i P_i = \kappa_i \lambda_i^2 P_i$. Multiplying by $(\boldsymbol{OO}^*)^{p-1}$ gives

$$(\boldsymbol{OO}^*)^p x_0 = \sum_{i=1}^{N} \kappa_i \lambda_i^p P_i \tag{31}$$

$$\begin{aligned} OO^* P_i &= \lambda_i P_i \\ O^* O G_i &= \lambda_i G_i \end{aligned} \tag{32}$$

As $p$ becomes large, the largest eigenvalue, $\lambda_l$, dominates the sum so that $(\boldsymbol{OO}^*)^p x_0 \sim \kappa_i \lambda_i^p P_i$. This method is called the *power method* [91]. After a few iterations, a close approximation to the principal eigenvector (the eigenvector with the greatest eigenvalue) is obtained. Consequently, the next guess for the pulse can be obtained by multiplying the previous guess for the pulse by $\boldsymbol{OO}^*$. The next guess for the gate can be obtained by multiplying the previous guess for the gate by $\boldsymbol{O}^*\boldsymbol{O}$. (For polarization-gate FROG, the absolute value of the result for the gate is taken.) While better approximations for the eigenvectors may be obtained by using these operators several times per iteration, once per iteration is adequate in practice [42]–[44], [69]. For real-time applications, this is the best approach [43], [44], [75] and generally does not affect the convergence.

Practically, the power method implementation of the PCGP algorithm (Figure 5) is extremely fast and quite robust. Indeed, the power method implementation can loop at a few thousand iterations/second on modern consumer computers. Good approximations for the pulse usually occur in about 40 iterations [43], [44].

The Principal Components Generalized Projections Algorithm (PCGPA) is an algorithm that can be used for the extraction of the functions forming a spectrogram or sonogram, namely the pulse and the gate, from its magnitude. In its most basic form, it makes no assumption about the relationship between the two functions. In the phase retrieval literature, such retrievals are known as “blind” because the pulse and the gate are independent. This has also been called “Twin Retrieval of Excitation Electric fields FROG” (TREEFROG) when applied specifically to FROG [92]. Because it makes no *a priori* assumptions about the relationships between the probe and the gate [42], we shall refer to this general case as blind-FROG when we are specifically referring to the PCGPA and blind-spectrogram when we are discussing blind retrievals independent of the algorithm. While contrary to the goal of creating a FROG retrieval algorithm where the gate is a function of the pulse, having the pulse and the gate independent allows the development of a generalized projections algorithm that does not require a minimization step, ultimately speeding algorithm execution and simplifying programming [44]. This formulation also provides a path for an operator-based development, further simplifying construction and implementation of the PCGPA. The next section discusses some of the applications of the blind PCGPA algorithm before moving on to constraining the PCGPA for FROG-type applications.

## VII. APPLICATIONS OF THE BLIND PCGP ALOGORITHM

For quite some time, blind-FROG was merely a way to develop the FROG algorithm. Interestingly, blind-FROG has found some applications and they are difficult to achieve using other ptychographic algorithms. That said, care must be taken because of the many ambiguities that can occur in these types of retrieval problems [42], [92], [93]. Two important causes that will be discussed here include the measurement of telecommunications pulses [76] where the measurement configuration required that the gate was independent of the pulse and the other has been the measurement of attosecond laser pulses [77] where the amplitude of the gate was fixed to be a constant (phase only).

In the case of telecommunications pulse measurement, the pulse to be measured is sent into an electroabsorptive modulator, which is phase-locked to the pulse train. Scanning the time delay between the pulse to be measured and the electroabsorptive gate is accomplished by adjusting the relative phase between the pulse train and the electroabsorptive modulation. Thus, 360 degrees of phase is equivalent to the time spacing between each pulse. Different temporal portions of the pulse are then gated out as the relative phase between the pulse train and the electroabsorptive modulation is changed [76].

The gated portion of the pulse is spectrally resolved using an optical spectrum analyzer. Because the exact gating function is not known, the resulting spectrogram could be inverted using the blind-FROG algorithm. An advantage of the technique is that it is a linear technique—no nonlinearities are driven by the pulse to be measured. Thus, very weak pulses can be measured (i.e., telecommunications pulses with only milliwatts of peak power). Another advantage is that there is no direction of time ambiguity as in the case of SHG-FROG [76].

Attosecond pulse generation represents a whole new facet to the field of ultrafast laser physics, and, of course, pulse measurement. One elegant solution was described in a paper by F. Quéré, et al. and is now called Frequency-Resolved Optical Gating for Complete Reconstruction of Attosecond Bursts (FROG CRAB) [77], [94], [95]. This method makes no assumptions about the XUV attosecond pulses so that the pulses can be completely arbitrary.

The gist of the technique is that an XUV pulse generates an electron wave-packet by photoionizing atoms. An IR femtosecond laser pulse can then phase modulate the electron wavepacket acting as a phase gate. The photoelectron spectrum is recorded as a function of time delay between the attosecond pulse and the IR femtosecond laser pulse. The resulting spectrogram of the photoelectron spectrum is inverted using the blind PCGP algorithm [77], [94], [95].

## VIII. Conversion of the PCGPA to a FROG Inversion Algorithm

The PCGPA is fundamentally a blind-FROG algorithm where the pulse and the gate are independent. Without additional constraints, conducting FROG inversions where the pulse and the gate are related through a nonlinear process, creates retrieval errors, which can produce erroneous results if ignored [42], [92], [93]. Therefore, constraints are required to facilitate FROG retrievals. FROG is self-referencing; the FROG geometry defines the relationship between the pulse and the gate. For second harmonic generation (SHG) FROG, the pulse and gate are equal. For polarization-gate (PG) FROG, the gate is the magnitude of the pulse squared.

For the first part of this discussion, we will focus on the SHG FROG algorithm. The next guess for the pulse and the gate is calculated from Equation 32; one equation for each vector. At this point, for SHG FROG, it would be possible to set the pulse and the gate equal; however, this has been attempted (months of frustration, quite honestly) and will cause algorithm stagnation. However, the task of coercing the algorithm to converge to a solution where the pulse and the gate are equal is still a requirement. The other location where the constraint can be added is in the outer product. Suppose we set the next guess for the pulse to be equal to the average between the calculated pulse and the calculated gate, or equivalently, their sum. The output product becomes:

$$\begin{aligned} O_k^{ij} &= \left(pulse_k^i + \overline{gate_k^j}\right)\overline{\left(\overline{pulse_k^i} + gate_k^j\right)} \\ O_k^{ij} &= pulse_k^i pulse_k^j + pulse_k^i \overline{gate_k^j} + gate_k^i pulse_k^j + \overline{gate_k^i}\,\overline{gate_k^j} \end{aligned} \tag{33}$$

where the bar accent is the notation for the complex conjugate to maintain consistency with the matrix vector formulation. We know that this outer product will stagnate, and we can ascertain (by judicious guessing) that the first and last terms of the expanded product are problematic because they force the algorithm to different solutions for the pulse and the gate. Removing them we obtain:

$$O_k^{ij} = pulse_k^i \overline{gate_k^j} + \overline{gate_k^i} pulse_k^j \quad (34)$$

forming the FROG trace from the sum of two outer products. Because only the principal outer product pair is used for the next estimate of the electric field, the two outer products are still forced to be equal. In practice, using this outer product works quite well for SHG FROG [43], [44].

SHG FROG is a special case, however; Equation 34 is valid only for SHG FROG and must be modified for other FROG geometries. We define Γ as the function that produces the gate from the pulse, E(t); its inverse, denoted $\Gamma^{-1}$, produces the probe from the gate. Thus, the $pulse = \Gamma(gate)$ and the $pulse = \Gamma^{-1}(gate)$. Rather than using the product of $pulse^i \overline{gate^j}$ to produce the next time domain FROG trace, the sum of the products $pulse_k^i \overline{gate_k^j}$ and $\Gamma^{-1}(gate_k)^i \overline{\Gamma(pulse_k)^j}$ is used so that the outer product on the next iteration is given by

$$O_k^{ij} = pulse_k^i \overline{gate_k^j} + \Gamma^{-1}(gate_k)^i \overline{\Gamma(pulse_k)^j} \quad (35)$$

where $O_k$ is the sum of two products for the *k*th iteration.

Equation 35 allows the PCGPA to be used with any FROG geometries where $\Gamma^{-1}$ exists. However, for most FROG interactions other than SHG, the inverse function does not exist. For example, the inverse of the gate function does not exist in PG FROG and a pseudo-inverse must be constructed. For PG FROG, the pseudo-inverse is constructed from the square root of the gate intensity and the phase of the pulse. Because small fluctuations in the wings of the gate can be produced by the square root whenever values are small and prone to noise, instabilities may occur in the algorithm. This can be remedied by applying the square root only to well-defined portions of the gate. Where the gate is not well defined (i.e., the intensity is near zero and dominated by noise), the intensity (and phase) of the pulse is used. To increase the robustness of the PG algorithm, the pseudo-inverse constraint is applied on alternate iterations [75]. The pseudo-inverse method works well for polarization-gate (PG) FROG on the synthetic test sets, converging to experimental error for 90% of the test pulses, and has worked well for experimental data [69]. However, this approach to converting the PCGPA to a self-diffraction geometry has not worked well. In the sections that follow, we develop a method to add additional constraints to the PCGPA, which helps to reduce issues with self-diffraction inversions. The first step in that development is the construction of the cross-correlation FROG (X-FROG) PCGPA.

## IX. PCGP Cross-correlation FROG (X-FROG) derivation

Cross-correlation FROG uses a known reference pulse or gate function to interrogate an unknown pulse in a cross-correlation configuration [54], [86], [88], [89]. Since the reference pulse is known *a priori* and is cross-correlated with the unknown pulse to create the X-FROG trace, extraction of the unknown pulse should be easier [54].

However, insertion of a known gate into the PCGP FROG algorithm is not straightforward because the PCGP algorithm is naturally a blind algorithm—the pulse and the gate are found independently and no *a priori* information about the gate is used. Indeed, insertion of the known gate can cause severe stagnation and retrieval problems. Using blind deconvolution for X-FROG retrievals will determine both the pulse and the gate, but this process is prone to errors and ambiguities especially in the case of similar pulse shapes and durations; therefore, it would be much better to use known gate ($g$) information [42]. Indeed, for years, it has not been known how to use a known gate in the PCGP algorithm to develop a PCGP X-FROG algorithm that does not stagnate[76], [87].

In the following discussion, we derive the X-FROG PCGPA, which will not only add to the repertoire of the PCGPA but will also provide a robust means to add additional constraints to the PCGPA while still maintaining the goal of an operator formalism. As in the case of the FROG PCGPA, the goal is to determine the best rank 1 approximation for $O$, the outer product form matrix. For fixed choices of unit eigenvectors $p$ (pulse) and $g$ (gate), a best rank 1 approximation for $O$ must be of the form $e^{i\theta}\sigma_1 p g^*$ for some $\theta$ [96]. The scalar factor $e^{i\theta}$ must be chosen to minimize

$$\|O - e^{i\theta}\sigma_1 p g^*\|_2^2$$
$$= \|O\|_2^2 - 2\sigma_1 \, Re[\, tr(e^{-i\theta} O (pg^*)^*)] + \sigma_1^2 \|p\|_2^2 \|g\|_2^2, \quad (36)$$

where $\theta$ is a phase constant, $|| \quad ||_2^2$ is the $l_2$ norm (sum of the elements squared) squared, and *Re* is the real part. Minimizing equation 36 above is equivalent to maximizing

$$Re[\, tr(e^{-i\theta} O (pg^*)^*)] = Re[\, e^{-i\theta} p^* O g] \quad (37)$$

where $tr$ is the trace of the matrix. Both *p* and *g* are unit eigenvectors, and $p^*Og$ needs to be maximized. Because $\sigma_1$ is the principal eigenvalue, it is the largest eigenvalue. In addition, the optimal scalar factor, $e^{i\theta}$, is given by $(p^*\boldsymbol{O}g)/\sigma_1$. Therefore, equation (37) is maximized when $\boldsymbol{O}g = e^{i\theta}\sigma_1 p$ [96]. In other words, ***O*** projects the gate, *g*, onto the unknown pulse, *p*, within a constant phase factor, which is the basis for the X-FROG algorithm. As a result, to convert the FROG PCGP algorithm to the X-FROG PCGP algorithm, the outer product form matrix ***O*** becomes the operator to obtain the best pulse for a known gate, or vice versa:

$$\boldsymbol{O}g = \sigma_1 p, \; \boldsymbol{O}^* p = \sigma_1 g \quad (38)$$

where *g* is the gate or reference pulse, and *p* is the unknown pulse. Depending on if *g* or *p* is the known pulse (either one can be the known or unknown quantity), only one of the equations in Equation 38 needs to be used in the X-FROG retrieval algorithm. For example, if *g* is known, then ***O*** obtains *p* from a known *w* and the outer product of *p* and *g* creates the next guess for ***O***. Also, like the FROG PCGP algorithm, equation 38 is applied every iteration, which greatly favors the vector with the largest eigenvalue. The outer product is constructed from a single outer product and is the same as the blind-FROG algorithm. The resulting algorithm is extremely robust [86], [97].

By developing an operator to obtain the best pulse given a known gate (and vice versa), we have developed a means to add additional constraints to the PCGPA to fully develop the Constrained PCGPA (C-PCGPA) discussed in the next section.

## X. Constrained PCGPA—The Final Step in the Operator Formalism of the PCGPA

The procedure to add constraints to the original formulation of the PCGPA [42]–[44], [69], [76], [87], led to stagnation and required significant algorithm modification. Using the X-FROG PCGPA, now have everything required to develop a complete *operator* formalism for the PCGPA. With this formulation, we can quickly add and remove constraints, both internal constraints (i.e., constraints based on relationships between the pulse and gate or a.k.a., FROG mathematical form constraints) and external constraints (e.g., spectral constraints, compact support, etc.) all while improving convergence and retrievals. By incorporating internal or external constraints into, for example, generating the next iteration of the probe or gate estimate, the PCGPA can provide better convergence and greater accuracy. Spectral constraints can be easily added; compact support or other known characteristics of the pulse (and gate) can be added as well. We refer to relationships or expressions of mathematical constraints associated with a particular FROG geometry as internal constraints. Constraints using known or measurable quantities about the pulse we will call external constraints. These constraints do not sacrifice the iteration speed, simplicity or the fast inversion speed and compact code that can run efficiently on typical computer processing units (CPU), and graphics processing units (GPU).

Like the FROG PCGPA, the constrained principal components generalized projections (C-PCGP) system uses an iterative method to find a desired vector pair that corresponds to the measured FROG trace and other constraints. Figure 8 shows the flow of the C-PCGPA. The illustrated algorithm uses the techniques, structures, and operators that have been developed in both the FROG PCGPA section and the X-FROG PCPGPA section.

*A. FROG Geometries*

Figure 8 shows that the algorithm is started with a random guess, which is like the random guess used to start the PCGPA usually taking the form of a noise modulated Gaussian. To construct the initial guess for the phase, a C-PCGPA constructs a FROG trace using vector pairs, one complex (pulse) vector and, in the case of polarization-gate FROG, one real (gate) vector, that are random noise modulated by a broad Gaussian pulse. The constrained output product matrix

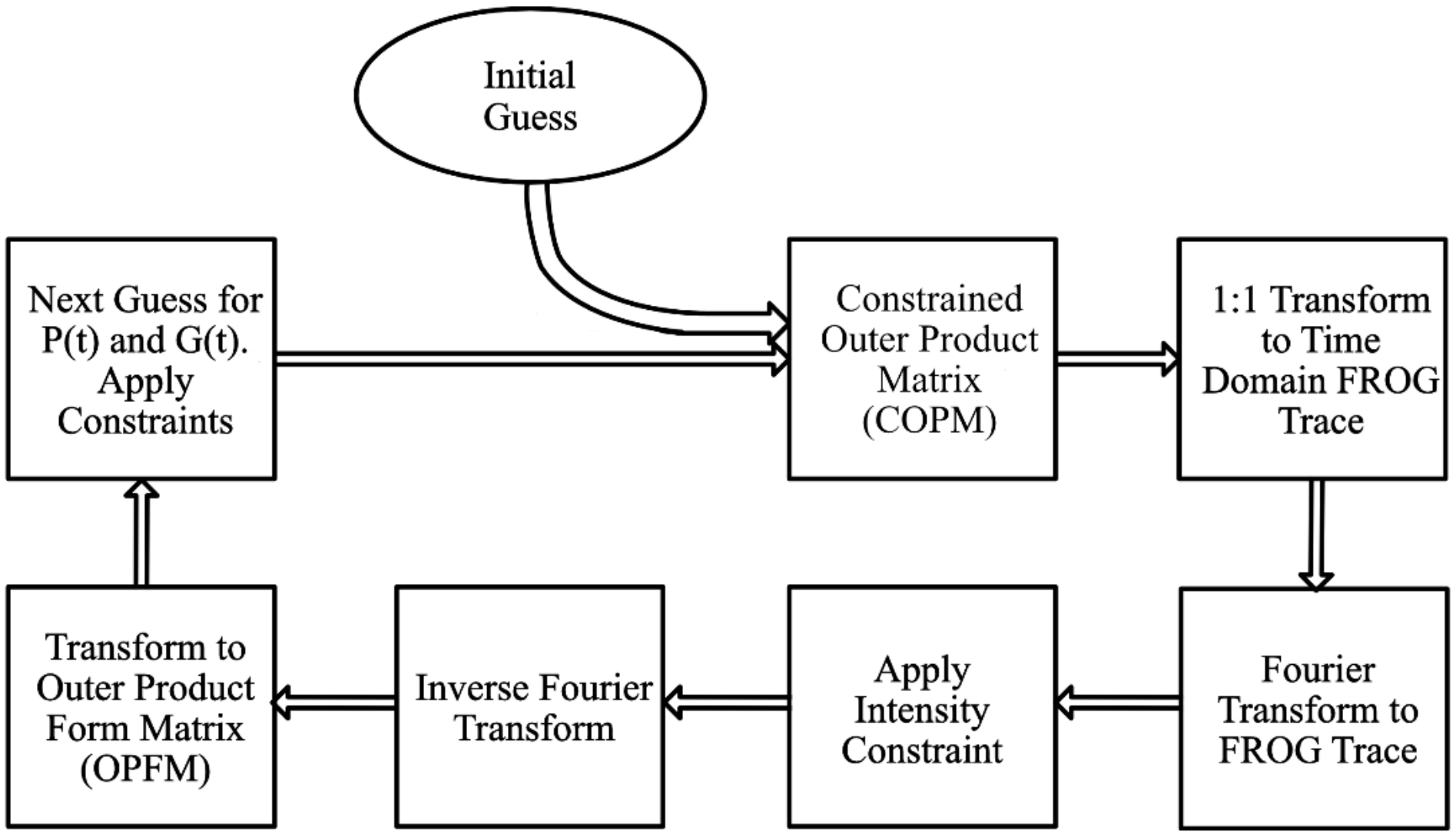

**Figure 8**. Flow chart of the C-PCGPA. The constraints are used in two places. The first is during the construction of the next guess, and the second is when the outer product form matrix is constructed. The rest of the algorithm is identical to the PCGPA.

(COPM) is constructed from both constrained and unconstrained outer products.

The COPM is transformed, through matrix row rotations, into the time domain FROG trace. The transform to convert the COPM to the time domain FROG trace is a 1:1 transform where the elements of the outer product matrix are rearranged to form the time domain FROG trace, identical to the transform discussed previously (Equations 23, 24 and 27). A Fourier transform of the columns of the time domain FROG trace transforms the time domain FROG trace to the frequency domain. The magnitude of the newly constructed FROG trace is replaced by the square root of the magnitude of the measured (experimental) FROG trace (i.e., intensity). That is, like the PCGPA, the FROG trace constraint for the C-PCGPA is applied in the conjugate domain (If the pulse starts in the time domain, then the intensity constraint is applied in the frequency domain and *vice versa*.) An inverse Fourier transform transforms the FROG trace back to the time domain. The process reverses the 1:1 transform by reversing the rearranging of matrix elements to convert the matrix back to the outer product form matrix (OPFM).

After the transform, the flow (Figure 8) generates a next estimate for the gate and probe pulse vectors that will be used for the next iteration. In addition, constrained versions of the pulse and the gate are also constructed. The constraints used depend on the available information and the geometry of the FROG system. To demonstrate, the first example is for the polarization gate geometry (single shot or scanning). The product of the OPFM, $O$, with its transpose $O^*$ transforms the probe pulse, $pulse_i$, into the next guess for the pulse, $pulse_{i+1}$ using the relation $pulse_{i+1} = OO^* pulse_i$. Similarly, the function $\boldsymbol{O^*O}$ transforms the gate, $gate_i$, into the next guess for the gate, $gate_{i+1}$. Because in the case of the polarization gate (PG) geometry, the gate is real, the magnitude is taken such that the next iteration of the gate, $gate_{i+1} = |O^* O gate_i|$. In summary,

$$
\begin{aligned}
pulse_{i+1} &= OO^* pulse_i \\
gate_{i+1} &= |O^* O gate_i|.
\end{aligned}
\tag{39}
$$

In equation 39, a multiplied constant related to the eigenvalue of the dominant vector is omitted, since it is a scaling factor. The functions $\boldsymbol{OO^*}$ and $\boldsymbol{O^*O}$ map a pulse and a gate, respectively, to a projection for the next iteration in the polarization gate geometry. Alternately, the new estimates for the gate and probe fields can be determined using the singular value decomposition (SVD) procedure [70].

If we were to stop here, the resulting inversion would be blind. No relationship between the pulse and the gate would be assumed. However, for PG-FROG, the gate is the $|pulse_{i+1}|^2$. Using the operators in the X-FROG implementation, we can add internal constraints to the inversion. For the ith iteration for PG-FROG they are:

$$pulse'_{i+1} = O|pulse_{i+1}|^2$$
$$gate'_{i+1} = |Ogate_{i+1}|^2 \quad (40)$$

The relations in Equation 40 include the relevant mathematical form relationship; a scaling factor related to the eigenvalue corresponding to ***O*** is omitted from both statements of the equation. The operator ***O***, an OPFM, is from the i-th iteration when calculating the value for the (i+1)-th pulse and gate iterations.

In the first of the Equation 40 relations for the polarization gate geometry, the function $|p_{i+1}|^2$ is close to the gate function. The operator *O* projects the $|pulse_{i+1}|^2$ function to the pulse space and generally corrects the $|pulse_{i+1}|^2$ for variations from the actual gate at that iteration. The operator *O* maps the $|pulse_{i+1}|^2$ function to the best estimate of the pulse (in a least squares sense) for that iteration. Similarly, the function $Ogate_{i+1}$ maps the gate to the best estimate of the pulse, which is then converted to the gate space by the absolute value squared function shown in Equation 40. To generate the constraints, $pulse'_{i+1}$ and $gate'_{i+1}$we typically use the next iteration of the pulse and gate used from equation 39.

Selection of the COPM for an implementation of the Figure 8 flow is intended to enhance convergence and the effectiveness of the applied constraints. An example of a constrained outer product matrix for the Figure 8 FROG retrieval is given by:

$$O_{i+1}^{jk} = pulse_{i+1}^{j}\overline{gate_{i+1}^{k}} + pulse'^{j}_{i+1}\overline{gate_{i+1}^{k}} + pulse_{i+1}^{j}\overline{gate'^{k}_{i+1}}. \quad (41)$$

$O_{i+1}$ is an outer product matrix formed as the sum of three outer product matrices where the bar accent is the notation for the complex conjugate to maintain consistency with the matrix vector formulation. In equation 41, $O_{i+1}$ is the outer product form matrix for the (i+1)-th iteration and is based on estimates obtained through using equations 39 and 40. Equation 40 produces the $pulse'_{i+1}$ and $gate'_{i+1}$ vectors in this example of the polarization gate geometry. Using equation 41 as the form of the constrained outer product matrix provides better convergence and immunity to a time center offset artifact.

However, when equation 41 is used on every iteration, stagnation can occur. Better convergence for the polarization gate implementation occurs when equation 41 is alternated with only the first outer product in equation 41 so that OPFM $O_{i+1}^{jk} = pulse_{i+1}^{j}\overline{gate_{i+1}^{k}}$. Testing shows that the PG C-PCGPA is very robust and converges nearly 100% of the time.

In the case of SHG FROG, the gate and the pulse are equal, and the pulse and the gate are focused into a nonlinear medium, usually a second harmonic generation crystal to generate the product of the pulse and the gate in the time domain (autocorrelation signal), which is spectrally resolved. Most of the Figure 8 flow for the SHG geometry is implemented in the same manner as discussed above in the context of the polarization gate. Only the differences are discussed here, beginning with determining the next estimates for the SHG geometry pulse and gate pulses. After the transform to the OPFM, the next estimate for the gate and pulse vectors is generated that will be used for the next iteration through the algorithm flow (Figure 8). The product of the OPFM, $O$, with its transpose $O^*$ transforms the pulse, $pulse_i$, into the next guess for the pulse $pulse_{i+1}$ using the relation $pulse_{i+1} = OO^*pulse_i$. Similarly, the function $O^*O$ transforms the gate pulse, $gate$, into the next guess for the gate, $gate_{i+1}$is given by $O^*Ogate_i$. In summary,

$$pulse_{i+1} = OO^{*}pulse_{i}$$
$$gate_{i+1} = O^{*}Ogate_{i} \tag{42}$$

where the constant scaling factors are omitted. Thus, operators $\boldsymbol{OO^{*}}$ and $\boldsymbol{O^{*}O}$ map a pulse and a gate pulse, respectively, to a projection for the next iteration.

The mathematical form relationship between the gate and pulse in the SHG geometry is that the pulse and gate are equal. (For the matrix notation used here, the gate is set equal to the complex conjugate of the pulse to make the outer product consistent with the SHG interaction.) In the C-PCGPA implementation of the SHG geometry, the next iteration's gate or pulse can also be taken as:

$$pulse'_{i+1} = Ogate_{i+1}$$
$$gate'_{i+1} = O^{*}pulse_{i+1} \tag{43}$$

where $gate_{i+1}$ and $pulse_{i+1}$ are defined in Equation 42. Equation 43 relations apply the mathematical form relationship by using the X-FROG operators defined above; the best next guess for the pulse, $pulse'_{i+1}$, assuming the next guess for the gate, $gate_{i+1}$, is correct, is found by operating on $gate_{i+1}$with $\boldsymbol{O}$. Likewise, the best next guess for the gate, $gate'_{i+1}$, assuming the pulse, $pulse_{i+1}$, is correct is found by operating on $pulse_{i+1}$with $\boldsymbol{O^{*}}$**.** Again, the scaling factors are ignored because they contain no information.

The operators $\boldsymbol{O^{*}}$ and $\boldsymbol{O}$ map a known pulse and gate to the best (in the least squares sense) gate and pulse, respectively. These operators $\boldsymbol{O^{*}}$ and $\boldsymbol{O}$ produce projections to the target (pulse or gate) space. The SHG implementation of the C-PCGPA uses the relationships listed in Equations 42 and 43 to form the outer product sum shown in Equation 41. This outer product sum is used in each iteration in the algorithm depicted in Figure 8. It is of course possible to use different permutations of the relationships shown in Equations 42 and 43. The constrained SHG FROG algorithm was found to converge 4% better for the random phase test and 13% better for the compound pulse test [43], [44]. 3-5X better convergence was found overall when the time center was shifted, which is problematic for the original SHG PCGPA [98].

As mentioned above, the constrained PCGP algorithm can also be used to retrieve pulses from self-diffraction FROG traces [99]. Like SHG and PG FROG, the next estimates for the pulse and gate vectors in the self-diffraction geometry are given by:

$$pulse_{i+1} = OO^{*} pulse_i$$
$$gate_{i+1} = O^{*} O gate_i \tag{44}$$

In this case, we define the $pulse = E_{in}^2$ and the $gate = E_{in}$ where $E_{in}$ is the measured field. The mathematical form constraint is calculated by first operating on (multiplying) ${gate_{i+1}}^2$ by $O*$ to obtain the best gate assuming assuming ${gate_{i+1}}^2$ is correct pulse:

$$gate'_{i+1} = O^{*} gate_{i+1}^2. \tag{45}$$

The self-diffraction geometry COPM is slightly different than the COPM used by SHG and PG implementations for the C-PCGPA. It contains only two outer products:

$$O_{i+1}^{jk} = pulse_{i+1}^{j} \overline{gate_{i+1}^{k}} + pulse_{i+1}^{j} \overline{gate'^{k}_{i+1}}. \tag{46}$$

The second term is used only in every other iteration. The SD FROG C-PCGPA is still being researched and is work in progress.

Table 1 shows a summary of the different constraints and COPMs used for the SHG, PG and SD FROG geometries.

**Table 1.** Table comparing the original PCGPA with the constrained PCGPA. SHG: second harmonic generation; PG: polarization gate; SD: self-diffraction.

| **Previous Algorithm** | **Constrained Algorithm** |
|---|---|
| **Outer Product** | **Constrained Outer Product** |
| $O_{i+1}^{jk} = pulse_{i+1}^{j}\overline{gate_{i+1}^{k}} + \Gamma^{-1}(gate_{i+1})^{j}\overline{\Gamma(pulse_{i+1})^{k}}$<br>where $gate = \Gamma(pulse)$ every other *i*-th iteration. | SHG, PG:<br>$O_{i+1}^{jk} = pulse_{i+1}^{j}\overline{gate_{i+1}^{k}} + pulse'^{j}_{i+1}\overline{gate_{i+1}^{k}} + pulse_{i+1}^{j}\overline{gate'^{k}_{i+1}}$<br>Terms 2 and 3 every other *i*-th iteration for PG.<br><br>SD:<br>$O_{i+1}^{jk} = pulse_{i+1}^{j}\overline{gate_{i+1}^{k}} + pulse_{i+1}^{j}\overline{gate'^{k}_{i+1}}$<br><br>second term every other *i*-th iteration |
| **Next Guess** | **Next Guess and Constraints** |
| SHG:<br>$pulse_{i+1} = OO^{*}pulse_{i}$<br>$gate_{i+1} = O^{*}Ogate_{i}$<br><br>PG:<br>$pulse_{i+1} = OO^{*}pulse_{i}$<br>$gate_{i+1} = \lvert O^{*}Ogate_{i}\rvert$<br><br>SD: N/A | SHG:<br>$pulse_{i+1} = OO^{*}pulse_{i}$<br>$gate_{i+1} = O^{*}Ogate_{i}$<br>$pulse'_{i+1} = Ogate_{i+1}$<br>$gate'_{i+1} = O^{*}pulse_{i+1}$<br>PG<br>$pulse_{i+1} = OO^{*}pulse_{i}$<br>$gate_{i+1} = \lvert O^{*}Ogate_{i}\rvert$<br>$pulse'_{i+1} = O\lvert pulse_{i+1}\rvert^{2}$<br>$gate'_{i+1} = \lvert Ogate_{i+1}\rvert^{2}$<br>SD:<br>$pulse_{i+1} = OO^{*}pulse_{k}$<br>$gate_{i+1} = O^{*}Ogate_{i}$<br>$gate'_{i+1} = O^{*}gate_{i+1}^{2}$ |

*B. External Constraints*

So far, we have discussed using constraints in the C-PCGPA to implement mathematical form constraints for different FROG geometries—internal constraints. The same strategies can also be applied to *a priori* pulse information such as the pulse spectrum or compact support. Constrained outer products obtained from such external information are called externally constrained outer products. Even using only internal constraints, such as a mathematical relationship between the gate and pulses, inversion performance can be significantly improved. When other, additional information, such as pulse spectra, is available, externally constrained outer products can be summed with other outer products to form a constrained outer product matrix (COPM).

For either the polarization gate or the SHG geometry, additional constraints, as needed, are applied to either the pulse or the gate and then using the respective equations 40 or 43 to determine the best complementary pulse or gate to use in further processing. For example, when the next guesses for the probe and gate are determined in Figure 8, a spectral constraint of an externally measured spectrum can be applied to either the pulse or the gate using the following equation:

$$probe_{constrained} = F^{-1}\left\{\frac{F\{probe\}}{|F\{probe\}|}\sqrt{I(\omega)}\right\}. \tag{47}$$

where probe indicates either the gate or the pulse.

In equation 47, $F\{\}$ is the Fourier transform, $F^{-1}\{\}$ is the inverse Fourier transform, and I(ω) is the measured intensity spectrum of the probe pulse. The operation of equation 47 is applied for each element in the current iteration of the probe vector. The best gate for the spectrally constrained pulse, $pulse_{\text{constrained}}$, is then obtained using the appropriate statement from equations 40 or 43. For example in the SHG application, Equation 43 is used and $gate_{\text{constrained}} = O^{*}pulse_{\text{constrained}}$. $gate_{\text{constrained}}$ is then used in the outer product directly rather than the intensity constrained pulse to help prevent stagnation. The reason is that the constrained pulse is not a projection solution to the best rank 1 approximation of $O$, but $gate_{\text{constrained}}$ is the best rank 1 approximation for the gate of $O$ assuming the constrained pulse is a solution. A constrained outer product matrix can be determined from the sum of outer products:

$$O_{i+1}^{jk} = pulse_{i+1}^{j}\overline{gate_{i+1}^{k}} + \overline{(O_i^{*}pulse_{i-constrained})^{j}}\,\overline{gate_{i+1}^{k}} + pulse_{i+1}^{j}\overline{(O_i gate_i)^{k}}, \tag{48}$$

where $i$ is the iteration number, $pulse_i$ is the $i$-th iteration of the probe pulse, $gate_i$ is the $i$-th iteration of the gate, $\boldsymbol{O_i}$ is the constrained outer product matrix (COPM) from the previous iteration, and $\boldsymbol{O}^{*}_{\text{i}}$ is its transpose. Thus, $pulse_{i\text{-}constrained}$ is the constrained pulse whereby $\boldsymbol{O}^*$ converts the constrained pulse to the constrained gate. Applying the spectral constraint in this manner reduces stagnation in the algorithm depicted in Figure 8.

Many different constraints can be used including combinations of internally and externally constrained outer products to form a constrained outer product matrix (COPM). Intensity constraints can be applied for both spectrogram and sonogram-based measurements. In the case of a sonogram-based measurement, where the wave is in the frequency domain, the intensity constraint can be applied as:

$$probe_{constrained} = F\left\{\frac{F^{-1}\{probe\}}{|F^{-1}\{probe\}|}\sqrt{I(t)}\right\}, \qquad (49)$$

where $I(t)$ is the time domain intensity and $F$ is the Fourier transform as discussed above and probe can be either the pulse or the gate. In the case of spectrogram measurements, an intensity constraint can be applied as:

$$probe_{constrained} = \frac{probe}{|probe|}\sqrt{I(t)}. \qquad (50)$$

An intensity constraint can also be applied in analyzing a spectrogram only to regions where the intensity is known. For example, in the case of compact support the intensity is zero in the wings. In this case, an intensity constraint can be applied as

$$\begin{aligned} probe_{constrained} &= probe \qquad a < t < b \\ probe_{constrained} &= 0 \qquad t < a \; and \; t > b \end{aligned} \qquad (51)$$

where the region between a and b defines where the probe (either the pulse or the gate) is known to be non-zero. Probes constrained in the Equation 51 for spectrogram analysis should be converted to the complementary wave identified in the Equation 43 relations before being used in the outer product sum. Other forms of external contraints that can be applied simlarly include knowledge of a pulse shape or the likelihood that such a pulse shape will occur, or knowledge of chirp. Depending on the constraints used, different combinations of constrained outer products and unconstrained outer products may be needed to develop a robust algorithm.

The constrained PCGPA can also be used to measure and analyze sonograms. In the case of a sonogram, the probe pulse is the spectral or frequency representation of the input pulse. The gate is then a frequency filter that is shifted in frequency and the output is the frequency filtered output pulse. The temporal profile of the output pulse is then measured, which may for example be done by cross-correlating the output pulse with the input pulse, provided the input pulse is much shorter than the filtered output pulse. Analysis can then use the constrained PCGP algorithm to add external constraints such as the pulse spectrum or the shape of the frequency filter. The constraints can be applied "softly" by combining both the direct outer product and the gate by using an externally constrained outer product:

$$O_{i+1}^{jk} = pulse_{i+1}^{j}\overline{gate_{i+1}^{k}} + (O_i gate_{i-constrained})^j\overline{gate_{i+1}^{k}}, \tag{52}$$

where the first outer product of the sum is the "blind" or direct outer product, which is not constrained, and the second externally constrained outer product assumes an external constraint applied to the gate.

While outer products can be summed to form the COPM for both spectrograms and sonograms, as shown in equation 52, the sum is not necessarily a projection because the sum is typically just an average of the outer products. Not using a projection can hurt convergence and stability of the processing and the measurement. Methods such as singular value decomposition (SVD) can be used to extract a projection instead of merely using a sum of outer products to produce the COPM [70].

A COPM that is a projection can also be obtained by using the power method on the outer products matrix. While this may not be as accurate as the full singular value decomposition, it is computationally much faster, and may result in faster convergence. In general, the sum of three outer products as shown in equation 41 seems to work well, without the need to find a projection.

# *XI.* Error Analysis

Error analysis is important for any measurement. One error analysis technique is the parametric bootstrap [123] where the noise is modeled and added to the retrieved FROG trace [42]. The pulse (and, if appropriate, the gate) are extracted from a noisy FROG trace. This process is repeated a statistically significant number of times, which we define here as $\frac{\sqrt{N}}{N} \ll 1$ where N is the number of data sets. Error bars are then determined for the intensity and phase. Another method, called the "jackknife" [123] removes points from the measured FROG trace, and the algorithm is run on this modified FROG trace [100], [101]. Because points are missing from the FROG trace, a modified algorithm is required to extract the pulse. The process is repeated a statistically significant number of times with points randomly removed. Error bars can then be determined from the set of extracted pulses for the intensity and phase.

## *A. Diagnoses of Errors*

The PCGP algorithm is not only useful for inverting spectrograms and sonograms, but it can also be helpful in analyzing the measurements obtained from these spectrograms because the process of applying the PCGP algorithm orthogonalizes the components of the measured spectrogram making them mathematically independent [102]. This analysis presented here can be used for any spectrogram, no matter how it was retrieved. The only issue is that the spectrogram must be resampled to the full Nyquist sampling.

Computing the SVD of an outer product form matrix obtained from a perfectly retrieved FROG trace (i.e., the phase has been retrieved) results in only one outer product pair; that is, the SVD into the U, Σ and V matrices results in only one vector in both the U and V matrices and only one non-zero diagonal term in the Σ matrix. However, FROG inversion is over-determined[42], [103], [104]. Thus, additional outer products resulting from additional column in the U and V matrices together from additional diagonal terms (which we refer to as "weights") in Σ are produced by an SVD of an experimentally obtained outer product form matrix even after algorithm convergence. Noise present in the detection system and distortions caused by geometrical effects, bandwidth limitations, etc. causes measured FROG traces to become non-ideal. In typical measurements in which the inversion algorithm is applied, this added information from the additional outer products is discarded [43], [44], [69]. However, this additional information, can provide clues as to how well the FROG measurement has been made, how well the FROG retrieval

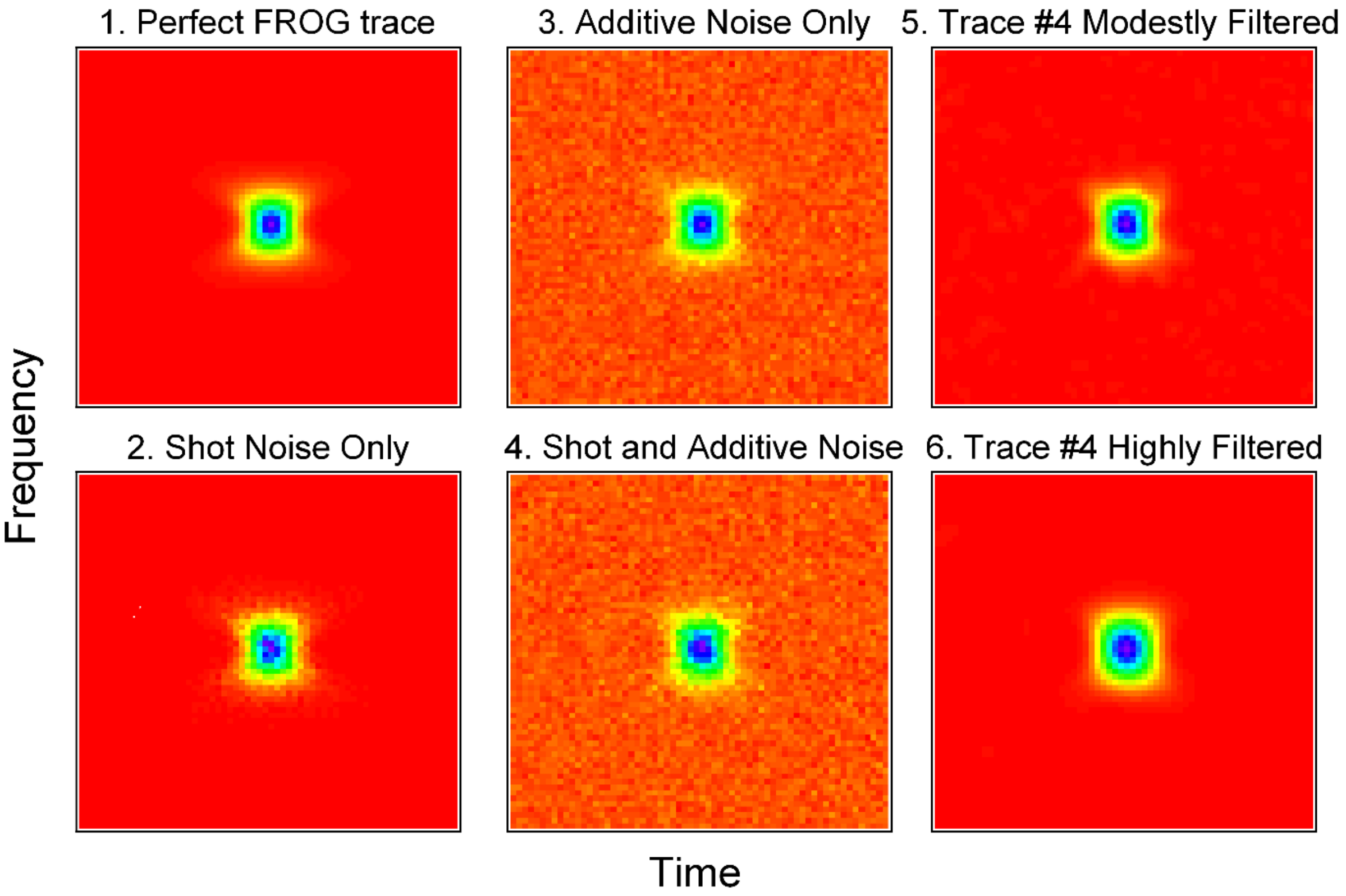

Figure 9. Series of synthetic SHG FROG traces (FT) (the square root of the intensity is shown versus frequency and time delay is presented to show more detail) with cubic spectral phase and various amounts of noise and distortion. 1. Perfect FT with no distortions. FT 2 has shot noise added (It has a per-pixel rms deviation of $12V^{1/2}$, where V is the pixel value, which varies from 0 to 62,500.). FT 3 is FT 1 with only additive noise added (per-pixel rms deviation of 2% of the maximum pixel value). FT 4 has both shot and additive noise. FT 5 is a modestly filtered version of FT 4. FT 6 is a filtered version of trace 4, but visibly distorted because of filtering. After pulse retrieval the FT error and the retrieved pulse error are 1, 0%, 0%; 2, 0.8%, 0.14%; 3, 2.0%, 3.5%; 4, 2.2%, 2.7%; 5, 0.44%, 3.9%; 6, 0.3%, 8.5%, respectively. Adapted with permission from [69]. © The Optical Society.

has been completed, and can also help to diagnose systematic errors that can occur. To demonstrate, six synthetic FROG traces with varying amounts of added noise and distortion are shown in Figure 9. FROG trace (FT) 1 is perfect, undistorted with no noise. FT 2 has only shot noise added; FT 3 has only additive noise added. FT 4 has both shot noise and additive noise. FT 5 and FT 6 are filtered by removing spatial high frequencies in a 2-D FFT. All the FROG traces shown in this figure were subsequently inverted using 200 iterations of the SHG-PCGP algorithm

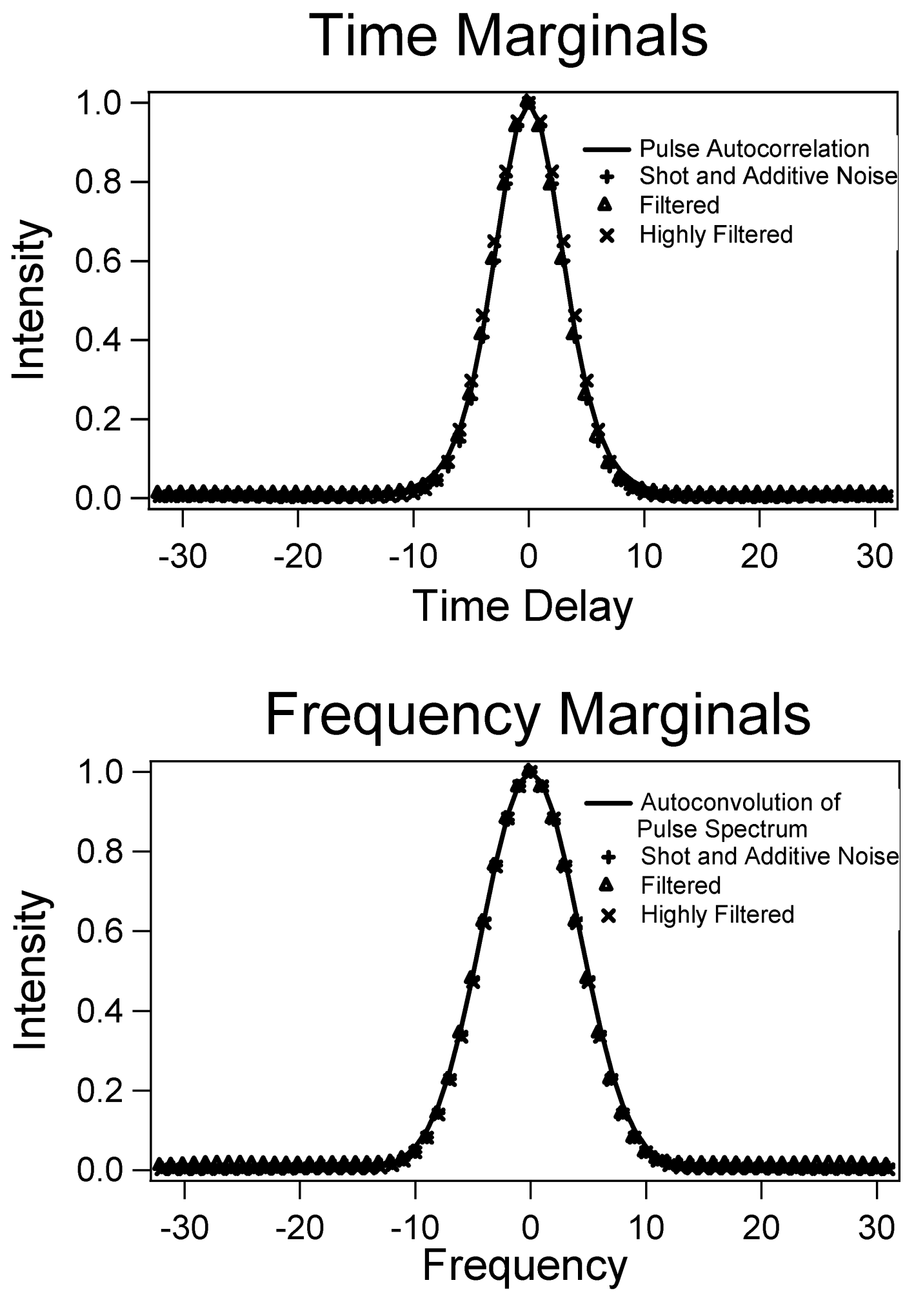

Figure 10. Time and frequency marginals for the pulses retrieved from the synthetic FROG traces shown above. FTs 4-6 are compared with the FT 1. Adapted with permission from [102]. © The Optical Society.

[43], [44]. (See Figure 9 caption for retrieval errors and errors of the retrieved pulses and a more

detailed description of the FTs.) The time and frequency marginals were calculated from the retrieved pulses from FT 4-6 and are compared to the marginals for FT 1 [15]. Agreement is excellent (Figure 10) even though the error of the retrieved pulse is as high as 8.5%.

Figure 11 shows outer product weight plots for FROG traces in Figure 9. FT 1, the perfect FROG trace, has only one non-zero outer product weight. All the other FROG traces depicted in Figure 9 show more than one non-zero outer product weight because added noise and distortions make the FROG trace less ideal. For example, when FT's 2-4, consisting of the original FROG trace corrupted with only noise are decomposed into outer products, the weights (which are always ordered from largest to smallest by the SVD), excluding the first weight, fall on a straight line. Further application of the inversion algorithm cannot change this because noise is uncorrelated to the FROG trace. Indeed, the outer products produced by the SVD are orthogonal to the retrieved pulse and gate pair; a single outer product can no longer describe the FROG trace.

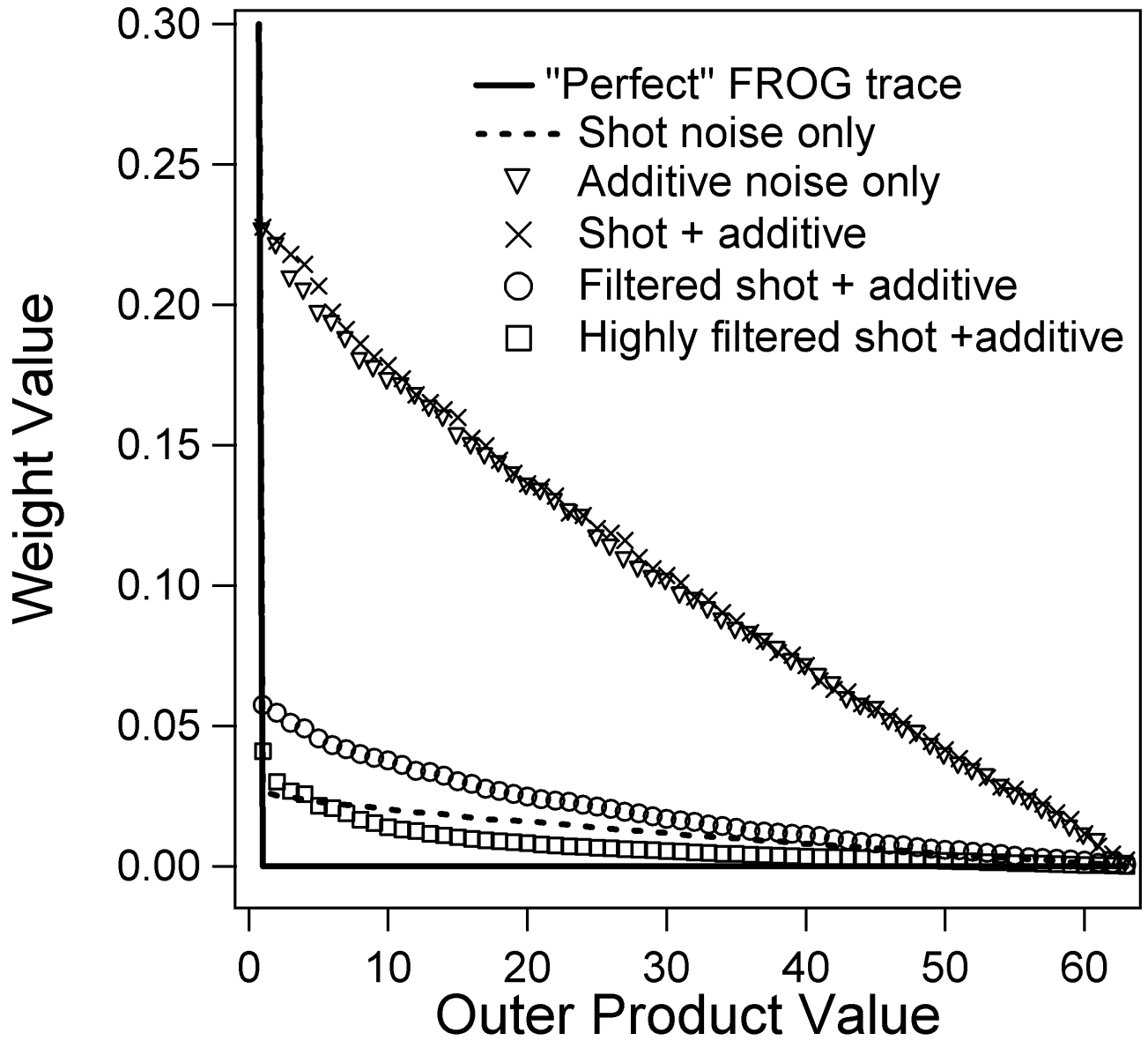

Figure 11. Outer product weight plot from the retrieved, phased FROG traces. The perfect FROG trace is produced from only one outer product. Additive noise adds a significant amount of "noise" outer products, much more than the shot noise. Filtering too aggressively produces the curved distortion signature in the retrieval. Adapted with permission from [102]. © The Optical Society.

Distortions present on the FROG trace are a different matter. Inversion algorithms struggle to fit the distortions with a single outer product pair resulting in a weight plot that is *curved* because the salient features of the FROG trace require multiple outer products to describe them. However, the distortions are correlated to the FROG trace, which means that fewer outer products are needed to reproduce the distortions in the FROG trace than the number that would be required to reproduce uncorrelated noise. This is shown in the weight plots of filtered FROG traces 5 and 6. Although the act of filtering FT 4 removes the "noise" from the FROG trace, it also distorts the FROG trace slightly. The distortion shows up as a curve in the weight plot. Noise can be distinguished from distortions because the weights of the outer product reproducing noise nearly fall on a straight line whereas distortions cause a curve in the weight plot in the region of the largest weights.

Because distortions have contributions spread over relatively few outer products, enticing retrieval algorithms to fit them (FT's 5,6), they contribute relatively little to the FROG trace error, but contribute significantly to the measured pulse error. Indeed, distortion-limited convergence can be quite deleterious, producing relatively low FROG trace errors, but large errors in the reconstructed pulse.

In the next example, we demonstrate the power of our technique by analyzing systematic experimental errors. Figure 12a shows an experimental second harmonic generation (SHG) FROG trace and Figure 12b shows the retrieved pulse and gate obtained by using the SHG-PCGP algorithm [42]–[44]. Because this algorithm uses a soft constraint to force the pulse and gate to be equal, slight asymmetries between the negative and positive time delays in the SHG FROG trace cause a slightly different pulse and gate to be returned. The FROG trace shown in Figure 12a is only slightly asymmetric; as a result, the differences between retrieved pulse and the retrieved gate are minimal (0.4 % difference). The weight plot from an SVD analysis shows a small amount of distortion in the retrieval (Figure 12c). The slight curve in the weight plot is caused by FROG trace asymmetry and some lost bandwidth due to group velocity mismatch in the doubling crystal. If we average the negative and positive time delays to symmetrize the FT in Figure 12a and divide the frequency axis by a Gaussian to enhance the frequency components in the wings and reapply the SHG-PCGP algorithm on the corrected FT to retrieve a new pulse, we see a straight weight plot indicative of noise-limited convergence (Figure 12c, dashed line), and the FROG trace error decreased from 0.4% to 0.25%. There are only negligible differences between the pulse retrieved from the original FT and the corrected FT so we are assured of an accurately reconstructed pulse.

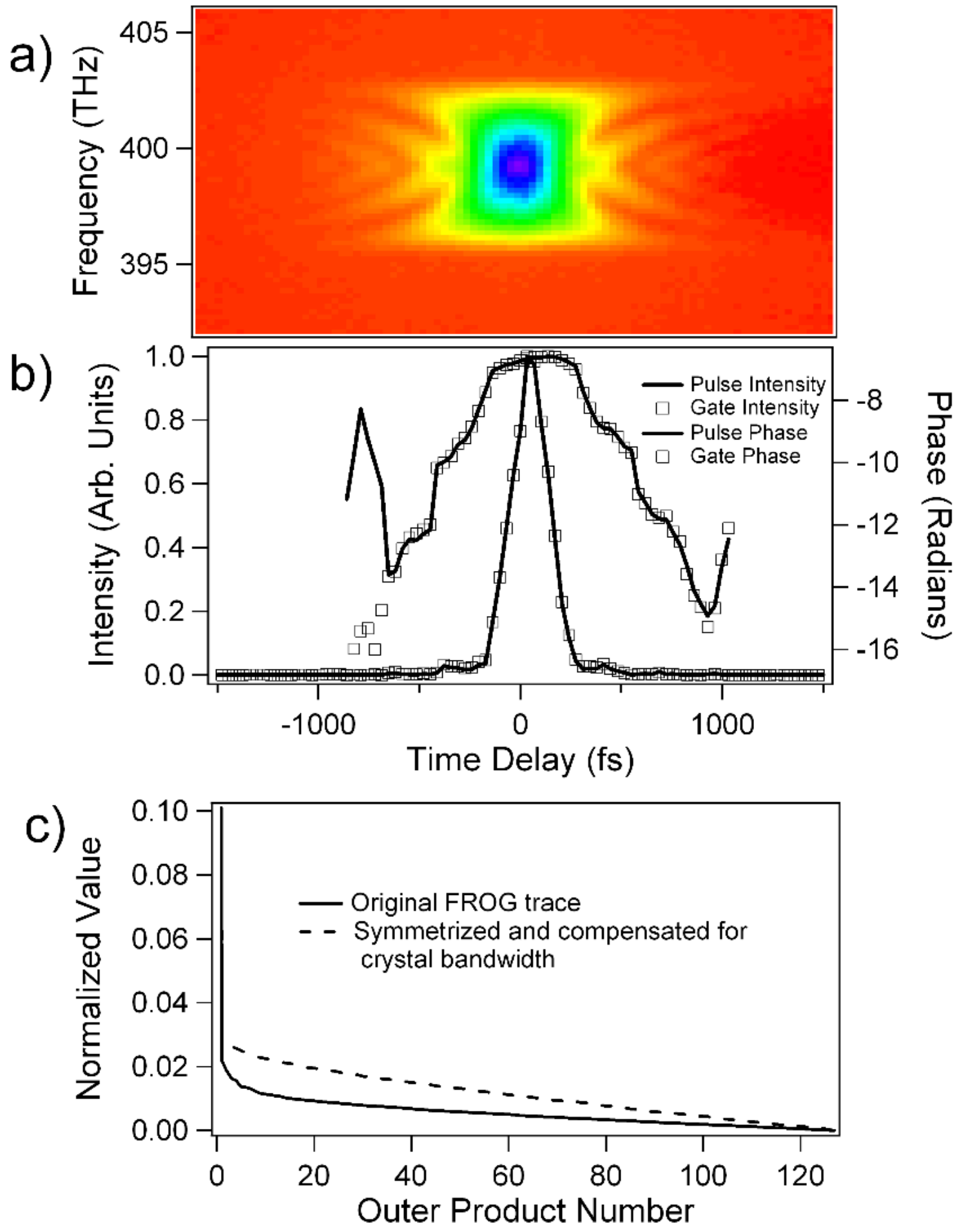

Figure 12. (a) Experimental SHG FROG trace (square root of the intensity). (b) Retrieved pulse. (c) SVD analysis. The slight curvature in the normalized weight plot indicates only a small amount of distortion. If we assume that the distortion is due only to the bandwidth limitation of the doubling crystal and compensate for it, nearly noise limited convergence is achieved. Adapted with permission from [69]. © The Optical Society.

This convergence analysis is quite useful; however, it is not fool proof. It is based on the PCGP algorithm, which requires the wings of the pulse intensity to approach zero. Analysis of FROG traces when the algorithm has not fully converged will also show distortion. Also, any distortion that converts a FROG trace into something that still appears to be a FROG trace cannot be found using this method. Examples include a linearly chirped SHG FROG trace that has been amplitude limited along the time coordinate by an invalid time delay calibration.

## XII. Comparison of the Ptychographic iterative engine and the PCGPA

### A. *Using a benchmark to compare algorithm performance*

In this section, we use a previously published benchmark test [44] to compare the PIE, ePIE, and FROG PCGP algorithms applied to ptychographic data. The benchmark consists of three different sets of test pulses constructed from filtered random noise, random chirp (including self-phase modulation), and compound (multiple) pulses [44]. At least 500 test pulses from each category were used on a 64 x 64 grid. The algorithm was run for 200 iterations for each test pulse, and convergence was assumed to be a FROG trace error of 1.5 x $10^{-3}$. (For the PIE and ePIE algorithms, one iteration was the application of all the spectra in one data set.) For each X-FROG test, an independent pulse and gate was constructed using the model under test. The algorithms were started using the known constructed gate, and an initial guess of a noise modulated Gaussian for the unknown pulse. Both the PIE algorithm and the PCGP X-FROG algorithms converged >99% of the time, which is much better than any FROG retrieval algorithm including the polarization-gate FROG algorithm [44].

The algorithms were all programmed in MATLAB. No attempt was made to optimize the algorithms for speed for these tests. The PIE algorithm iteration speed improved as spectra were dropped.

The PIE algorithm performance depended on the alpha parameter used. With $\alpha = 1$, the PIE had 100% convergence for all test cases while the PCGP X-FROG algorithm converged 100% of the time for both the filtered random phase test and the compound pulse test and 99% of the time for the random polynomial chirp test. When $\alpha$ was varied randomly between 0.05 and 0.1, the PIE algorithms performance dropped to 94% convergence on the random polynomial chirp test. The dependence on the alpha parameter is an important consideration for the PIE algorithm as we shall see in the sections that follow. When the PIE (and ePIE) algorithms were compared directly to the PCGP algorithm, alpha parameters producing the best performance for the PIE and ePIE algorithms were used.

### *B. Second Harmonic Generation FROG*

Because the pulse and the gate are equal in SHG, it is a special case. Thus, a blind inversion is not required, and the ePIE algorithm in this case can be thought of as a cascaded PIE algorithm where the pulse and the gate are set to be equal. Like the PIE algorithm, the X-FROG algorithm can be used in succession for SHG FROG, and a reasonably good SHG inversion algorithm can be obtained, albeit slightly less robust than the original SHG PCGP algorithm.

The SHG PCGP algorithm significantly outperformed the PIE/ePIE algorithm overall for SHG FROG retrievals when using full FROG data sets. Using unoptimized MATLAB code, the PCGP SHG algorithm ran approximately twice as fast as the PIE/ePIE SHG algorithm when a full data set was used. The performance of the PIE/ePIE algorithm was found to be dependent on the alpha parameter with the best performance found to randomly vary between 0.05 and 0.1. Overall, for the 3 test cases, the ePIE algorithm was found to have a convergence rate of 68.0%. The convergence rate of the cascaded X-FROG algorithm and the original SHG algorithm was found to be 65.3% and 73.6%, respectively.

Interestingly, the ePIE algorithm slightly out-performed the PCGP algorithms in the multiple pulse test where its convergence was 64.5% and the cascaded X-FROG convergence was 59.2% and the original SHG FROG's convergence rate was 58.9%. The ePIE algorithm’s performance on the filtered random noise retrieval test improved to 71.9% when the alpha parameter was set to vary randomly between 0.1 and 0.5; however, the overall convergence rate dropped to 64.1%. The PCGP based algorithms performed the best in the polynomial chirp test, converging 71.2 % and 90.2% for the cascaded X-FROG algorithm and the original SHG algorithm, respectively. The ePIE algorithm converged 69.7% of the time when the alpha parameter was optimized. Table 1 summarizes these results. Test case 1 is the filtered random noise test; test case 2 is the random chirp test, and test case 3 is the multiple (compound) pulse test. (Note: the random number generator used for test 1 was the MATLAB rand() function that generated random numbers uniformly between 0 and 1. The random chirp test and the multiple pulse test used the MATLAB random number function randn(), which produces a normal distribution of random numbers with a mean of 0 and a standard deviation of 1. These functions are significantly different than the Python random number generators.) Percentages in parentheses are for ($\alpha$ = 0.1 - 0.5).

**Table 1. Second Harmonic Generation Reconstruction Performance**

| SHG Retrieval Performance (Percent Success)<br>Best Case ($\alpha$ = 0.05 - 0.1 and ($\alpha$ = 0.1 - 0.5) randomly varied for ePIE)<br>Full FROG Data Set | | |
|---|---|---|
| Algorithm | Total Percentage | Test Case Success |
| Sequential/cascaded PIE (ePIE) | 68.0% (64.1%) | 1) 69.8% (74.8%)<br>2) 69.7% (62.5%)<br>3) 64.5% (55.1%) |
| Cascaded X-FROG | 65.3% | 1) 65.6%<br>2) 71.2%<br>3) 59.2% |
| Original SHG | 73.6% | 1) 71.9%<br>2) 90.2%<br>3) 58.9% |

*C. The effect of dropping spectra*

The greatest strength of the ePIE and PIE algorithms is that retrieval does not require a complete FROG set of data while the original PCGP algorithm does. However, dropping spectra does reduce conversion rates. When one of the inputs is known, the algorithm is surprisingly robust, converging over 94% of the time when only the center 25% of the data was used. The PIE algorithm with a known gate still worked as well as the X-FROG PCGP algorithm when using only half the data (every other spectrum). The same was not true in the SHG case, however. The algorithm performance dropped quickly when data was omitted. Still, in some cases, the algorithm could be useful with judicious data selection, but never as robust as the SHG PCGP algorithm when using a full FROG data set for general purpose reconstructions.

Table 2 shows the performance of the PIE algorithm (known, fixed gate) as spectral data is removed for both $\alpha$ randomly varied between 0.05 and 0.1 and $\alpha$=1. An entire FROG data set was constructed in each data set. Spectra were selected out for the algorithm, but the entire data set was compared to better compare with a complete data set.

**Table 2. Effect of dropping spectral data from known gate – unknown pulse retrievals.**

(Reprinted with permission from [97])

| Ptychography PIE performance percent success<br>Reduced Data: Known Gate – Unknown Pulse | | |
|---|---|---|
| | $\alpha$ = 1 | $\alpha$ = 0.05 - 0.1 |
| Every other spectrum | 100% | 94% |
| Every 4$^{th}$ spectrum | 99.7% | 67% |
| Every 8$^{th}$ spectrum | 32% | 8.3% |
| Every 16$^{th}$ spectrum | 2% | 0% |
| Center 3 | 5.7% | --- |
| Center 7 | 61.3% | --- |
| Center 17 | 98.3% | --- |

Table 3 shows the comparative performance of the algorithms when data are removed for an SHG retrieval (pulse and the gate). The ePIE algorithm is used for the ptychography test. For the SHG FROG algorithm, any unknown data is linearly interpolated from surrounding data. The robustness of the PCGP SHG algorithm bests the ePIE algorithm only when data are regularly placed, and a large amount is present. The PCGP algorithm has difficulties when data becomes sparse and are randomly removed. The alpha parameter for the SHG ptychography algorithm requires tuning and may depend not only on the amount of data present, but also on the pulse.

**Table 3. Comparison of the SHG Ptychography and the PCGP SHG algorithm with data removed. (Reprinted with permission from [97])**

| SHG Ptychography $\alpha$ = 0.05 - 0.1 ($\alpha$ = 0.1 - 0.5) vs. PCGP SHG using reduced data* Percent Success all Test Cases | | |
|---|---|---|
| | Cascaded PIE (ePIE) | PCGP SHG |
| Every other spectrum | 32.3% (38%) | 56.0% |
| Every 4th spectrum | 22.0% (28%) | 26.7% |
| Every 8th Spectrum | 2.0% | 0% |
| Every 16th Spectrum | 0.0% | 0% |
| Random 10% | 2.0% (3.2%) | 0% |
| Random 25% | 16.0% (20.5%) | 5.67% |
| Random 33% | 23.3% (27.8%) | 6.67% |
| Random 50% | 30.3% (43.1%) | 26.7% |

*missing data for the PCGP algorithm is linearly interpolated. (Reprinted with permission from [107])

*D. Noise Performance*

Noise performance is tricky to evaluate in statistically meaningful ways; i.e., methods that are not anecdotal. Three types of noise were added to the FROG trace magnitude. The first was random multiplicative noise with a mean of 4 % of the value of the pixel, and normally distributed with a full-width half maximum of 1. The second was shot noise at 8% of the square root of the value of the pixel, with the same distribution. The third was additive noise with a mean of about 0.25% of the maximum. The added noise was not meant to be an extreme level, but rather about the level noise expected from a good quality 8-bit camera used at full dynamic range.

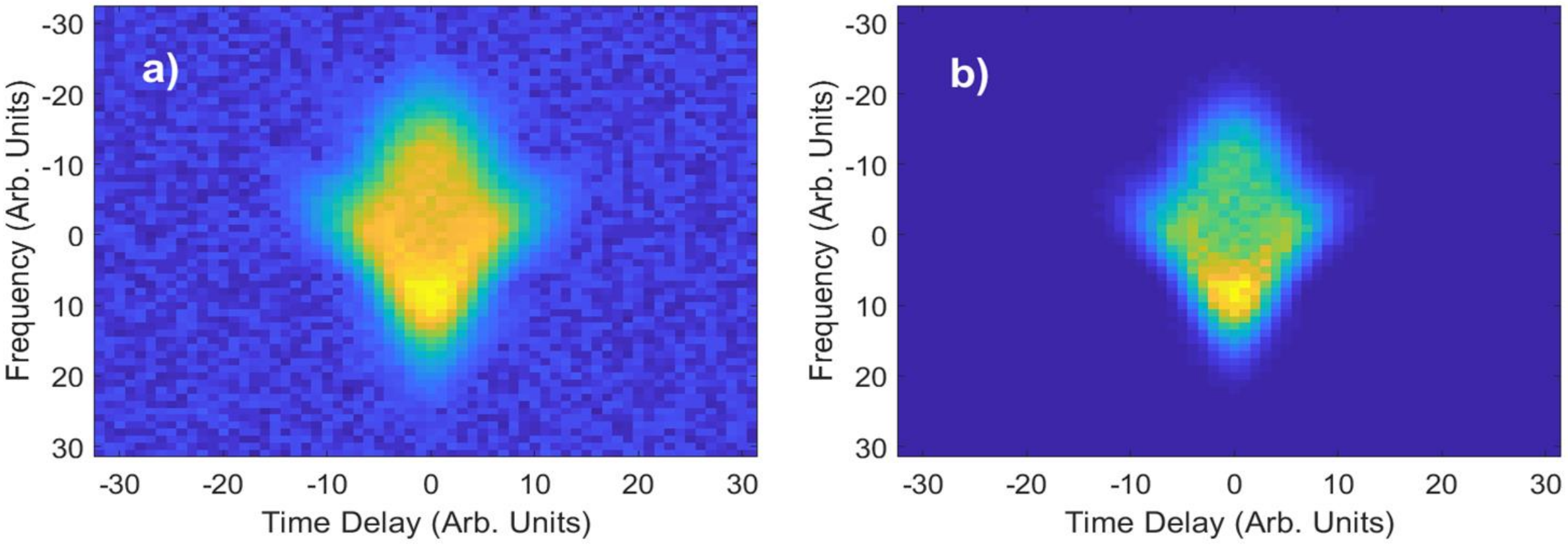

Figure 13. Figure showing an example noisy FROG trace representative of the noise level used in the comparison. a) shows the magnitude while b) is shows the intensity.

Figure 13 shows an example of "noisy" FROG data used in the comparison. Figure 13a shows the magnitude while Figure 13b shows the intensity. In the intensity display, noise in the background is difficult to see; however, since the algorithm input is the square root of the intensity, the magnitude is more indicative of the input into the algorithm. Figure 14 shows the pulse retrieved from the FROG data shown in Figure 13. The convergence was monitored, and the 1% and 2% in the retrieved temporal intensity are compared to the actual pulse. Part a) shows the temporal intensity, part b) the temporal phase, part c) the spectral intensity, and part d) the spectral phase. Even with only 1% temporal intensity error, the pulse is slightly longer and the spectrum is slightly narrower. However, both the temporal and spectral phase agree reasonably well. The difference between the 2% intensity and the actual pulse is significant in terms of intensity. For purposes of this comparison, we deem an average rms error/pixel of 1% in the intensity as the highest tolerable error. In many cases, 2% might be acceptable depending on pulse shape, but in general, anything significantly greater than 1% for these data sets are not useable. Shown in Figure 15 are plots of the percent convergence as a function of the retrieved pulse intensity error for the PCGP X-FROG algorithm and the PIE algorithm (known gate). To benchmark the performance, the percent retrieved as a function of pulse intensity error is plotted. Intensity temporal overlap was optimized and because SHG FROG has a time reversal ambiguity, both temporal directions were compared with the original pulse. The lowest reported error was used. No benchmark for phase was attempted. Interestingly, while both algorithms show virtually 100% convergence at errors of less than 1%, the PIE algorithm has better than 80% convergence for pulse intensity errors of less than 0.5% while the PCGP X-FROG algorithm has a convergence in this error range of only about 40-50%.

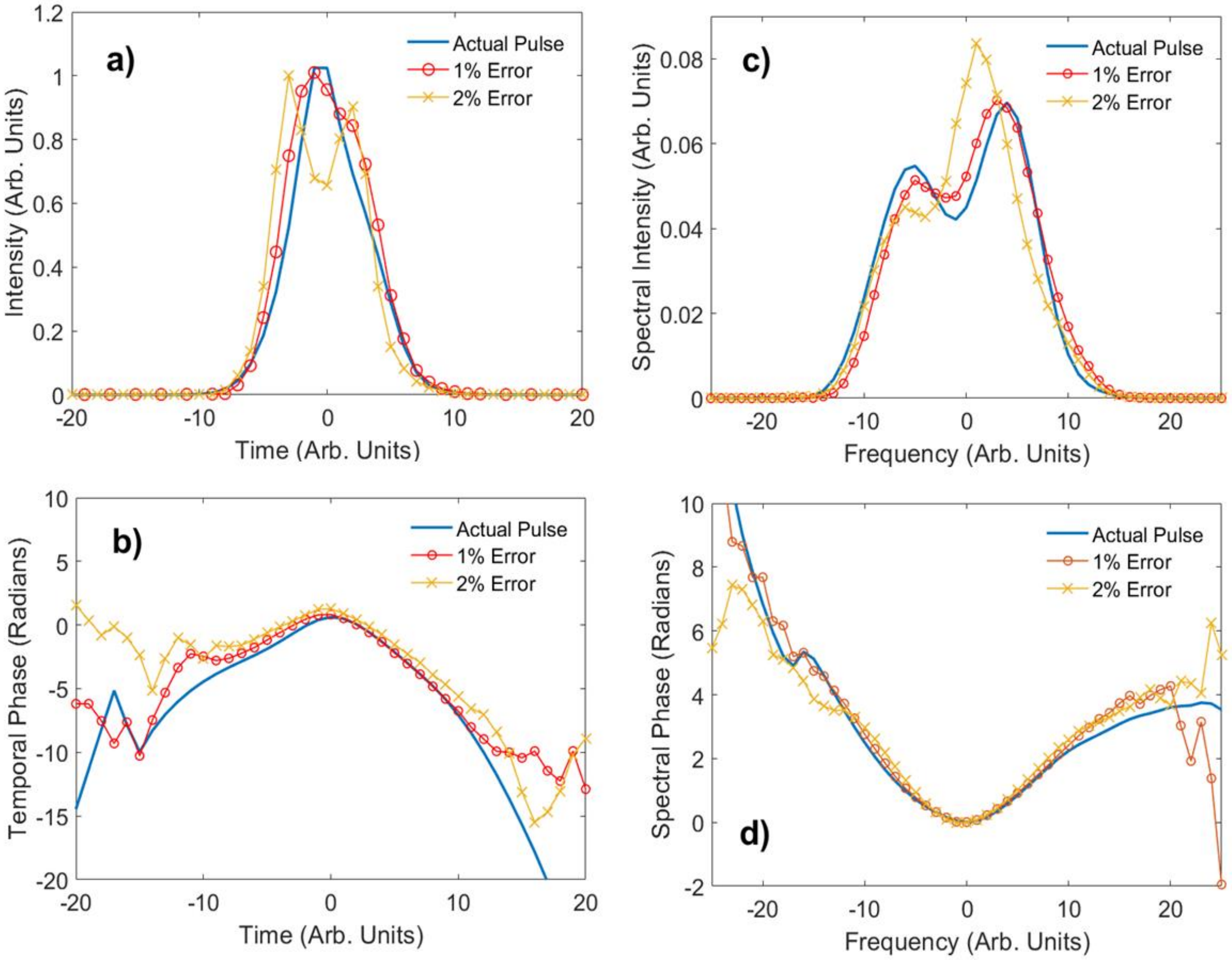

Figure 14. Figure comparing the actual pulse with partially retrieved pulses with 1% and 2% intensity errors in an example pulse. a) compares the temporal intensity. b) compares the retrieved temporal phase. c) compares the spectral intensity and d) compares the spectral phase. While the temporal intensity with 1% is slightly longer than the actual pulse, the temporal and spectral phases are very close to the actual pulse. The temporal intensity and the spectral intensity of the pulse with 2% intensity error is highly distorted.

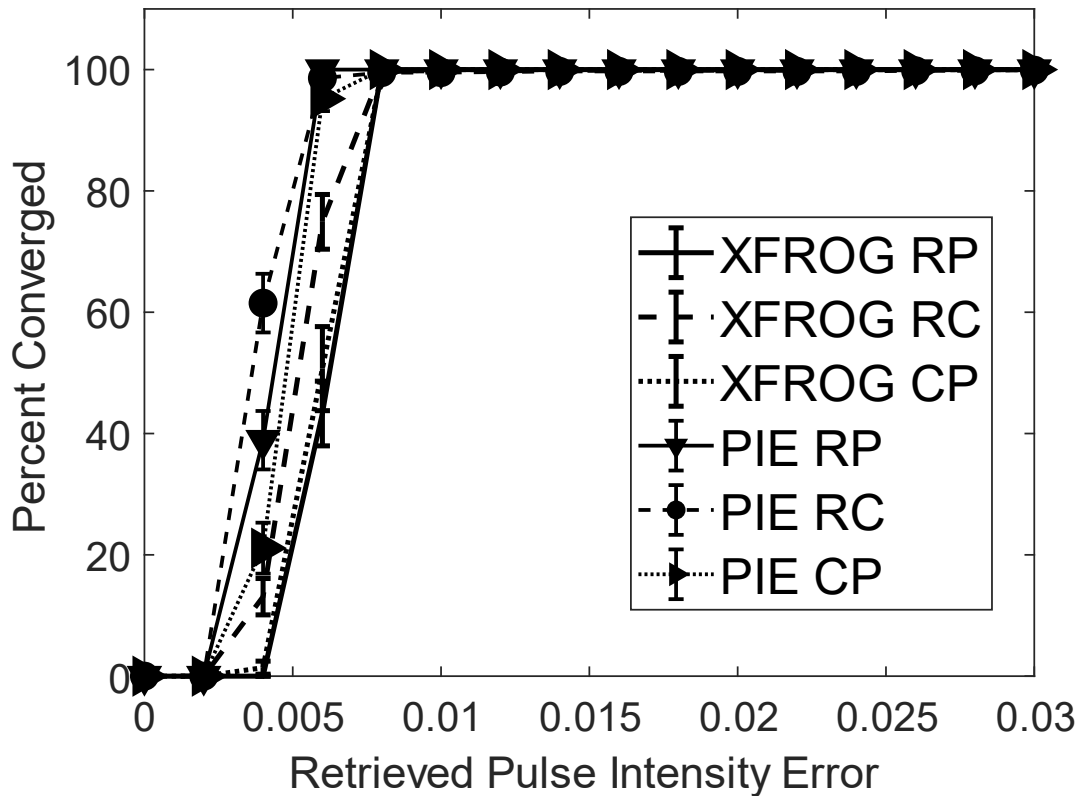

**Figure 15.** Comparison of retrieval accuracy between the X-FROG PCGP algorithm and the PIE (known gate) algorithm. While nearly all the test cases reached at least 1% error, many more of the PIE retrievals reached a temporal intensity error of 0.5% or better. Reprinted with permission from [97].

Figure 16 shows the noise performance for the PIE (cross-correlation retrievals where one of the fields is known) when less than a full FROG data set is used. Figure 16a show the noise performance when every $N^{th}$ spectrum is used. Performance does not degrade significantly until every $8^{th}$ spectrum is used on a 64-point retrieval (7 out of 8 spectra are dropped). For every $8^{th}$ spectrum and every $16^{th}$ spectrum, no pulses were retrieved at an acceptable error level.

Figure 16b shows the performance using only the center portion of the FROG trace when noise is added. We can see that the performance degrades from a full data set and from even every other spectrum in a full data set when the center 17 spectra are used. Performance deteriorates rapidly as fewer center spectra are used. When very few center spectra are used, the wings of the FROG trace are not sampled, which may be reason for poorer convergence [103].

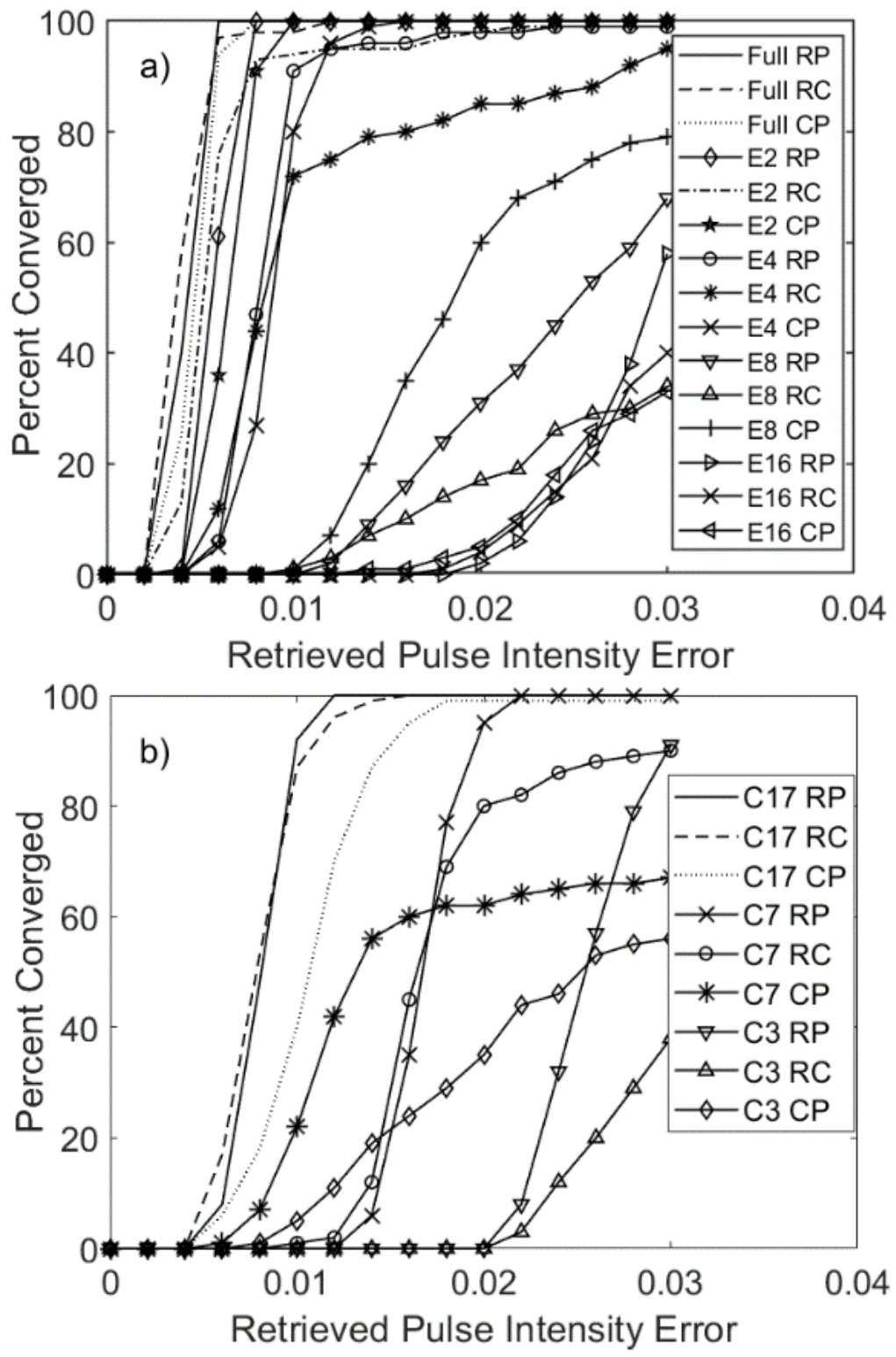

Figure 16. Plots showing noise performance of the PIE when data is removed. a) are plots of percent retrieved as a function of error for regularly spaced reduced data sets comparing full data sets (Full), every other spectrum (E2), every 4th spectrum (E4), every 8th spectrum (E8), and every 16th spectrum (E16) for each of the data sets (RP= Random pulse, RC = Random chirp, and CP = Compound pulse). b) are plots of noise performance when only the center 17 (C17), center 7 (C7), and center 3 (C3) spectra are used in the 3 different test cases. Reprinted with permission from [97].

Even though an $\alpha$=1 works best for no noise and partial data sets for the PIE (known gate), an $\alpha$ randomly varied between 0.05 and 0.1 works best when partial data sets centered on the FROG trace are used when noise is added.  For partial noisy data sets using every N[th] spectrum, choice of α was not critical.

Shown in Figure 17 are plots of the retrieved pulse temporal intensity error for the three different pulse test cases for the original SHG PCGP algorithm (a), a cascaded X-FROG algorithm (b), ePIE algorithm used for SHG retrieval $\alpha \in [0.05, 0.1]$, and an SHG ePIE with $\alpha \in [0.1, 0.5]$.

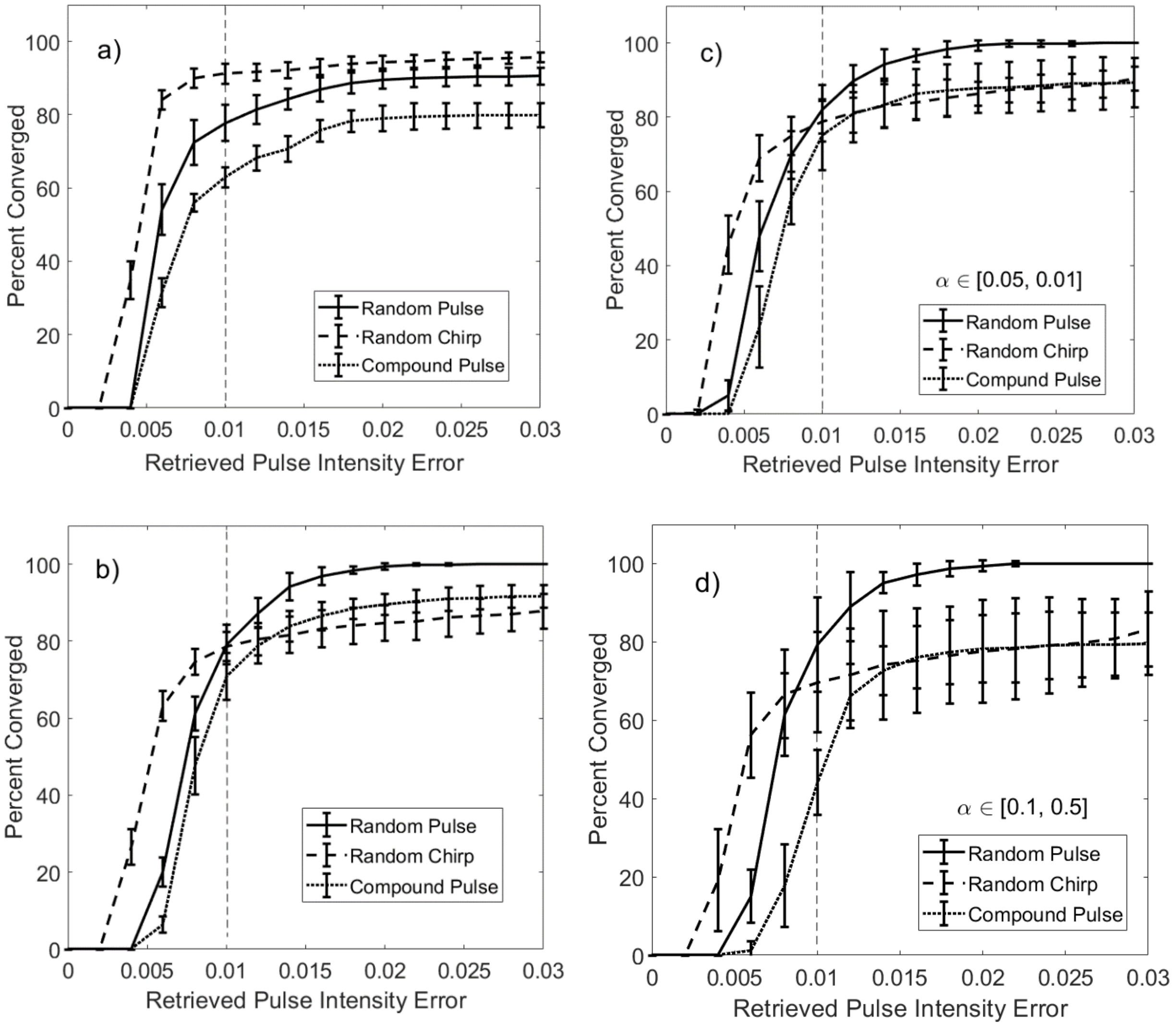

Figure 17. Plots showing the noise performance of the original SHG PCGP algorithm (a), a cascaded X-FROG algorithm set up for SHG inversion (b), the ePIE SHG algorithm with α = [0.05, 0.1] (c), and the ePIE SHG algorithm with α = [0.1, 0.5] (d).  The percent retrieved is plotted as a function of intensity error in the retrieved pulse.  At 1% error, the pulse intensity is still a good, quantitative representation of the actual pulse. These tests utilize a full FROG data set.

The original SHG algorithm test results (Figure 17a) are a good benchmark because this algorithm is used successfully in commercial FROG devices with virtually no stagnation problems. At 1% pulse temporal intensity error, for the original SHG PCGP algorithm, over 90% of the random chirp pulses were retrieved at better than 1% intensity error. About 78% of the random pulses were retrieved and nearly 70% of the chirped, compound pulses were retrieved with intensity errors of 1% or less. The error bars show the standard deviation of the percent converged for a given retrieved temporal intensity error.

In Figure 17b, the results for the cascaded X-FROG SHG algorithm are shown to work well. This is probably because it is a cascaded generalized projections algorithm, which are predicted to work well in general [105], [106]. Figure 17c and d show the noise converge of the ePIE SHG to be close to the SHG PCGPA. When $\alpha \in [0.05, 0.1]$, the results are nearly identical to the cascaded X-FROG algorithm. Indeed, an optimized ePIE SHG algorithm overall performs nearly as well as the PCGP algorithm.

The ePIE SHG algorithm performed better for the lower values of $\alpha$ in the random chirp and the compound pulse test sets when a full FROG data set was used. In general, we noticed that the ePIE tends to perform slightly better on incomplete data sets when larger values of $\alpha$ are used. It is also interesting to note that the structure of the corrected ePIE SHG algorithm noise performance plot (a) is remarkably like the cascaded X-FROG algorithm, which becomes mathematically similar to the ePIE near convergence. Tuning of the $\alpha$ parameter appears desirable in real world applications.

It is instructive to observe that no difference in the performance of the ePIE SHG algorithm is seen between applying intensity constraints sequentially versus applying them randomly if only test sets 1 and 2 were used. Thus, we conclude that benchmarks, provided they test a wide range of types of pulses, can be helpful for algorithm development, comparisons (relative), and, of course, debugging of algorithms [107]. Absolute algorithm testing using this benchmark is difficult because different programming languages use different definitions for random number generators, and performance also depends on the initial guess.

## *XIII.* TIME RESOLVED OPTICAL GATING – SONOGRAM RETRIEVALS

Ptychography will not just work for time gated (spectrographic) data acquisition. Frequency gated data acquisition can be retrieved as well. In this work, we refer to this type of data acquisition as a sonogram. Spectrograms and sonograms are closely related. Rather than displaying the arrival time of frequency filtered pulses (as a sonogram does), a spectrogram displays the frequency content of time slices of a pulse. Thus, the spectrogram and the sonogram contain the same information about the constructing functions. The difference results from how the data is taken. Figure 18 shows conceptually how sonogram data is taken. The input is frequency filtered, and the time profile of the result is recorded as a function of the center frequency of the filter. The resulting data is the form of:

$$S_{\tau\omega}^{sonogram}(\tau,\omega) = \left|\int_{-\infty}^{\infty} \mathrm{P}(\Omega)\mathrm{G}(\omega-\Omega)e^{-i\Omega\tau}d\Omega\right|^2, \tag{53}$$

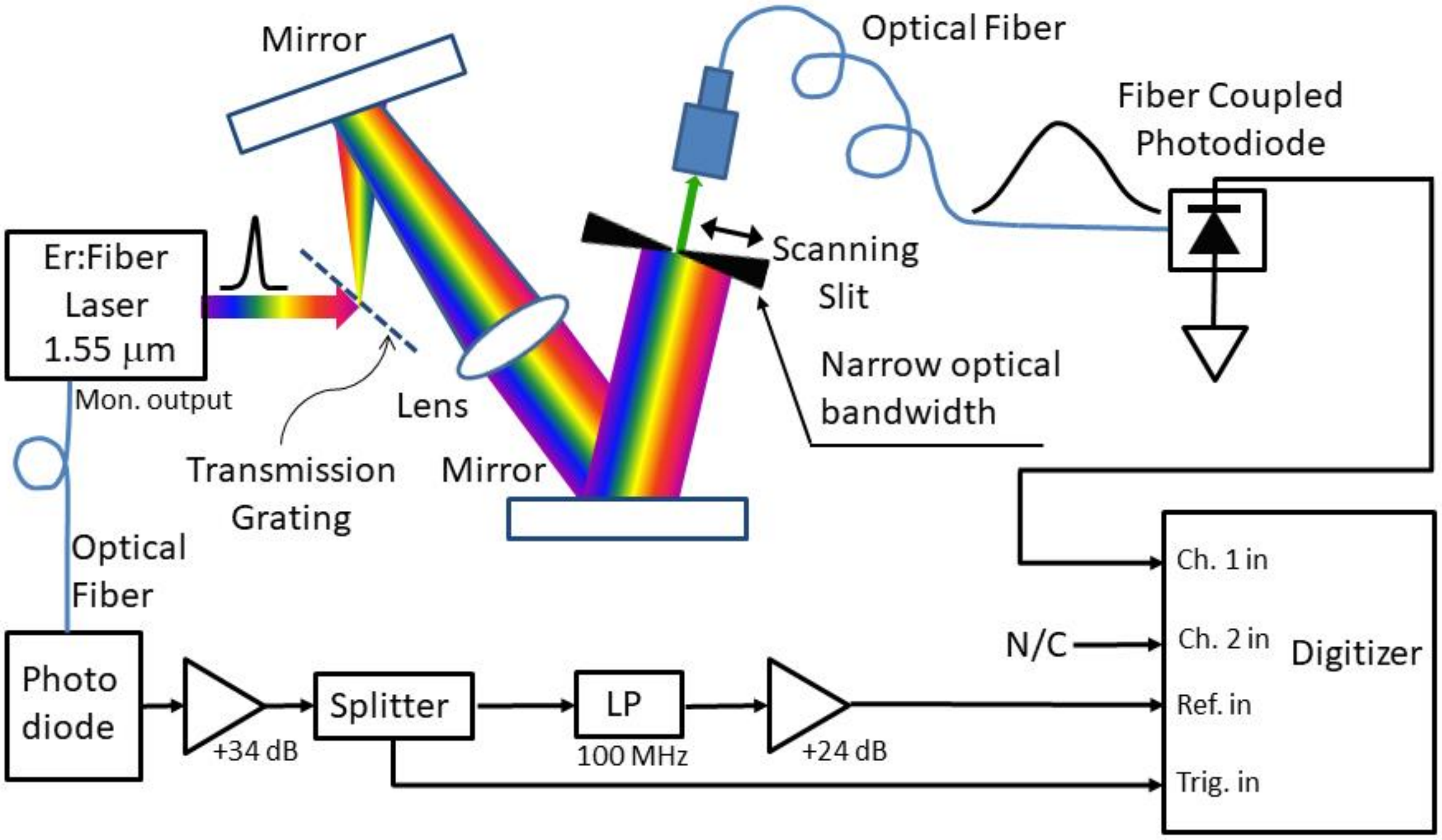

Figure 18. Experimental schematic of a time resolved optical gating experimental apparatus using eletronic detection.

The equation is like, but not identical to Equation 10, which reverses the direction of the gate. Because a square-law detector is used, the measured intensity is given by:

$$I_{TROG}(\tau,\omega) = \int |G_{DET}(t,\tau)|^2 \left(\left|\int_{-\infty}^{\infty} \mathrm{P}(\Omega)\mathrm{G}(\omega-\Omega)e^{-i\Omega\tau}d\Omega\right|^2\right) dt \tag{54}$$

In the case of Chilla and Martinez [13], the frequency filter is a slit such that

$$G(\omega) = \begin{cases} 1, \text{when } |\omega-\Omega| \le \frac{1}{2}\Omega_0 \\ 0, \text{when } |\omega-\Omega| > \frac{1}{2}\Omega_0 \end{cases} \tag{55}$$

where the input pulse is cross correlated with the frequency filtered pulse to determine the time arrival of the frequency filtered pulse. This technique is known in the literature as frequency-domain phase measurement (FDPM). As long as the frequency filter, *G(ω)* is narrow enough such that the frequency filtered pulse is much longer than the input pulse, then $|G_{DET}(t,\tau)|^2 \rightarrow \delta(t-\tau)$, and the detected signal is given by Equation 53. Because Chilla and Martinez used the time arrival of the frequency filtered pulse, the phase across the frequency filter slit is assumed to be linear; therefore, a narrower slit can improve phase reconstruction [13]. Wong and Walmsley showed that a phase retrieval algorithm could be used to relax the slit requirements [64].

An important variant of TROG was developed by Koumans and Yariv called dispersion propagation (DP) TROG, or DP-TROG [34], [35]. A schematic diagram is shown in Figure 19. The setup consists of a phase stationary filter (the disperser) followed by an amplitude nonstationary filter (an autocorrelator). The disperser can be any method of dispersion that is

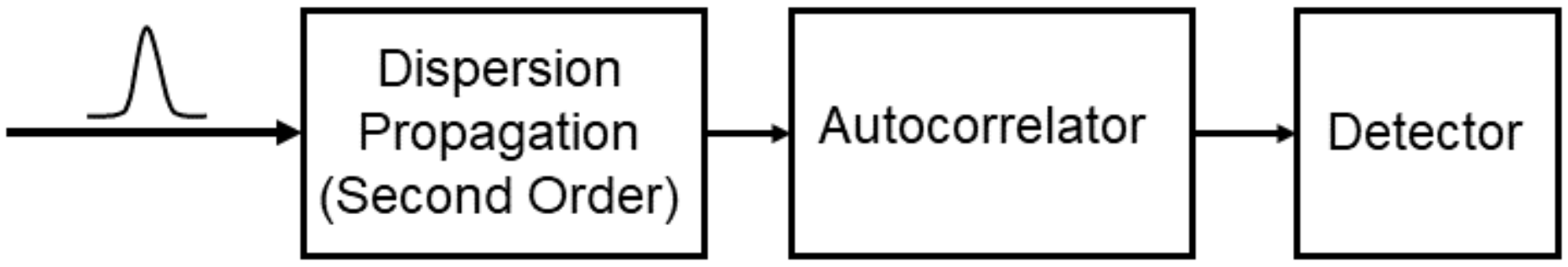

Figure 19. Schematic of the steps for a dispersion propagation time resolved optical gating apparatus. An autocorrelation is measured for varying amounts of positive and negative dispersion added the pulse. The autocorrelation is Fourier transformed and remapped to a sonogram.

primarily second order. The gate function, G(ω) is then a function of both ω and β, the dispersion parameter:

$$G(\omega, \beta) = e^{i\beta\omega^2}. \tag{56}$$

Through the use of the convolution theorem for the Fourier transforms, Koumans and Yariv show that the autocorrelation data as a function of dispersion can be Fourier transformed and remapped to a sonogram where the P(ω) is the frequency domain representation of the pulse and the gate, G(ω) is the complex conjugate of P(ω). This technique has been shown to work quite well, and errors induced by $3^{rd}$ order material dispersion have been explored [66], [67].

Like spectrograms, the gate and the pulse can be arbitrary and have similar ambiguities. Thus, blind retrievals can be problematic (see section XIV below). For sonogram measurements where the gate is known and the X-FROG or PIE retrieval are used, reconstruction quality is a weak function of the gate shape as long as the gate is well characterized.

## A. *Single-Shot Time Resolved Optical Gating with Electronics*

Ultrafast laser pulse measurements are largely optical domain methods. However, optics can be quite limiting when moving from the idea of pulse measurement (short duration) to waveform measurement when large time durations (>1 ns) need to be measured at high bandwidths. This is largely a result from practical considerations such as the use of imaging and spectrographic optics in the optical implementation of pulse measurement. On the other hand, electronics provides a means to measure arbitrarily long waveforms. However, even modern electronics, with bandwidths up to over 120 GHz, is still far from being able to measure optical pulses of any duration shorter than a few picoseconds. Ideally, combining the speed/bandwidth advantage of optical methods with the long-time duration (low bandwidth) of electronics would be ideal.

While typical optical frequencies are in the petahertz region, with bandwidths in the terahertz region, optical structures such as spectrometers and telecommunications parts can reduce optical bandwidths to less than 1 terahertz. The 100 GHz speed of electronics is an especially important milestone. At this point, electronic bandwidths are commensurate with optical bandwidths that can be achieved using physical optics. Spectrometers can be made with off the shelf components that routinely have spectral resolutions of 0.5 nm at 1550 nm, which has a bandwidth of ~60 GHz, well within the 100 GHz electronics bandwidth available.

We demonstrate a hybrid optical and electronic system for measuring optical waveforms up to several nanoseconds in duration at bandwidths up to about 2 THz at single-shot rates up to 1 MHz. Indeed, we demonstrate that electronics have the required precision and accuracy to measure pulse characteristics that are far beyond the limits of the electronic bandwidth—more in line with the analog-to-digital convertor's jitter specification. The technique uses sonogram decomposition (time arrival of frequency components of the pulses) of the waveform to deconstruct the waveform into multiple frequency bins that span about 4 THz. Each bandpass reduces the bandwidth to less than 40 GHz - levels measurable with current high-speed electronics. This work uses a 20 GHz oscilloscope to measure the filtered waveform. The filtered waveform from each frequency component was correlated with an average filtered waveform to determine the relative delay between each filtered frequency bin (group delay). A delay measurement precision for each frequency of about 1 ps was obtained in a single shot. Using this device, we measured shaped sub-400 fs optical pulses. The group delay calculated from the phase applied by the pulse shaper matched the measured group delay.

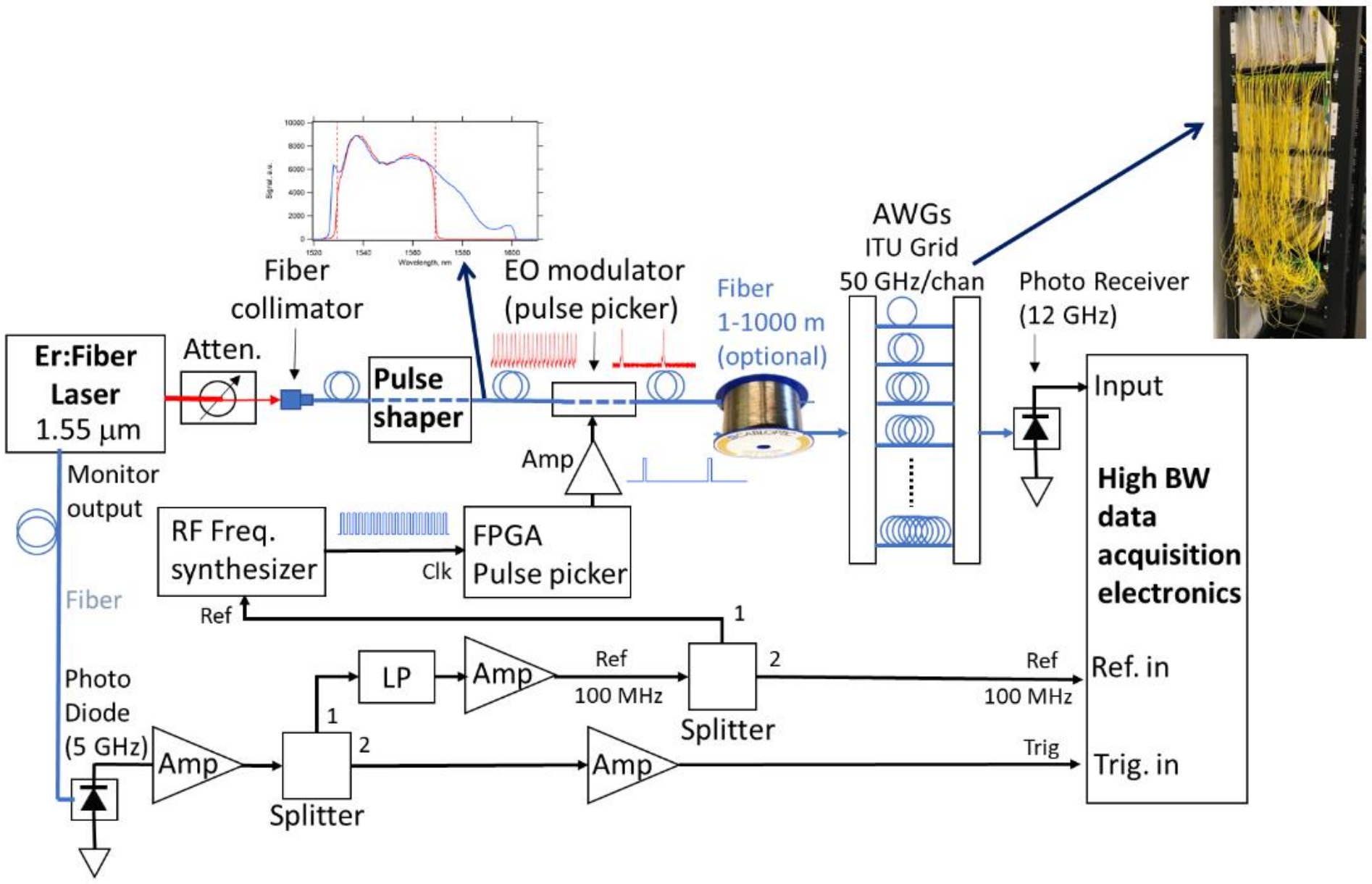

**Figure 20.** Schematic diagram of the high-speed opto-electronic pulse measurement system. The Er:fiber laser is the optical source and acts as the master clock. A pulse shaper is used to change the intensity and phase of the optical pulse. The monitor output of the laser is used to synchronize the data acquisition. The insets (left) show the spectral section of the mode-locked laser used and (right) a photograph of the fiber multiplexer.

Figure 20 shows a schematic diagram of the experimental setup. A mode locked Er:fiber laser (Menlo Systems TC-1550) generates ~80 fs pulses at a repetition rate of 100 MHz. The laser output is attenuated (using a combination of half-wave plate and linear polarizer), coupled to a single mode fiber and passed to a fiber coupled Finisar WaveShaper 1000A programmable optical filter (pulse shaper) and a pulse picker. The pulse shaper is used to modify the intensity and spectral phase of the laser pulse. The home-made pulse picker is used to eliminate the overlap of the pulse trains coming out from the AWG's that can be spread in time over hundreds of nanoseconds. The output of the pulse picker (typically passing one out of 48 laser pulses coming at a 100 MHz rate) can optionally be coupled to an SMF-28 single mode optical fiber cable (length 1–1000 m) to introduce additional dispersion. The light is then passed to a pair of 96-channel 50 GHz athermal array waveguide gratings (AWG's). The first AWG spectrally disperses the incoming optical signal into up to 96 channels that are 50 GHz apart. The individual channel outputs are connected to the corresponding channels of the second AWG with single-mode fibers of different lengths with 1 m increments thus introducing different optical delays that are ~5 ns apart. The second AWG recombines the spectral channels into a single output comprising a train of pulses separated by ~5 ns that correspond to each individual spectral channel. With 96 channels implemented, the pulse train is almost 500 ns long. The AWG's output is passed to a 12 GHz photoreceiver (Newport 1544) and is digitized using a 20 GHz 50 GSPS digital serial analyzer/oscilloscope (Tektronix DSA 72004C).

Synchronization with the laser is done by monitoring the monitor output of the Er:fiber laser with a 5 GHz photodiode. The output is amplified and split. One leg is filtered to provide a 100 MHz reference, which can be used to synchronize the RF generator with the oscilloscope. The unfiltered portion is used as a trigger for the oscilloscope.

Figure 21 shows the transmission spectrum of the fiber multiplexer. The peaks correspond to the International Telecommunication Union (ITU) frequency grid for 50 GHz spacing as selected from an Athermal Array Waveguide Grating (AAWG). The optical spectrum was separated into these channels, and each channel had an increasing amount of delay. The temporal waveform (b) is shown in the plot just below the transmission spectrum (a). The nominal time spacing between each peak is about 5 ns. The bottom plot (c) shows an expended view of the temporal waveform at the start and end. Because only the first pulse of the temporal waveform triggers the oscilloscope, after passing through 100 m of fiber, the pulses are shifted to early times (blue trace) as compared to going through no fiber (red trace). (Optical fiber has negative dispersion at 1550 nm.)

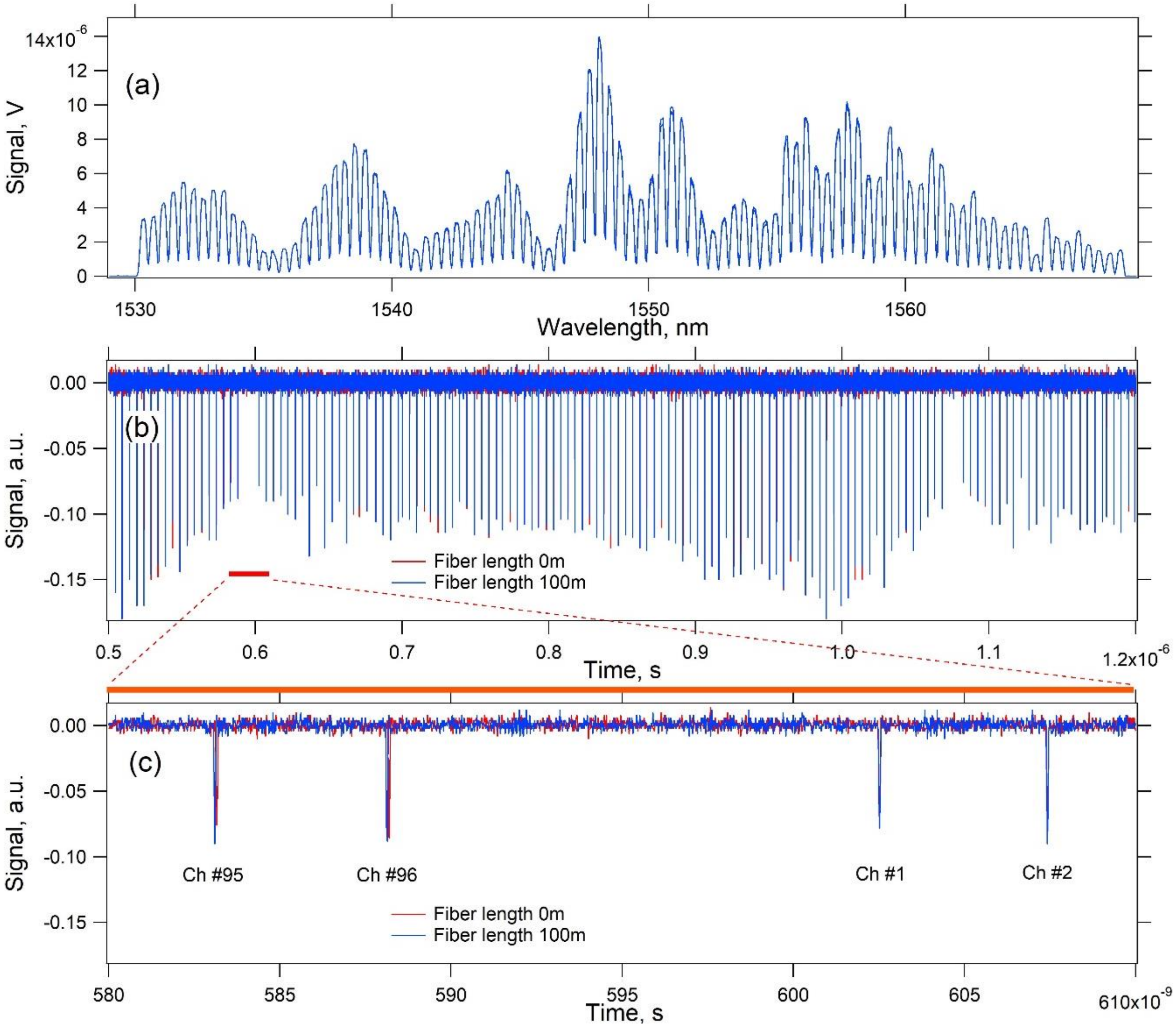

**Figure 21**. Transmission spectrum (a) and the recorded waveform (b). An expanded view of the waveform (c) shows the waveform when dispersed by no fiber (red) and 100 m (blue).

To determine the arrival time of each pulse, the impulse response is constructed using multiple waveforms sampled at different times. The result is an impulse response effectively sampled at 1 ps time intervals (Figure 22). This impulse response is correlated with every pulse in the temporal waveform to determine the arrival time. Because each delay can be slightly different, each time arrival must be calibrated by inputting a transform limited pulse to the pulse measurement system.

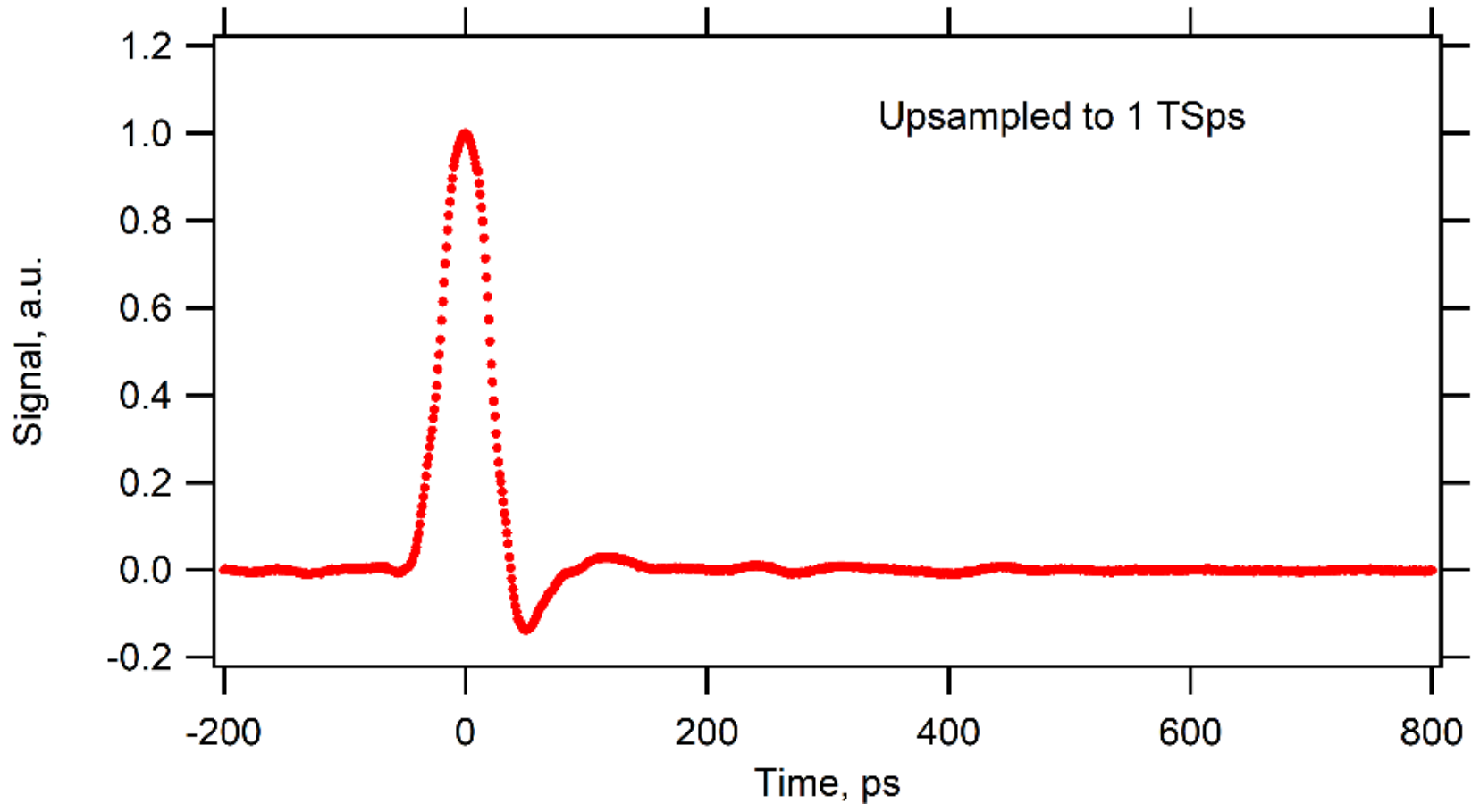

**Figure 22**. Figure showing the upsampled impulse response. Pulses with slightly different timing are interleaved to form the upsampled impulse response.

Figure 23 shows the measured group delay for different lengths of SMF-28 the pulse was sent through. 2 m of fiber was about the shortest length we could reliably measure, which corresponds to about 40,000 $fs^2$ of dispersion at 1550 nm. The total bandwidth of the pulse was about 2 THz. The error of the slope was consistently about 150 fs/THz.

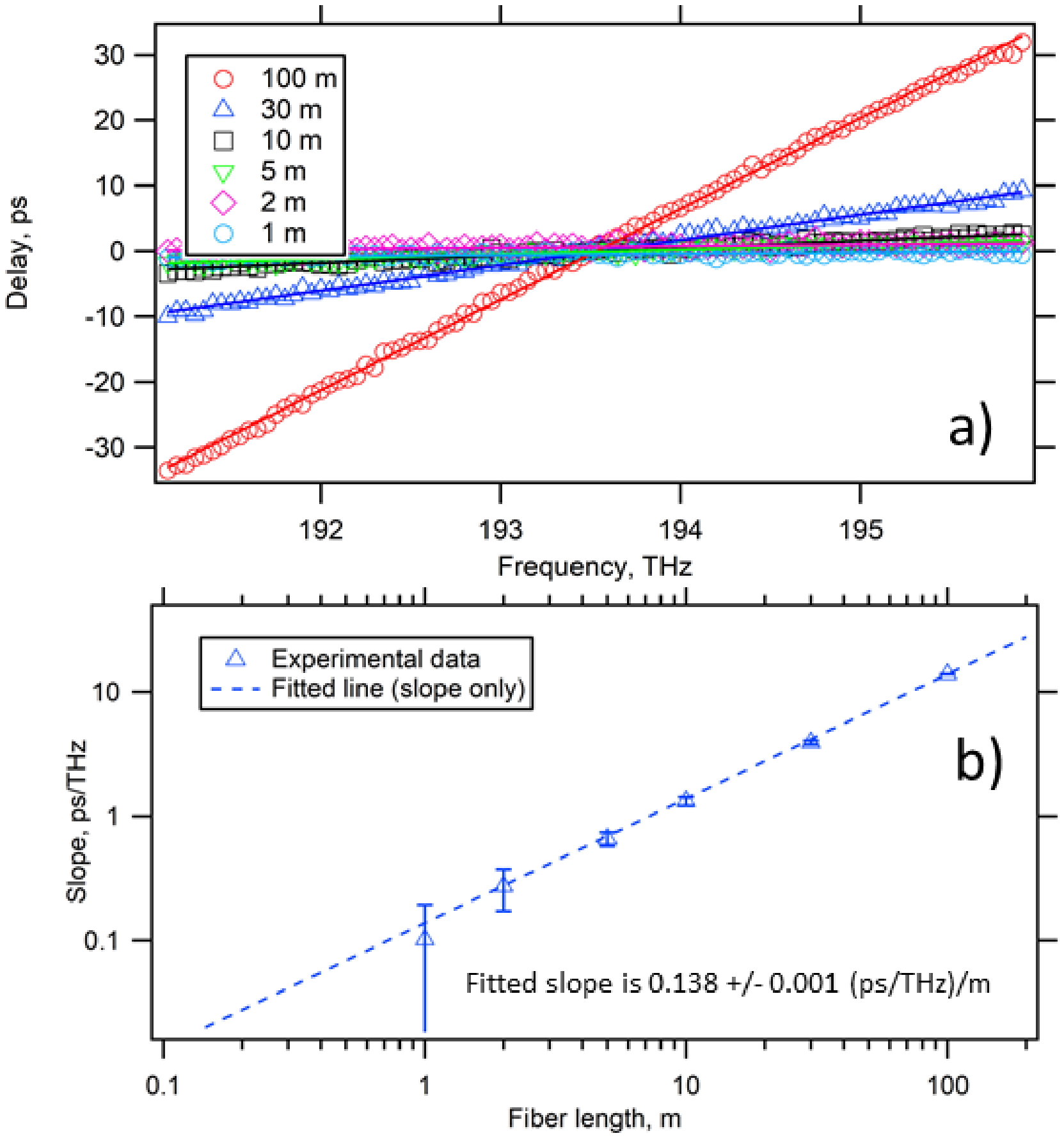

**Figure 23.** (a) Representative single shot measurements of the delay as a function of optical frequency performed using different lengths of SMF-28 fiber. Points are color coded according to the fiber length, as indicated in the legend. Lines are linear fits of the experimental data. (b) Slopes of the data sets shown in (a) plotted as a function of the fiber length.

Figure 24 shows the group delay measurement obtained when quadratic phase is applied to the pulse using a pulse shaper. The measured group delay compares very well with the expected values. Figure 24(c) shows the plot of the slope of the group delay vs. the dispersion coefficient. From this plot, the minimum amount of dispersion that can be measured is about 44,000 $fs^2$, which compares well with the fiber data.

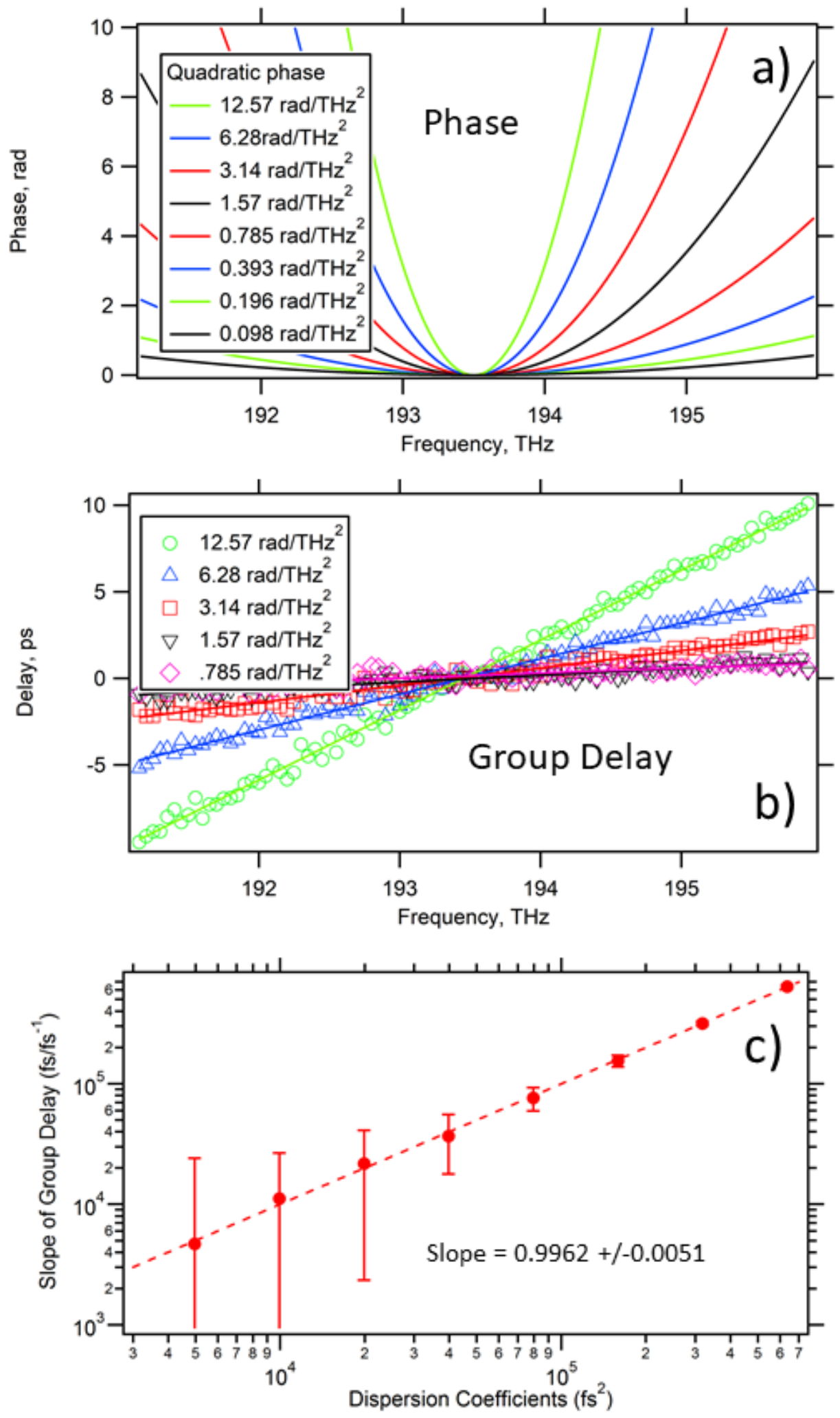

**Figure 24.** Measured group delay for quadratic phase added by the pulse shaper. (a) shows the plots of the applied phase. (b) shows the measured group delay. (c) compares the slope of the group delay to the dispersion coefficients. The slope is close to 1.

Figure 25 shows group delay measurements obtained when the pulse shaper is used to apply a sinusoidal phase to the pulse: $\phi = \phi_0 \sin(2\pi f-\omega_0)$, where $\phi_0$ is the phase amplitude, f is the frequency component of the pulse and $\omega_0$ is an arbitrary constant. Part a) shows eight levels of phase modulation applied: 0.08, 0.16, 0.32, 0.64, 1.28, 2.56, 5.12 and 10.24 radians. Since group delay is the derivative of the phase with respect to angular frequency, the expected group delay can be expressed as $\tau = \tau_0 \cos(2\pi f-\omega_0)$. We note that, with this particular phase waveform, when $\phi$, f and $\tau$ are expressed in radians, THz and ps, respectively, there is a one-to-one correspondence between the numerical values of the applied phase and the resulting group delay, e.g. a spectral phase shift of 1 radian corresponds to a 1 ps group delay. Therefore, group delay measurement precision provides an immediate estimate of the phase measurement precision. Figure 25b) shows the best sinusoidal fits of the group delay for different phase waveforms applies. Figure 25 c) shows the fitted amplitudes for both a three-parameter fit, with $\tau_0$, f and $\omega_0$ fitted (blue points), and single-parameter fit, with $\tau_0$ fitted and f and $\omega_0$ fixed (green points). The fitted amplitudes match the expected values quite well. Using these data, we estimate a phase measurement accuracy of about 0.2 radians.

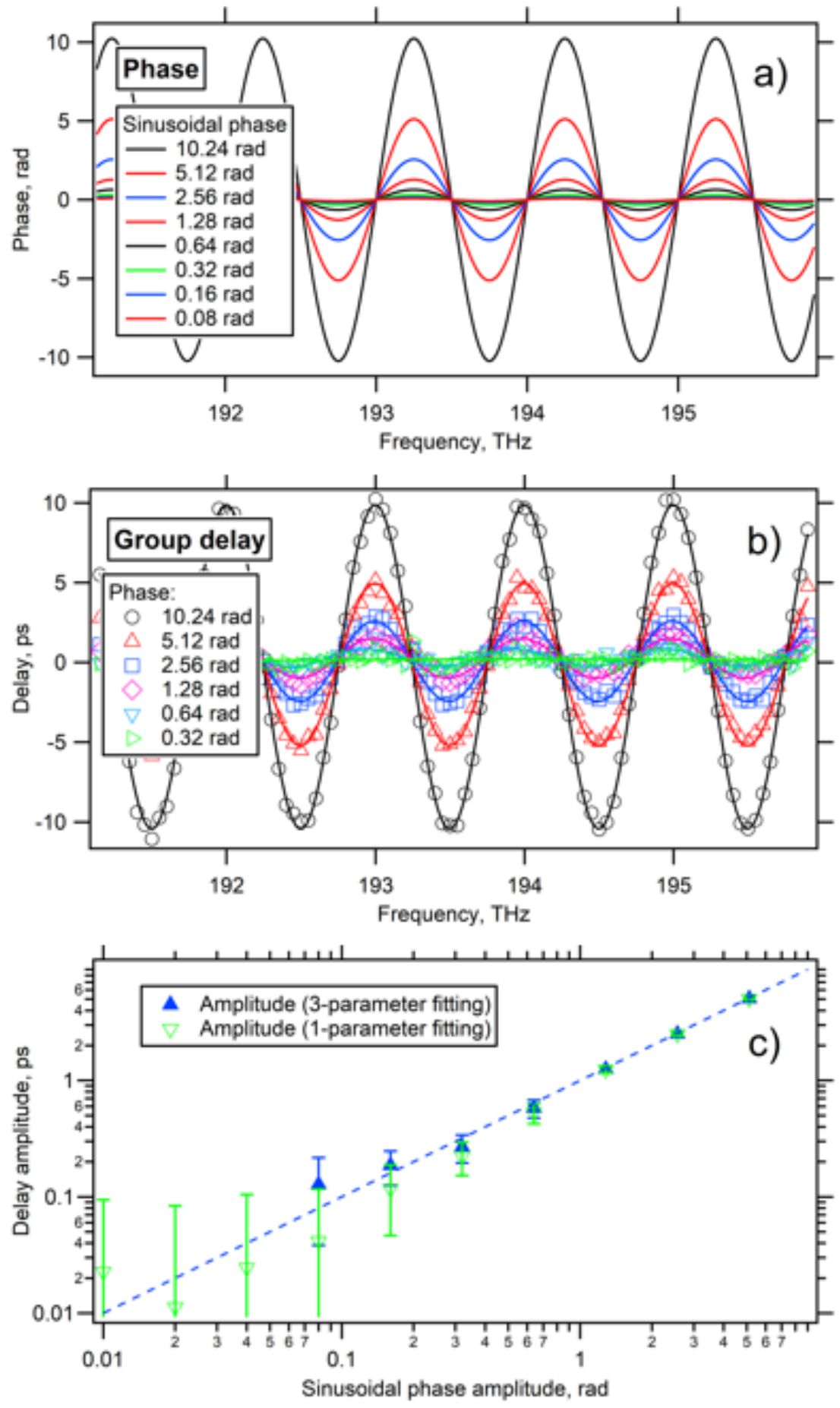

**Figure 25.** (a) Sinusoidal phase waveforms of different amplitudes generated using the Finisar pulse shaper. The phase amplitudes are shown in the legend. (b) Points: representative single shot measurements of the group delay as a function of the optical frequency for the sinusoidal phase waveforms. Lines: 3-parameter fits. (c) Log-log plot of the fitted group delay amplitude vs. phase amplitude. The dashed line has a slope of 1. Blue triangles show the amplitude if all three parameters are fitted while the green points represent the amplitude if only the amplitude is fitted.

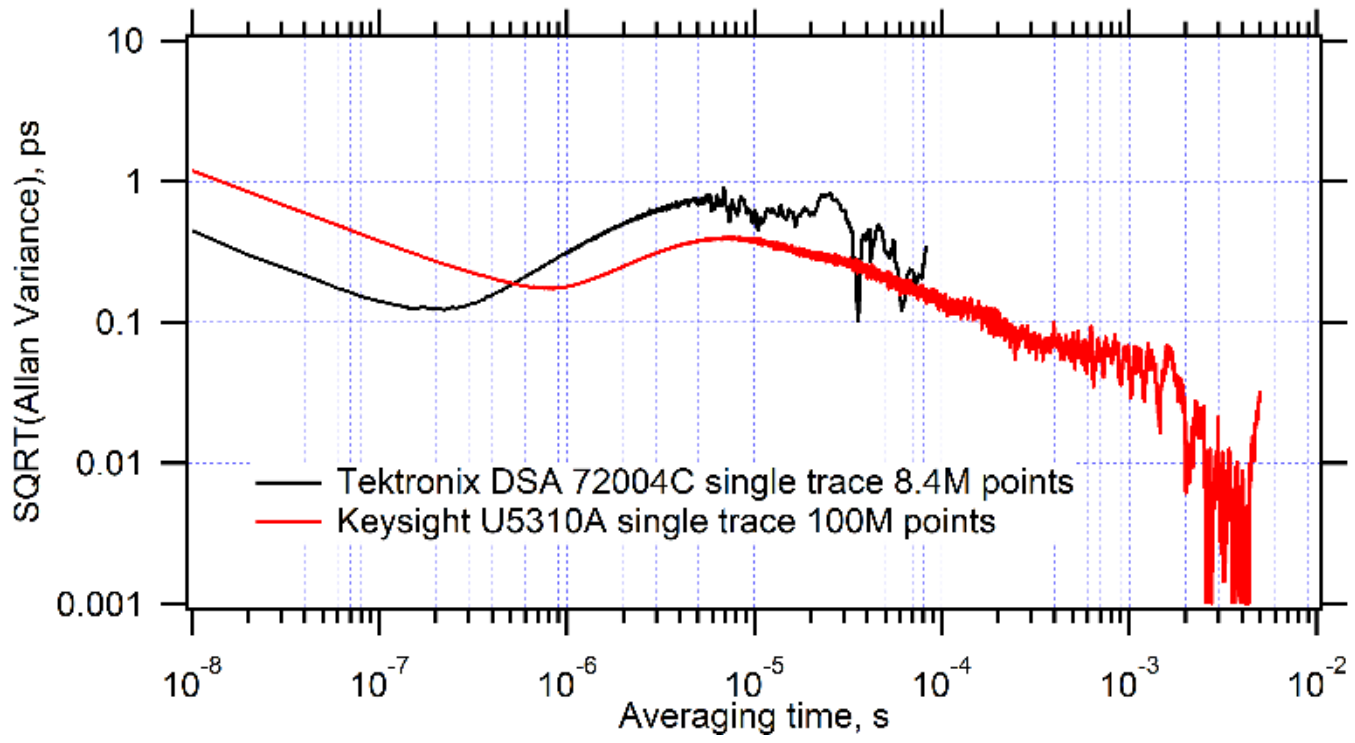

**Figure 26**. Allan variance of the jitter. A single shot has an aperture jitter of about 1 ps, but averaging rapidly improves the jitter performance demonstrating that it should be possible to measure optical pulses a few 10's of fs in duration using high-speed electronics.

Figure 26 shows the Allan variance of the jitter. The raw data sets for Allan variance analysis are single oscilloscope waveforms containing ~$10^4$–$10^7$ laser pulses arriving at a 100 MHz repetition rate. Each pulse in the waveform is correlated with the impulse response. As averaging time is increased, the jitter, and hence, the timing accuracy improves indicating that the precision of the measurement is determined by the A/D aperture jitter (sampling uncertainty) rather than the bandwidth. Indeed, the black line on the Allan variance plot is for an oscilloscope with a bandwidth of 20 GHz while the red line is for a data acquisition card with a bandwidth of only about 2.5 GHz. This is not that unexpected because the data acquisition card uses a newer technology, and it has a 10-bit digitizing depth. Aperture jitter limits both the bandwidth and the bit-depth [108].

In conclusion, we demonstrate a hybrid optical/electronic system for high-speed pulse measurement. Precision of the measured time delays and data is much better than would be expected from the electronics bandwidth. Averaging can improve the group delay measurement precision because the electronic jitter is reduced, allowing pulses on the order of 10's of femtoseconds to be measured.

## XIV. Uniqueness in ptychographic pulse measurements

When we discuss uniqueness, we are referring to a given data set returning a single solution. This is strictly rarely true [109]. There are almost always trivial ambiguities such as phase offsets and time or frequency translations that in the strict sense are ambiguities, but do not change the utility of the result in a significant way. Indeed, spectrogram retrieval cannot, without a reference, extract the absolute phase, or the absolute time arrival or the absolute frequency of an optical waveform unless the experiment is specifically designed to measure these quantities [110]–[112]. Normally, only terms that exist in the second and higher derivatives can be determined. Complete discussions of both trivial and nontrivial ambiguities can be found elsewhere [15], [93], [103], [104].

### *A. 1-D uniqueness*

It is well known that 1-D phase retrieval using only the autocorrelation and the spectrum will not yield unique results [57]. A recent series of studies have been conducted on 1-D phase retrieval to determine how much information is required for essentially unique retrievals [57], [58], [113]. Constraints in both domains are required, and if the intensity is known in both domains, the phase retrieval is almost always unique up to a constant phase factor.

### *B. Spectrogram retrieval*

Uniqueness of the spectrogram phase retrieval has been discussed for some time. In the original work [19], it was argued that converting a 1-D phase retrieval problem into a 2-D phase retrieval algorithm makes the problem unique because 2-D phase retrieval problems are almost always unique. Later work has shown this to be true for at least second harmonic generation (SHG)

FROG (aside from SHG FROG's well-known direction of time ambiguity). Indeed, in this work, it was determined that only about 3 cross-correlation spectra are required to uniquely determined the pulse [103].

Work on blind-FROG has shown that there are significant ambiguities both trivial and nontrivial [42], [93]. Generally, the trivial ambiguities include opposite linear shifts in frequencies between the pulses, exchanging the pulse and gate fields when $I_{FROG}(\omega,\tau) = I_{FROG}(\omega,-\tau)$, or changing the pulse pair to the complex pulse pair when $I_{FROG}(\omega,\tau) = I_{FROG}(\omega,-\tau)$. The work of Seifert et al. [93] also showed that a necessary, but not sufficient condition for blind FROG uniqueness is that the spectral intensities of the pulse and gate are distinguishable. More recent work on the uniqueness of blind spectrogram retrievals utilized a different, more general proof, demonstrated that the inversion is unique if the spectra of both pulses are known [104]. However, this proof did not address the special cases described in [93]. Likewise, if both temporal magnitudes are known, then a blind sonogram retrieval is unique. Indeed, because a spectrogram can be written as a sonogram, we can deduce that if either the temporal profiles are known or the spectra are known, then a blind spectrogram or sonogram retrieval are unique. (Except for special cases [93].)

## XV. OTHER ALGORITHMS/TECHNIQUES

### *A. Very Advanced Method for Phase and Intensity Retrieval of E-fields (VAMPIRE)*

VAMPIRE is a self-referencing technique for measuring amplitude and phase of ultrashort laser pulses [114]. Two different arbitrary pulses are cross-correlated in an SFG X-FROG configuration. One pulse is unchanged while the other is sent through an unbalanced Michelson interferometer where the pulse replicas are time delayed relative to each other and one is sent through a dispersive material. The resulting spectrogram from the SFG X-FROG configuration is retrieved by decomposing the phase retrieval into a series of 1-D phase retrievals using the spectra as an added constraint to remove ambiguities present in SHG FROG and blind-FROG.

### *B. Direct reconstruction of ultrashort pulse from FROG spectrograms*

Hidalgo-Rojas et al. [115] have recently published a method to directly reconstruct the pulses constructing a spectrogram. The method solves the differential equation:

$$\left.\frac{\partial I_\Psi(t)}{\partial \Psi}\right|_{\Psi=0} = \frac{\partial}{\partial t}\left[I(t)\frac{\partial \varphi_E(t)}{\partial t}\right]$$

where I(t) is the spectrogram field, $\varphi$ is the time domain phase, and $\Psi$ is a group delay where the spectrogram's spectral phase is offset by $\phi(\omega) = \frac{1}{2}\Psi\omega^2$. This differential equation comes from differentiating the time marginal of the Wigner distribution [22], [23], [72], [115]. The solution is obtained experimentally by measuring two spectrograms except that for one spectrogram, one of the pulses has a phase of $\frac{1}{2}\Psi\omega^2$ applied to it. The difference between the spectrograms gives:

$$\left.\frac{\partial I_\Psi(\Omega,\tau)}{\partial \Psi}\right|_{\Psi=0} \approx [I_\Psi(\Omega,\tau) - I(\Omega,\tau)]/\Psi.$$

Integration obtains $\varphi(\Omega,\tau)$, solving the phase of the spectrogram, which determines the fields.

### *C. Differential Evolution*

Differential evolution can be utilized for pulse characterization methods where it is difficult or even impossible to use generalized projections or steepest decent algorithms [116]–[119]. The advantage of differential evolution is that it can be easily adapted to completely different characterization methods. Only the forward equation is required to compute the respective spectrogram from the pulse amplitude and phase. In most work, differential evolution will always converge globally for data with reasonable signal-to-noise ratio and complexity. The primary disadvantage is that it is slow, taking 1-2 minutes to converge.

Although first published for FROG retrievals, differential evolution can be applied to other pulse measurement techniques such as d-scan [120].

### D. *Common Phase Retrieval Algorithm (COPRA)*

A recent work provides an elegant framework for all pulse measurement techniques utilizing a parameterized nonlinear process coupled with an iterative phase retrieval algorithm to determine the pulse complex field [121]. This class of techniques includes FROG, interferometric-FROG (iFROG), d-scan, ptychographic retrieval, and MIIPS. Within the framework, each technique can be reduced to a variation in measurement scan vector and choice of nonlinear interaction; these are implemented as variations in algorithm operators. The authors put forth an algorithm named the Common Phase Retrieval Algorithm (COPRA) for solving the general problem of the class in an efficient manner. COPRA is designed to yield the Maximum Likelihood Estimate (MLE) under conditions of white Gaussian measurement noise, such that a least-squares optimization can be used to yield the MLE. To achieve this, the algorithm employs well-known projection methods with a gradient-descent in a local 'pre-processing' step, then switches to a purely gradient-descent method for the main processing step. In addition to its generality, a key feature of the COPRA algorithm is its accuracy in the presence of noise. It is also efficient, typically achieving convergence in 10's of iterations.

### E. *Multimode Phase Retrieval for D-Scan*

D-scan is a single beam pulse measurement method that maps changes in dispersion to changes in the second harmonic spectrum. However, d-scan is not a true ptychographic method. While it does use the product of the field and a second function that is translated (dispersion) it is not the intensity of the product field that is propagated to another plane; it is the second harmonic that is propagated. Because the second harmonic does not have a clean inverse, the construction of inversion algorithms for d-scan has been challenging. Recent work has adapted the ptychographic iterative engine to invert d-scan traces by utilizing a form of Newton's method as a solution to the square root problem. This new ptychographic inversion algorithm retrieves both single-mode and multi-mode d-scan traces [122].

## XVI. Conclusions

Ptychography is a rich and powerful technique for ultrafast laser pulse measurement. There is a wide variety of experimental strategies that can be used, and a variety of algorithms available for the analysis of the experimental data. The development of the operator formalism of the PCGPA provides added versatility including the ability to easily add constraints containing *a priori* information about the pulse to improve convergence and reduce ambiguities. These constraints can be added without significant changes to the algorithm, allowing complete software control of the inversion process. External constraints can also be combined with mathematical form constraints to add external constraints to FROG retrievals.

The PIE and ePIE [45]–[49] provide ways to extract measurements from reduced data sets providing convenient ways to measure highly chirped pulses that are difficult to measure using standard FROG approaches. Recent uniqueness work has shown that at least some FROG measurements are unique [104].

We have also demonstrated that modern electronics are now at a level that can be successfully interfaced with TROG to extend both the temporal window and the electronics bandwidth, which effectively increases the bandwidth of electronics for measuring optical signals by 40 times. Indeed, sub-picosecond optical pulses can now be measured using only optical filters and electronics. This method also increases the speed of the measurement. Single-shot measurements can be obtained in a single oscilloscope sweep making measurements rates as fast as 1 MHz. Interestingly, the limits of the electronic detection, at least for group delay measurement, seems to lie more in the jitter of the digitization step rather than the bandwidth.

Work on new algorithms continues, and new approaches are constantly solving past problems and adapting ptychographic measurement techniques to new measurement needs. New algorithms and approaches continue to blur the lines of measurement methodologies, simultaneously shrinking and expanding the idea of ptychography.

This material is based upon work supported by the U.S. Department of Energy, Office of Science, Office of Fusion Energy Sciences, under Award Number DE-SC0013866.

The authors would like to thank Keith Wernsing for his helpful suggestions.

The following authors have affiliations with organizations with direct or indirect financial interest in the subject matter discussed in the manuscript:

Daniel J. Kane Affiliation: Mesa Photonics, LLC owner and employee

Andrei B. Vakhtin Affiliation: Mesa Photonics, LLC employee

# XVII. References

[1] G. Taft *et al.*, "Ultrashort optical waveform measurements using frequency-resolved optical gating," *Opt. Lett.*, vol. 20, no. 7, p. 743, Apr. 1995, doi: 10.1364/OL.20.000743.

[2] I. P. Christov, M. M. Murnane, H. C. Kapteyn, and V. D. Stoev, "Sub-10-fs operation of Kerr-lens mode-locked lasers," *Opt. Lett.*, vol. 21, no. 18, p. 1493, Sep. 1996, doi: 10.1364/OL.21.001493.

[3] J. D. Harvey, P. F. Curley, C. Spielmann, J. M. Dudley, and F. Krausz, "Coherent effects in a self-mode-locked Ti:sapphire laser," *Opt. Lett.*, vol. 19, no. 13, p. 972, Jul. 1994, doi: 10.1364/OL.19.000972.

[4] B. Kohler, V. V. Yakovlev, K. R. Wilson, J. Squier, K. W. DeLong, and R. Trebino, "Phase and intensity characterization of femtosecond pulses from a chirped-pulse amplifier by frequency-resolved optical gating," *Opt. Lett.*, vol. 20, no. 5, pp. 483–485, Mar. 1995, doi: 10.1364/OL.20.000483.

[5] C. Dorrer *et al.*, "Characterization of chirped-pulse amplification systems with spectral phase interferometry for direct electric-field reconstruction," *Appl. Phys. B*, vol. 70, no. 1, pp. S77–S84, Jun. 2000, doi: 10.1007/s003400000334.

[6] C. Dorrer *et al.*, "Single-shot real-time characterization of chirped-pulse amplification systems by spectral phase interferometry for direct electric-field reconstruction," *Opt. Lett.*, vol. 24, no. 22, pp. 1644–1646, Nov. 1999, doi: 10.1364/OL.24.001644.

[7] G. Stibenz, N. Zhavoronkov, and G. Steinmeyer, "Self-compression of millijoule pulses to 7.8 fs duration in a white-light filament," *Opt. Lett.*, vol. 31, no. 2, pp. 274–276, Jan. 2006, doi: 10.1364/OL.31.000274.

[8] A. Baltuška, M. S. Pshenichnikov, and D. A. Wiersma, "Amplitude and phase characterization of 4.5-fs pulses by frequency-resolved optical gating," *Opt. Lett.*, vol. 23, no. 18, pp. 1474–1476, Sep. 1998, doi: 10.1364/OL.23.001474.

[9] V. Wong and I. A. Walmsley, "Linear filter analysis of methods for ultrashort-pulse-shape measurements," *JOSA B*, vol. 12, no. 8, pp. 1491–1499, Aug. 1995, doi: 10.1364/JOSAB.12.001491.

[10] I. A. Walmsley and V. Wong, "Characterization of the electric field of ultrashort optical pulses," *JOSA B*, vol. 13, no. 11, pp. 2453–2463, Nov. 1996, doi: 10.1364/JOSAB.13.002453.

[11] J.-C. M. Diels, J. J. Fontaine, I. C. McMichael, and F. Simoni, "Control and measurement of ultrashort pulse shapes (in amplitude and phase) with femtosecond accuracy," *Appl. Opt.*, vol. 24, no. 9, pp. 1270–1282, May 1985, doi: 10.1364/AO.24.001270.

[12] E. B. Treacy, "Measurement and Interpretation of Dynamic Spectrograms of Picosecond Light Pulses," *J. Appl. Phys.*, vol. 42, no. 10, pp. 3848–3858, Sep. 1971, doi: 10.1063/1.1659696.

[13] J. L. A. Chilla and O. E. Martinez, "Direct determination of the amplitude and the phase of femtosecond light pulses," *Opt. Lett.*, vol. 16, no. 1, pp. 39–41, Jan. 1991, doi: 10.1364/OL.16.000039.

[14] D. J. Kane and R. Trebino, "Characterization of arbitrary femtosecond pulses using frequency-resolved optical gating," *IEEE J. Quantum Electron.*, vol. 29, no. 2, pp. 571–579, Feb. 1993, doi: 10.1109/3.199311.

[15] R. Trebino *et al.*, "Measuring ultrashort laser pulses in the time-frequency domain using frequency-resolved optical gating," *Rev. Sci. Instrum.*, vol. 68, no. 9, pp. 3277–3295, Sep. 1997, doi: 10.1063/1.1148286.

[16] I. A. Walmsley and C. Dorrer, "Characterization of ultrashort electromagnetic pulses," *Adv. Opt. Photonics*, vol. 1, no. 2, pp. 308–437, Apr. 2009, doi: 10.1364/AOP.1.000308.

[17] D. J. Kane, A. J. Taylor, R. Trebino, and K. W. DeLong, "Single-shot measurement of the intensity and phase of a femtosecond UV laser pulse with frequency-resolved optical gating," *Opt. Lett.*, vol. 19, no. 14, pp. 1061–1063, Jul. 1994, doi: 10.1364/OL.19.001061.

[18] D. J. Kane and R. Trebino, “Single-shot measurement of the intensity and phase of an arbitrary ultrashort pulse by using frequency-resolved optical gating,” *Opt. Lett.*, vol. 18, no. 10, pp. 823–825, May 1993, doi: 10.1364/OL.18.000823.
[19] R. Trebino and D. J. Kane, “Using phase retrieval to measure the intensity and phase of ultrashort pulses: frequency-resolved optical gating,” *JOSA A*, vol. 10, no. 5, pp. 1101–1111, May 1993, doi: 10.1364/JOSAA.10.001101.
[20] C. Iaconis and I. A. Walmsley, “Spectral phase interferometry for direct electric-field reconstruction of ultrashort optical pulses,” *Opt. Lett.*, vol. 23, no. 10, pp. 792–794, May 1998, doi: 10.1364/OL.23.000792.
[21] J.-K. Rhee, T. S. Sosnowski, A.-C. Tien, and T. B. Norris, “Real-time dispersion analyzer of femtosecond laser pulses with use of a spectrally and temporally resolved upconversion technique,” *JOSA B*, vol. 13, no. 8, pp. 1780–1785, Aug. 1996, doi: 10.1364/JOSAB.13.001780.
[22] M. Beck, M. G. Raymer, I. A. Walmsley, and V. Wong, “Chronocyclic tomography for measuring the amplitude and phase structure of optical pulses,” *Opt. Lett.*, vol. 18, no. 23, pp. 2041–2043, Dec. 1993, doi: 10.1364/OL.18.002041.
[23] C. Dorrer and I. Kang, “Complete temporal characterization of short optical pulses by simplified chronocyclic tomography,” *Opt. Lett.*, vol. 28, no. 16, pp. 1481–1483, Aug. 2003, doi: 10.1364/OL.28.001481.
[24] C. Bennett and B. Kolner, “Principles of parametric temporal imaging. I. System configurations,” *IEEE J. Quantum Electron.*, 2000, doi: 10.1109/3.831018.
[25] C. Bennett and B. Kolner, “Principles of parametric temporal imaging. II. System performance,” *IEEE J. Quantum Electron.*, 2000, doi: 10.1109/3.845718.
[26] M. T. Kauffman, W. C. Banyai, A. A. Godil, and D. M. Bloom, “Time-to-frequency converter for measuring picosecond optical pulses,” *Appl. Phys. Lett.*, vol. 64, no. 3, pp. 270–272, Jan. 1994, doi: 10.1063/1.111177.
[27] L. Kh. Mouradian, F. Louradour, V. Messager, A. Barthelemy, and C. Froehly, “Spectro-temporal imaging of femtosecond events,” *IEEE J. Quantum Electron.*, vol. 36, no. 7, pp. 795–801, Jul. 2000, doi: 10.1109/3.848351.
[28] J. Azaña, N. K. Berger, B. Levit, and B. Fischer, “Spectral Fraunhofer regime: time-to-frequency conversion by the action of a single time lens on an optical pulse,” *Appl. Opt.*, vol. 43, no. 2, pp. 483–490, Jan. 2004, doi: 10.1364/AO.43.000483.
[29] A. A. Godil, B. A. Auld, and D. M. Bloom, “Time-lens producing 1.9 ps optical pulses,” *Appl. Phys. Lett.*, vol. 62, no. 10, pp. 1047–1049, Mar. 1993, doi: 10.1063/1.108790.
[30] J. E. Rothenberg and D. Grischkowsky, “Measurement of optical phase with subpicosecond resolution by time-domain interferometry,” *Opt. Lett.*, vol. 12, no. 2, pp. 99–101, Feb. 1987, doi: 10.1364/OL.12.000099.
[31] K. C. Chu *et al.*, “Direct measurement of the spectral phase of femtosecond pulses,” *Opt. Lett.*, vol. 20, no. 8, pp. 904–906, Apr. 1995, doi: 10.1364/OL.20.000904.
[32] K. C. Chu, J. P. Heritage, R. S. Grant, and W. E. White, “Temporal interferometric measurement of femtosecond spectral phase,” *Opt. Lett.*, vol. 21, no. 22, pp. 1842–1844, Nov. 1996, doi: 10.1364/OL.21.001842.
[33] S. Prein, S. Diddams, and J.-C. Diels, “Complete characterization of femtosecond pulses with all-electronic detection,” in *Conference on Lasers and Electro-Optics (1996), paper CTuX3*, Jun. 1996, p. CTuX3. Accessed: Jun. 23, 2021. [Online]. Available: https://www.osapublishing.org/abstract.cfm?uri=CLEO-1996-CTuX3
[34] R. G. M. P. Koumans and A. Yariv, “Time-resolved optical gating based on dispersive propagation: a new method to characterize optical pulses,” *IEEE J. Quantum Electron.*, vol. 36, no. 2, pp. 137–144, Feb. 2000, doi: 10.1109/3.823457.
[35] R. G. M. P. Koumans and A. Yariv, “Pulse characterization at 1.5 μm using time-resolved optical gating based on dispersive propagation,” *IEEE Photonics Technol. Lett.*, vol. 12, no. 6, pp. 666–668, Jun. 2000, doi: 10.1109/68.849078.
[36] V. V. Lozovoy, I. Pastirk, and M. Dantus, “Multiphoton intrapulse interference. IV. Ultrashort laser pulse spectral phase characterization and compensation,” *Opt. Lett.*, vol. 29, no. 7, pp. 775–777, Apr. 2004, doi: 10.1364/OL.29.000775.
[37] B. Xu, J. M. Gunn, J. M. D. Cruz, V. V. Lozovoy, and M. Dantus, “Quantitative investigation of the multiphoton intrapulse interference phase scan method for simultaneous phase measurement and compensation of femtosecond laser pulses,” *JOSA B*, vol. 23, no. 4, pp. 750–759, Apr. 2006, doi: 10.1364/JOSAB.23.000750.

[38] Y. Coello *et al.*, “Interference without an interferometer: a different approach to measuring, compressing, and shaping ultrashort laser pulses,” *JOSA B*, vol. 25, no. 6, pp. A140–A150, Jun. 2008, doi: 10.1364/JOSAB.25.00A140.

[39] V. V. Lozovoy, B. Xu, Y. Coello, and M. Dantus, “Direct measurement of spectral phase for ultrashort laser pulses,” *Opt. Express*, vol. 16, no. 2, pp. 592–597, Jan. 2008, doi: 10.1364/OE.16.000592.

[40] A. M. Allende Motz, J. A. Squier, C. G. Durfee, and D. E. Adams, “Spectral phase and amplitude retrieval and compensation technique for measurement of pulses,” *Opt. Lett.*, vol. 44, no. 8, p. 2085, Apr. 2019, doi: 10.1364/OL.44.002085.

[41] M. Miranda, T. Fordell, C. Arnold, A. L’Huillier, and H. Crespo, “Simultaneous compression and characterization of ultrashort laser pulses using chirped mirrors and glass wedges,” *Opt. Express*, vol. 20, no. 1, p. 688, Jan. 2012, doi: 10.1364/OE.20.000688.

[42] D. J. Kane, G. Rodriguez, A. J. Taylor, and T. S. Clement, “Simultaneous measurement of two ultrashort laser pulses from a single spectrogram in a single shot,” *J. Opt. Soc. Am. B*, vol. 14, no. 4, p. 935, Apr. 1997, doi: 10.1364/JOSAB.14.000935.

[43] D. J. Kane, “Real-time measurement of ultrashort laser pulses using principal component generalized projections,” *IEEE J. Sel. Top. Quantum Electron.*, vol. 4, no. 2, pp. 278–284, Apr. 1998, doi: 10.1109/2944.686733.

[44] D. J. Kane, “Recent progress toward real-time measurement of ultrashort laser pulses,” *IEEE J. Quantum Electron.*, vol. 35, no. 4, pp. 421–431, Apr. 1999, doi: 10.1109/3.753647.

[45] D. Spangenberg, E. Rohwer, M. H. Brügmann, and T. Feurer, “Ptychographic ultrafast pulse reconstruction,” *Opt. Lett.*, vol. 40, no. 6, p. 1002, Mar. 2015, doi: 10.1364/OL.40.001002.

[46] D.-M. Spangenberg, M. Brügmann, E. Rohwer, and T. Feurer, “All-optical implementation of a time-domain ptychographic pulse reconstruction setup,” *Appl. Opt.*, vol. 55, no. 19, pp. 5008–5013, Jul. 2016, doi: 10.1364/AO.55.005008.

[47] P. Sidorenko, O. Lahav, Z. Avnat, and O. Cohen, “Ptychographic reconstruction algorithm for frequency-resolved optical gating: super-resolution and supreme robustness,” *Optica*, vol. 3, no. 12, pp. 1320–1330, Dec. 2016, doi: 10.1364/OPTICA.3.001320.

[48] P. Sidorenko, O. Lahav, Z. Avnat, and O. Cohen, “Ptychographic reconstruction algorithm for frequency resolved optical gating: super-resolution and extreme robustness: erratum,” *Optica*, vol. 4, no. 11, pp. 1388–1389, Nov. 2017, doi: 10.1364/OPTICA.4.001388.

[49] A. M. Maiden and J. M. Rodenburg, “An improved ptychographical phase retrieval algorithm for diffractive imaging,” *Ultramicroscopy*, vol. 109, no. 10, pp. 1256–1262, Sep. 2009, doi: 10.1016/j.ultramic.2009.05.012.

[50] J. R. Fienup, “Re: 1D Ptychography,” May 11, 2016.

[51] M. Guizar-Sicairos and J. R. Fienup, “Phase retrieval with transverse translation diversity: a nonlinear optimization approach,” *Opt. Express*, vol. 16, no. 10, p. 7264, May 2008, doi: 10.1364/OE.16.007264.

[52] R. G. Paxman, T. J. Schulz, and J. R. Fienup, “Joint estimation of object and aberrations by using phase diversity,” *JOSA A*, vol. 9, no. 7, pp. 1072–1085, Jul. 1992, doi: 10.1364/JOSAA.9.001072.

[53] B. H. Dean and C. W. Bowers, “Diversity selection for phase-diverse phase retrieval,” *JOSA A*, vol. 20, no. 8, pp. 1490–1504, Aug. 2003, doi: 10.1364/JOSAA.20.001490.

[54] D. Griffin and J. Lim, “Signal estimation from modified short-time Fourier transform,” *IEEE Trans. Acoust. Speech Signal Process.*, vol. 32, no. 2, pp. 236–243, Apr. 1984, doi: 10.1109/TASSP.1984.1164317.

[55] P. Thibault, M. Dierolf, O. Bunk, A. Menzel, and F. Pfeiffer, “Probe retrieval in ptychographic coherent diffractive imaging,” *Ultramicroscopy*, vol. 109, no. 4, pp. 338–343, Mar. 2009, doi: 10.1016/j.ultramic.2008.12.011.

[56] J. M. Rodenburg *et al.*, “Hard-X-Ray Lensless Imaging of Extended Objects,” *Phys. Rev. Lett.*, vol. 98, no. 3, p. 034801, Jan. 2007, doi: 10.1103/PhysRevLett.98.034801.

[57] R. Beinert and G. Plonka, “Ambiguities in One-Dimensional Discrete Phase Retrieval from Fourier Magnitudes,” *J. Fourier Anal. Appl.*, vol. 21, no. 6, pp. 1169–1198, Dec. 2015, doi: 10.1007/s00041-015-9405-2.

[58] R. Beinert and G. Plonka, “One-Dimensional Discrete-Time Phase Retrieval,” p. 25.

[59] R. Beinert, “Ambiguities in one-dimensional phase retrieval from Fourier magnitudes,” p. 258.

[60] J. R. Fienup, “Reconstruction of a complex-valued object from the modulus of its Fourier transform using a support constraint,” *JOSA A*, vol. 4, no. 1, pp. 118–123, Jan. 1987, doi: 10.1364/JOSAA.4.000118.

[61] J. R. Fienup, “Phase retrieval algorithms: a personal tour [Invited],” *Appl. Opt.*, vol. 52, no. 1, p. 45, Jan. 2013, doi: 10.1364/AO.52.000045.

[62] J. R. Fienup, “Phase retrieval algorithms: a comparison,” *Appl. Opt.*, vol. 21, no. 15, p. 2758, Aug. 1982, doi: 10.1364/AO.21.002758.
[63] M. J. Humphry, B. Kraus, A. C. Hurst, A. M. Maiden, and J. M. Rodenburg, “Ptychographic electron microscopy using high-angle dark-field scattering for sub-nanometre resolution imaging,” *Nat. Commun.*, vol. 3, no. 1, p. 730, Jan. 2012, doi: 10.1038/ncomms1733.
[64] V. Wong and I. A. Walmsley, “Ultrashort-pulse characterization from dynamic spectrograms by iterative phase retrieval,” *J. Opt. Soc. Am. B*, vol. 14, no. 4, p. 944, Apr. 1997, doi: 10.1364/JOSAB.14.000944.
[65] D. T. Reid, B. C. Thomsen, J. M. Dudley, and J. D. Harvey, “Sonogram characterisation of picosecond pulses at 1.5 /spl mu/m using waveguide two photon absorption,” *Electron. Lett.*, vol. 36, no. 13, pp. 1141–1142, Jun. 2000, doi: 10.1049/el:20000820.
[66] I. G. Cormack, W. Sibbett, D. T. Reid, and J. F. Allen, “Measurement of femtosecond optical pulses using time-resolved optical gating,” in *Technical Digest. Summaries of papers presented at the Conference on Lasers and Electro-Optics. Postconference Technical Digest (IEEE Cat. No.01CH37170)*, Baltimore, MD, USA, 2001, p. 201. doi: 10.1109/CLEO.2001.947705.
[67] I. G. Cormack, W. Sibbett, and D. T. Reid, “Practical measurement of femtosecond optical pulses using time-resolved optical gating,” *Opt. Commun.*, vol. 194, no. 4, pp. 415–424, Jul. 2001, doi: 10.1016/S0030-4018(01)01281-0.
[68] R. A. Altes, “Detection, estimation, and classification with spectrograms,” *J. Acoust. Soc. Am.*, vol. 67, no. 4, pp. 1232–1246, Apr. 1980, doi: 10.1121/1.384165.
[69] D. J. Kane, “Principal components generalized projections: a review [Invited],” *J. Opt. Soc. Am. B*, vol. 25, no. 6, p. A120, Jun. 2008, doi: 10.1364/JOSAB.25.00A120.
[70] D. J. Kane, “Method and apparatus for determining wave characteristics using constrained interactions of waves,” US10274378B1, Apr. 30, 2019 Accessed: Jun. 03, 2021. [Online]. Available: https://patents.google.com/patent/US10274378B1/en?oq=10%2c274%2c378
[71] T. Franck, “Colliding Pulse Mode-Locked Lasers and Duobinary Tranmitters for Optical Communication Systems,” Technical University of Denmark, Lyngby, Denmark, 1997.
[72] L. Cohen, *Time-frequency analysis*. Englewood Cliffs, N.J: Prentice Hall PTR, 1995.
[73] K. W. DeLong, R. Trebino, J. Hunter, and W. E. White, “Frequency-resolved optical gating with the use of second-harmonic generation,” *JOSA B*, vol. 11, no. 11, pp. 2206–2215, Nov. 1994, doi: 10.1364/JOSAB.11.002206.
[74] J. Paye, M. Ramaswamy, J. G. Fujimoto, and E. P. Ippen, “Measurement of the amplitude and phase of ultrashort light pulses from spectrally resolved autocorrelation,” *Opt. Lett.*, vol. 18, no. 22, p. 1946, Nov. 1993, doi: 10.1364/OL.18.001946.
[75] D. J. Kane, J. Weston, and K.-C. J. Chu, “Real-time inversion of polarization gate frequency-resolved optical gating spectrograms,” *Appl. Opt.*, vol. 42, no. 6, pp. 1140–1144, Feb. 2003, doi: 10.1364/AO.42.001140.
[76] C. Dorrer and I. Kang, “Real-time implementation of linear spectrograms for the characterization of high bit-rate optical pulse trains,” *IEEE Photonics Technol. Lett.*, vol. 16, no. 3, pp. 858–860, Mar. 2004, doi: 10.1109/LPT.2004.823692.
[77] F. Quéré, Y. Mairesse, and J. Itatani, “Temporal characterization of attosecond XUV fields,” *J. Mod. Opt.*, vol. 52, no. 2–3, pp. 339–360, Jan. 2005, doi: 10.1080/09500340412331307942.
[78] K. W. DeLong and R. Trebino, “Improved ultrashort pulse-retrieval algorithm for frequency-resolved optical gating,” *JOSA A*, vol. 11, no. 9, pp. 2429–2437, Sep. 1994, doi: 10.1364/JOSAA.11.002429.
[79] K. W. DeLong, D. N. Fittinghoff, R. Trebino, B. Kohler, and K. Wilson, “Pulse retrieval in frequency-resolved optical gating based on the method of generalized projections,” *Opt. Lett.*, vol. 19, no. 24, pp. 2152–2154, Dec. 1994, doi: 10.1364/OL.19.002152.
[80] W. H. Press, H. William, S. A. Teukolsky, W. T. Vetterling, A. Saul, and B. P. Flannery, *Numerical recipes 3rd edition: The art of scientific computing*. Cambridge university press, 2007.
[81] K. W. DeLong, C. L. Ladera, R. Trebino, B. Kohler, and K. R. Wilson, “Ultrashort-pulse measurement using noninstantaneous nonlinearities: Raman effects in frequency-resolved optical gating,” *Opt. Lett.*, vol. 20, no. 5, pp. 486–488, Mar. 1995, doi: 10.1364/OL.20.000486.
[82] T. Jones *et al.*, “Measuring an ultrashort, ultraviolet pulse in a slowly responding, absorbing medium,” *Opt. Express*, vol. 29, no. 8, pp. 11394–11405, Apr. 2021, doi: 10.1364/OE.417293.
[83] D. J. Kane, “Method and apparatus for determining wave characteristics using interaction with a known wave,” US9423307B2, Aug. 23, 2016 Accessed: Jun. 01, 2021. [Online]. Available: https://patents.google.com/patent/US9423307B2/en

[84] C. M. Kewish *et al.*, “Reconstruction of an astigmatic hard X-ray beam and alignment of K-B mirrors from ptychographic coherent diffraction data,” *Opt. Express*, vol. 18, no. 22, pp. 23420–23427, Oct. 2010, doi: 10.1364/OE.18.023420.

[85] D. J. Kane, “Method and apparatus for determining wave characteristics from wave phenomena,” US6219142B1, Apr. 17, 2001 Accessed: Jun. 13, 2021. [Online]. Available: https://patents.google.com/patent/US6219142B1/en?q=(method+and+apparatus+for+determining+wave+characteristics+wave+phenomenon;)&inventor=Daniel+J+Kane&oq=(method+and+apparatus+for+determining+wave+characteristics+wave+phenomenon;)+inventor:(Daniel+J+Kane)

[86] D. J. Kane, “New, simplified algorithm for cross-correlation frequency resolved optical gating,” in *Frontiers in Ultrafast Optics: Biomedical, Scientific, and Industrial Applications XIII*, Mar. 2013, vol. 8611, p. 86110Q. doi: 10.1117/12.2005148.

[87] D. T. Reid, P. Loza-Alvarez, C. T. A. Brown, T. Beddard, and W. Sibbett, “Amplitude and phase measurement of mid-infrared femtosecond pulses by using cross-correlation frequency-resolved optical gating,” *Opt. Lett.*, vol. 25, no. 19, pp. 1478–1480, Oct. 2000, doi: 10.1364/OL.25.001478.

[88] J. M. Dudley *et al.*, “Cross-correlation frequency resolved optical gating analysis of broadband continuum generation in photonic crystal fiber: simulations and experiments,” *Opt. Express*, vol. 10, no. 21, pp. 1215–1221, Oct. 2002, doi: 10.1364/OE.10.001215.

[89] S. Linden, H. Giessen, and J. Kuhl, “XFROG — A New Method for Amplitude and Phase Characterization of Weak Ultrashort Pulses,” *Phys. Status Solidi B*, vol. 206, no. 1, pp. 119–124, Mar. 1998, doi: 10.1002/(SICI)1521-3951(199803)206:1<119::AID-PSSB119>3.0.CO;2-X.

[90] A. K. Jain, *Fundamentals of digital image processing*. Englewood Cliffs, NJ: Prentice Hall, 1989.

[91] Howard Anton, *Elementary Linear Algebra*, Second. John Wiley & Sons, 1977.

[92] K. W. DeLong, R. Trebino, and W. E. White, “Simultaneous recovery of two ultrashort laser pulses from a single spectrogram,” *JOSA B*, vol. 12, no. 12, pp. 2463–2466, Dec. 1995, doi: 10.1364/JOSAB.12.002463.

[93] B. Seifert, H. Stolz, and M. Tasche, “Nontrivial ambiguities for blind frequency-resolved optical gating and the problem of uniqueness,” *J. Opt. Soc. Am. B*, vol. 21, no. 5, p. 1089, May 2004, doi: 10.1364/JOSAB.21.001089.

[94] Y. Mairesse and F. Quéré, “Frequency-resolved optical gating for complete reconstruction of attosecond bursts,” *Phys. Rev. A*, vol. 71, no. 1, p. 011401, Jan. 2005, doi: 10.1103/PhysRevA.71.011401.

[95] I. Thomann *et al.*, “Temporal characterization of attosecond wave forms in the sub-optical-cycle regime,” *Phys. Rev. A*, vol. 78, no. 1, p. 011806, Jul. 2008, doi: 10.1103/PhysRevA.78.011806.

[96] R. A. Horn and C. R. Johnson, *Matrix analysis*. Cambridge university press, 2012.

[97] D. J. Kane, “Comparison of the Ptychographic Inversion Engine to Principal Components Generalized Projections,” *IEEE J. Sel. Top. Quantum Electron.*, vol. 25, no. 4, pp. 1–8, Jul. 2019, doi: 10.1109/JSTQE.2019.2904414.

[98] D. J. Kane, “Improved principal components generalized projections algorithm for frequency resolved optical gating,” in *Conference on Lasers and Electro-Optics (2017), paper STu3I.4*, May 2017, p. STu3I.4. doi: 10.1364/CLEO_SI.2017.STu3I.4.

[99] T. S. Clement, D. J. Kane, and A. J. Taylor, “Single-shot measurement of the amplitude and phase of ultrashort laser pulses in the violet,” *Opt. Lett.*, vol. 20, no. 1, p. 70, Jan. 1995, doi: 10.1364/OL.20.000070.

[100] Z. Wang, E. Zeek, R. Trebino, and P. Kvam, “Determining error bars in measurements of ultrashort laser pulses,” *JOSA B*, vol. 20, no. 11, pp. 2400–2405, Nov. 2003, doi: 10.1364/JOSAB.20.002400.

[101] Z. Wang, E. Zeek, R. Trebino, and P. Kvam, “Beyond error bars: Understanding uncertainty in ultrashort-pulse frequency-resolved-optical-gating measurements in the presence of ambiguity,” *Opt. Express*, vol. 11, no. 26, pp. 3518–3527, Dec. 2003, doi: 10.1364/OE.11.003518.

[102] D. J. Kane, F. G. Omenetto, and A. J. Taylor, “Convergence test for inversion of frequency-resolved optical gating spectrograms,” *Opt. Lett.*, vol. 25, no. 16, pp. 1216–1218, Aug. 2000, doi: 10.1364/OL.25.001216.

[103] T. Bendory, P. Sidorenko, and Y. C. Eldar, “On the Uniqueness of FROG Methods,” *IEEE Signal Process. Lett.*, vol. 24, no. 5, pp. 722–726, May 2017, doi: 10.1109/LSP.2017.2690358.

[104] T. Bendory, D. Edidin, and Y. C. Eldar, “On signal reconstruction from FROG measurements,” *Appl. Comput. Harmon. Anal.*, p. S1063520318300915, Oct. 2018, doi: 10.1016/j.acha.2018.10.003.

[105] E. Yudilevich, A. Levi, G. J. Habetler, and H. Stark, “Restoration of signals from their signed Fourier-transform magnitude by the method of generalized projections,” *JOSA A*, vol. 4, no. 1, pp. 236–246, Jan. 1987, doi: 10.1364/JOSAA.4.000236.

[106] Y. Yang, N. P. Galatsanos, and H. Stark, “Projection-based blind deconvolution,” *JOSA A*, vol. 11, no. 9, pp. 2401–2409, Sep. 1994, doi: 10.1364/JOSAA.11.002401.

[107] D. J. Kane, “Corrections to ‘Comparison of the Ptychographic Inversion Engine to Principal Components Generalized Projections,’” *IEEE J. Sel. Top. Quantum Electron.*, vol. 25, no. 4, pp. 1–2, Jul. 2019, doi: 10.1109/JSTQE.2019.2925721.
[108] B. Brannon and A. Barlow, “Aperture Uncertainty and ADC System Performance,” *Analog Devices Appl. Note*, vol. AN-501, p. 4, 2006.
[109] A. H. Barnett, C. L. Epstein, L. F. Greengard, and J. F. Magland, “Geometry of the phase retrieval problem,” *Inverse Probl.*, vol. 36, no. 9, p. 094003, Sep. 2020, doi: 10.1088/1361-6420/aba5ed.
[110] C. W. Siders, A. J. Taylor, and M. C. Downer, “Multipulse interferometric frequency-resolved optical gating: real-time phase-sensitive imaging of ultrafast dynamics,” *Opt. Lett.*, vol. 22, no. 9, p. 3, May 1984.
[111] C. W. Siders, J. L. W. Siders, F. G. Omenetto, and A. J. Taylor, “Multipulse interferometric frequency-resolved optical gating,” *IEEE J. Quantum Electron.*, vol. 35, no. 4, pp. 432–440, Apr. 1999, doi: 10.1109/3.753648.
[112] G. Rodriguez, C. W. Siders, C. Guo, and A. J. Taylor, “Coherent ultrafast MI-FROG spectroscopy of optical field ionization in molecular H/sub 2/, N/sub 2/, and O/sub 2/,” *IEEE J. Sel. Top. Quantum Electron.*, vol. 7, no. 4, pp. 579–591, Jul. 2001, doi: 10.1109/2944.974229.
[113] R. Beinert and G. Plonka, “Enforcing uniqueness in one-dimensional phase retrieval by additional signal information in time domain,” *ArXiv160404493 Cs Math*, Apr. 2016, Accessed: Jan. 12, 2020. [Online]. Available: http://arxiv.org/abs/1604.04493
[114] B. Seifert and H. Stolz, “A method for unique phase retrieval of ultrafast optical fields,” *Meas. Sci. Technol.*, vol. 20, no. 1, p. 015303, Jan. 2009, doi: 10.1088/0957-0233/20/1/015303.
[115] D. Hidalgo-Rojas, R. Rojas-Aedo, R. Alastair Wheatley, L. Gence, and B. Seifert, “Direct reconstruction of two ultrashort pulses based on non-interferometric frequency-resolved optical gating,” *Opt. Express*, vol. 29, no. 4, p. 5166, Feb. 2021, doi: 10.1364/OE.411597.
[116] P. Rocca, G. Oliveri, and A. Massa, “Differential Evolution as Applied to Electromagnetics,” *IEEE Antennas Propag. Mag.*, vol. 53, no. 1, pp. 38–49, Feb. 2011, doi: 10.1109/MAP.2011.5773566.
[117] J. Hyyti, E. Escoto, and G. Steinmeyer, “Pulse retrieval algorithm for interferometric frequency-resolved optical gating based on differential evolution,” *Rev. Sci. Instrum.*, vol. 88, no. 10, p. 103102, Oct. 2017, doi: 10.1063/1.4991852.
[118] J. Hyyti, E. Escoto, and G. Steinmeyer, “Third-harmonic interferometric frequency-resolved optical gating,” *JOSA B*, vol. 34, no. 11, pp. 2367–2375, Nov. 2017, doi: 10.1364/JOSAB.34.002367.
[119] J. Hyyti, E. Escoto, G. Steinmeyer, and T. Witting, “Interferometric time-domain ptychography for ultrafast pulse characterization,” *Opt. Lett.*, vol. 42, no. 11, pp. 2185–2188, Jun. 2017, doi: 10.1364/OL.42.002185.
[120] E. Escoto, A. Tajalli, T. Nagy, and G. Steinmeyer, “Advanced phase retrieval for dispersion scan: a comparative study,” *JOSA B*, vol. 35, no. 1, pp. 8–19, Jan. 2018, doi: 10.1364/JOSAB.35.000008.
[121] N. C. Geib, M. Zilk, T. Pertsch, and F. Eilenberger, “Common pulse retrieval algorithm: a fast and universal method to retrieve ultrashort pulses,” *Optica*, vol. 6, no. 4, pp. 495–505, Apr. 2019, doi: 10.1364/OPTICA.6.000495.
[122] A. M. Wilhelm, D. D. Schmidt, D. E. Adams, and C. G. Durfee, “Advanced Multi-Mode Phase Retrieval For Dispersion Scan,” *ArXiv210400136 Phys.*, Mar. 2021, Accessed: Jun. 26, 2021. [Online]. Available: http://arxiv.org/abs/2104.00136
[123] Bradley Efron, Robert J. Tibshirani, An Introduction to the Bootstrap, 1st ed., Springer Science+Business Media, Berlin/Heidelberg, Germany, 1993.